\documentclass[11pt,twoside]{article}

\usepackage[margin=1in]{geometry}
\usepackage[T1]{fontenc}
\usepackage[utf8]{inputenc}

\usepackage{mathptmx}
\usepackage[scaled=0.9]{helvet}   
\usepackage{microtype}            

\usepackage{booktabs}
\usepackage{array}
\usepackage{multirow}
\usepackage{tabularx}
\usepackage{makecell}
\usepackage[table]{xcolor}
\usepackage{graphicx}   
\usepackage[font=small,labelfont=bf,labelsep=period,justification=justified]{caption}
\usepackage[labelfont=bf]{subcaption} 
\graphicspath{{figs/}}
\usepackage{pdflscape}  
\usepackage{float}
\usepackage{placeins}   
\usepackage{tikz}
\usetikzlibrary{positioning,arrows.meta,fit,calc}
\newsavebox{\wffigbox}


\usepackage{dirtree}

\usepackage{amsmath}
\usepackage{enumitem}
\usepackage{setspace}

\usepackage[hidelinks]{hyperref}
\usepackage{cleveref}

\usepackage{fancyhdr}
\fancypagestyle{plain}{%
  \fancyhf{}%
  \fancyfoot[C]{\small\thepage}%
}

\newcolumntype{R}{>{\raggedleft\arraybackslash}X}
\newcolumntype{L}{>{\raggedright\arraybackslash}X}
\newcolumntype{C}{>{\centering\arraybackslash}X}
\newcolumntype{T}{>{\centering\arraybackslash}X}

\definecolor{grayrow}{gray}{0.9}

\definecolor{wfbluefill}{HTML}{E8F1FB}
\definecolor{wfblue}{HTML}{5E8FC7}
\definecolor{wforangefill}{HTML}{FCE6C9}
\definecolor{wforange}{HTML}{D99A5E}
\definecolor{wfgreenfill}{HTML}{DFF0E2}
\definecolor{wfgreen}{HTML}{6AA875}
\definecolor{wfpurplefill}{HTML}{EFE8F6}
\definecolor{wfpurple}{HTML}{8C79B5}
\definecolor{wfgrayfill}{HTML}{F4F5F5}
\definecolor{wfgray}{HTML}{6F7780}
\definecolor{wfline}{HTML}{4F5B66}

\tikzset{
  wfbox/.style={
    rectangle, rounded corners=2pt, draw, line width=0.5pt,
    inner sep=4pt, align=center, font=\scriptsize, minimum height=8.5mm
  },
  wfblue/.style ={wfbox, draw=wfblue,    fill=wfbluefill,    text=black},
  wforange/.style={wfbox, draw=wforange, fill=wforangefill, text=black},
  wfgreen/.style ={wfbox, draw=wfgreen,  fill=wfgreenfill,  text=black},
  wfpurple/.style={wfbox, draw=wfpurple, fill=wfpurplefill, text=black},
  wfgraybox/.style={wfbox, draw=wfgray,  fill=wfgrayfill,  text=black},
  wfband/.style ={font=\scriptsize\itshape, text=wfgray},
  wfarr/.style  ={-{Latex[length=1.6mm,width=1.4mm]}, draw=wfline, line width=0.5pt},
  wfsoftcontainer/.style={
    draw=wfline, densely dashed, rounded corners=3pt, line width=0.5pt,
    fill=wfgrayfill, fill opacity=0.18, inner sep=4mm
  }
}

\newcommand{\compacttable}{\scriptsize\setlength{\tabcolsep}{3pt}\renewcommand{\arraystretch}{1.0}}
\newcommand{\widetable}{\scriptsize\setlength{\tabcolsep}{0pt}\renewcommand{\arraystretch}{1.0}}
\newcommand{\standardtable}{\scriptsize\setlength{\tabcolsep}{2pt}\renewcommand{\arraystretch}{1.0}}

\title{\vspace{-0.75cm}WiFiSpectralJam: A Large-Scale Open Wi-Fi Spectral Scan Dataset with Controlled RF Jamming}
\author{Dania Herzalla\textsuperscript{1}\textsuperscript{*}, Govind Singh\textsuperscript{1}, Willian T. Lunardi\textsuperscript{1}, Martin Andreoni\textsuperscript{2}\\[0.35em]
\small \textsuperscript{1}Technology Innovation Institute, Abu Dhabi, United Arab Emirates\\
\small \textsuperscript{2}Khalifa University, Abu Dhabi, United Arab Emirates\\
\small \textsuperscript{*}Corresponding author}
\date{}

\begin{document}
\maketitle
\thispagestyle{plain}
\begin{center}
    \vspace{-1.0em}
    {\small\bfseries Data Descriptor}
\end{center}
\vspace{0.5em}

\begin{abstract}
WiFiSpectralJam is a Wi-Fi spectral-scan dataset comprising 14.52\,GB, 96{,}090 CSV files, and 522{,}771{,}130 ordered spectral observations using commodity Wi-Fi sensing hardware. Measurements were acquired with a Raspberry Pi Compute Module~4 equipped with a Qualcomm Atheros QCA9880 802.11ac network interface and the Linux ath10k spectral-scan interface. The dataset spans active and passive scan modalities across the 2.4 and 5\,GHz bands and includes real-world benign background captures, benign RF-chamber floor captures, and controlled RF-jamming captures generated with a HackRF One. Jamming conditions vary by transmit power, target channel, and, in the active subset, waveform type. The release provides the raw spectral-scan records together with a file-level metadata manifest, derived spectral-summary features, validation outputs, and reproducible benchmark protocols. These resources support reuse in RF interference characterisation, jamming detection, spectrum monitoring, distribution-shift evaluation, and machine-learning studies using commodity-NIC spectral measurements. The dataset is publicly available at: \url{https://www.kaggle.com/datasets/daniaherzalla/radio-frequency-jamming/data}.
\end{abstract}

\section{Background \& Summary}

Benchmark datasets have played a central role in advancing data-driven security research. In Wi-Fi jamming detection, however, open datasets containing physical-layer RF measurements remain scarce, making it difficult to reproduce and compare methods that use spectral or signal-strength features. This challenge is compounded by the accessibility of RF jamming: an attacker can disrupt connectivity without network association or credential compromise, and low-cost software-defined radios have made intentional interference increasingly practical~\cite{pelechrinis2011dos,xu2005feasibility,pirayesh2022survey}. Without common datasets collected under documented interference conditions, reported detection results remain difficult to reproduce, compare, or extend across experimental settings.

Public RF dataset releases span diverse wireless domains, hardware platforms, and signal representations; Table~\ref{tab:related_datasets_grouped} surveys selected adversarial-jamming datasets and experimental benchmarks most relevant to the present work. The comparison includes vehicular jamming traces~\cite{punal2012crawdad,punal2014crawdad}, IoT and UAV jamming datasets~\cite{hussain2022edgeaijamming,li2023uavjammingdetection}, and indoor raw-I/Q jamming measurements~\cite{alhazbi2023indoorjammingdataset}, together with Wi-Fi-specific jamming studies and datasets~\cite{punal2014jamming,davaslioglu2019deepwifi,ali2024rfjammingdataset,ali2022rfjammingdataport,panitsas2025jamshield}. Adjacent spectrum-sensing, signal-classification, and incumbent-detection datasets are useful for broader RF-sensing context, but they are not adversarial jamming benchmarks and are therefore not included in the survey table.

Within the Wi-Fi domain specifically, existing jamming datasets and experimental studies occupy distinct measurement regimes. DeepWiFi provides a simulated 802.11ac I/Q benchmark for distinguishing idle, Wi-Fi, and jammer conditions~\cite{davaslioglu2019deepwifi}; Panitsas \textit{et al.}\ release cross-layer PHY/MAC/network KPI measurements collected with a consumer access point and USRP jammer~\cite{panitsas2025jamshield}; and Ali \textit{et al.}\ provide a gated commodity-NIC spectral-scan archive comprising 6{,}750 scans collected with QCA9880-based hardware and controlled RF jamming~\cite{ali2024rfjammingdataset,ali2022rfjammingdataport}. Our dataset, WiFiSpectralJam, is most directly related to the latter sensing regime because it also uses commodity-NIC spectral scanning. The present release is, however, a separate collection campaign rather than a scaled-up repackaging of the earlier Ali \textit{et al.}\ deposit. Beyond its substantially larger scale, our dataset adds dual-band coverage, a passive scanning modality, active-scan waveform variation, RF-chamber benign floor captures, scan-mode-specific real-world backgrounds, and explicit matched-distribution, held-out-condition, chamber-control, and cross-scan-mode benchmark protocols.

WiFiSpectralJam is a large-scale commodity-NIC spectral-scan dataset comprising 14.5~GB across 96{,}090 CSV files and 522{,}771{,}130 spectral-scan observations collected with a Raspberry Pi Compute Module~4 (CM4) and Qualcomm Atheros QCA9880 802.11ac NIC, with a HackRF One used as the jammer. Figure~\ref{fig:workflow_schematic} summarises the collection setup, acquisition environments, processing pipeline, and benchmark protocols. Captures span the 2.4 and 5~GHz bands under active and passive scanning, with real-world benign backgrounds, RF-chamber benign floor captures, and controlled jamming across multiple powers, target channels, and waveform conditions. The release also includes a file-level metadata manifest, precomputed aggregate features, and documented benchmark splits for reproducible reuse.

\newcommand{\gcell}[1]{\cellcolor{gray!10}{#1}}

\section{Methods}

This section provides an overview of the collection hardware, environments, and downstream processing pipeline, as outlined in Figure~\ref{fig:workflow_schematic}.

\sbox{\wffigbox}{%
\begin{tikzpicture}[node distance=4mm and 8mm]

  \node[wforange, text width=35mm] (hackrf)
    {\textbf{HackRF One SDR}\\ jammer, dual-band SMA antenna, 3\,dBi};

  \node[wfblue, text width=35mm, right=36mm of hackrf] (nic)
    {\textbf{QCA9880 NIC}\\ CM4 host, ath10k driver};

  \node[wforange, text width=35mm, below=of hackrf] (jamrf)
    {\textbf{JamRF framework}\\ waveform, power, channel};

  \node[wfblue, text width=35mm, below=of nic] (scan)
    {\textbf{Spectral-scan interface}\\ debugfs FFT export};

  \node[
    wfsoftcontainer,
    fit=(hackrf)(jamrf)(nic)(scan),
    label={[font=\tiny,text=wfgray]above:Candela CT840A RF-shielded chamber}
  ] (chamber) {};

  \draw[wfarr] (hackrf) -- (jamrf);
  \draw[wfarr] (nic) -- (scan);

  \draw[densely dashed, draw=wfline, line width=0.5pt]
    (hackrf.east) --
    node[above=0.3mm, font=\tiny, text=wfgray] {20\,cm LOS}
    (nic.west);

  \node[wfgreen, text width=33mm, below=13mm of chamber.south west, anchor=north west] (floor)
    {\textbf{Chamber floor}\\ controlled baseline};

  \node[wfgreen, text width=33mm, right=7mm of floor] (rw)
    {\textbf{Real-world indoor}\\ 2--3 locations};

  \node[wforange, text width=33mm, right=7mm of rw] (jam)
    {\textbf{Chamber jamming}\\ Gaussian / single tone};

  \node[wfband, below=0.35mm of floor.south] {Benign};
  \node[wfband, below=0.35mm of rw.south] {Benign};
  \node[wfband, below=0.35mm of jam.south] {Malicious};

  \draw[wfarr] (chamber.south -| floor) -- (floor.north);
  \draw[wfarr] (chamber.south -| rw)    -- (rw.north);
  \draw[wfarr] (chamber.south -| jam)   -- (jam.north);

  \node[wfblue, text width=54mm, below=12mm of rw.south] (raw)
    {\textbf{96{,}090 raw CSV files}\\ 8 spectral fields, 522\,M rows};

  \node[wfgraybox, text width=54mm, below=of raw] (mani)
    {\textbf{Metadata manifest}\\ labels, conditions, validation};

  \node[wfgraybox, text width=54mm, below=of mani] (feat)
    {\textbf{File-level feature table}\\ 63 spectral-summary features};

  \coordinate (merge) at ($(raw.north)+(0,5.5mm)$);
  \coordinate (floor_elbow) at ($(floor.south |- merge)$);
  \coordinate (jam_elbow)   at ($(jam.south   |- merge)$);

  \draw[draw=wfline, line width=0.5pt] (floor.south) -- (floor_elbow);
  \draw[wfarr] (floor_elbow) -- (merge);

  \draw[wfarr] (rw.south) -- (raw.north);

  \draw[draw=wfline, line width=0.5pt] (jam.south) -- (jam_elbow);
  \draw[wfarr] (jam_elbow) -- (merge);

  \draw[draw=wfline, line width=0.5pt] (merge) -- (raw.north);

  \draw[wfarr] (raw) -- (mani);
  \draw[wfarr] (mani) -- (feat);

  \node[wfpurple, text width=31mm, below=11mm of feat.south] (p2)
    {\textbf{Held-out}\\ power, waveform, loc.};

  \node[wfpurple, text width=29mm, left=9mm of p2] (p1)
    {\textbf{Random splits}\\ file-level};

  \node[wfpurple, text width=29mm, right=9mm of p2] (p3)
    {\textbf{Cross-mode}\\ active\,$\leftrightarrow$\,passive};

  \coordinate (split) at ($(feat.south)+(0,-4.5mm)$);

  \draw[draw=wfline, line width=0.5pt] (feat.south) -- (split);

  \draw[wfarr] (split) -- (p2.north);
  \draw[wfarr] (split) -| (p1.north);
  \draw[wfarr] (split) -| (p3.north);

\end{tikzpicture}%
}
\begin{figure}[!htbp]
    \centering
    \usebox{\wffigbox}
    \caption{Dataset collection and processing workflow. Inside the Candela CT840A RF-shielded chamber, a HackRF One SDR jammer controlled by the JamRF framework and a QCA9880 receiver on a Raspberry Pi Compute Module~4 (CM4) running the ath10k spectral-scan driver are separated by 20\,cm line-of-sight. The three collection environments produce benign (chamber floor and real-world indoor) and malicious (chamber jamming) captures. Raw CSV files flow through a metadata manifest and file-level feature aggregation into the benchmark protocol families.}
    \label{fig:workflow_schematic}
\end{figure}

\begin{figure}[!htbp]
    \centering
    \includegraphics[width=0.85\linewidth]{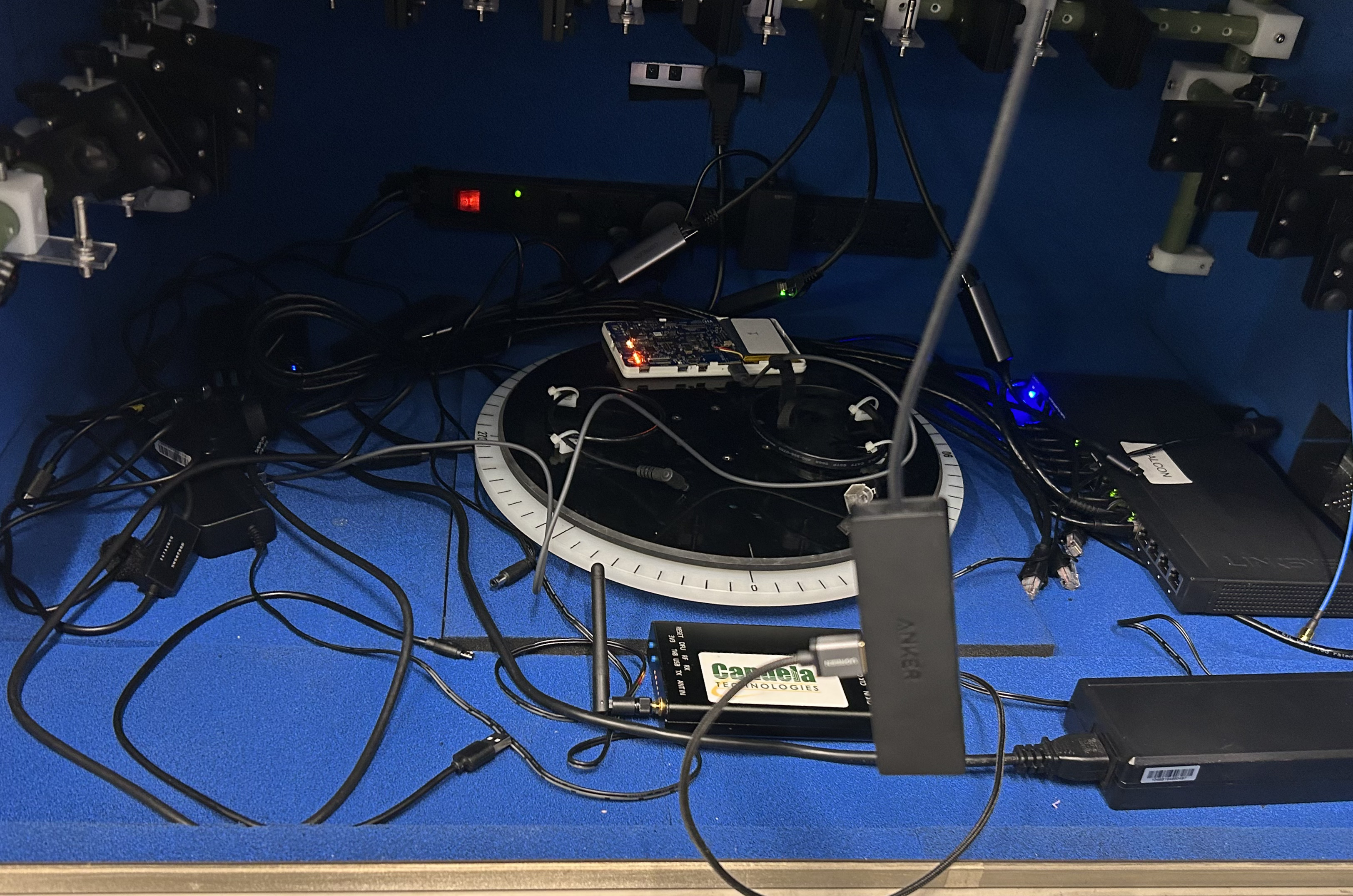}
    \caption{WiFiSpectralJam collection testbed inside the Candela CT840A RF-shielded chamber. The image shows the Raspberry Pi Compute Module~4 (CM4) spectral-sensing host with the Qualcomm Atheros QCA9880 NIC and the HackRF One SDR jammer placed approximately 20\,cm apart with clear line of sight inside the chamber. The laptop is connected to and controls the HackRF One using the JamRF toolkit during jamming collection.}
    \label{fig:testbed_photo}
\end{figure}

\subsection{Experimental Setup}

The receiving system comprised a Raspberry Pi Compute Module~4 (CM4) host equipped with a Qualcomm Atheros QCA9880 802.11ac NIC operating under the Linux ath10k driver. Spectral measurements were retrieved from the driver's debugfs interface using an open-source spectral scan utility released with the collection tooling.\footnote{Spectral-scan collection utility: \url{https://github.com/daniaherzalla/ath-spectral-scan}.} The QCA9880 reports FFT-derived spectral fields through this interface, making the receiver a representative commodity spectrum monitor deployable without specialised RF instrumentation.

Jamming signals were generated by a HackRF One software-defined radio equipped with a dual-band SMA Wi-Fi antenna with 3\,dBi gain, positioned 20\,cm from the CM4 receiver with an unobstructed line-of-sight path.
The HackRF One was controlled via the JamRF framework~\cite{ali2022jamrf}, a HackRF/GNU Radio toolkit for programmable Wi-Fi jamming waveforms, transmit powers, and target frequencies. The fixed 20\,cm separation provides a repeatable jammer--receiver geometry for comparing transmit-power settings. Figure~\ref{fig:testbed_photo} shows the chamber testbed, including the CM4 receiver, HackRF One jammer, and the control laptop used for acquisition.

Benign floor and all malicious captures were collected inside a Candela CT840A RF-shielded chamber,\footnote{Candela CT840A RF-shielded chamber product page: \url{https://www.candelatech.com/ct840a_product.php}.} providing a controlled RF environment. All intentional jamming transmissions were confined to the chamber, and no separate legitimate Wi-Fi association or traffic stream was operated during these captures. The benign floor condition therefore contains the receiver alone, whereas the malicious condition contains the same receiver with the transmitting HackRF One. These labels represent controlled jammer-absent versus jammer-present spectral conditions, not end-to-end link degradation. Real-world backgrounds were collected separately at indoor locations with varying ambient 2.4 and 5\,GHz activity (Section~\ref{sec:collection_design}). 

\subsection{Active and passive spectral scanning}\label{sec:active_passive_scanning}

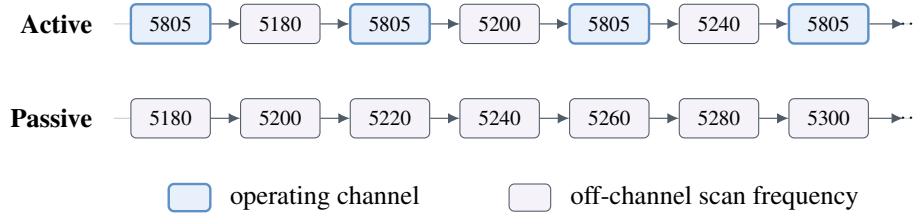
\begin{figure}[!htbp]
    \centering
    \begin{tikzpicture}[x=1cm,y=1cm,>=Latex,font=\scriptsize]
        \tikzstyle{seqbox}=[draw=wfline, rounded corners=2pt, line width=0.4pt,
            minimum width=1.05cm, minimum height=0.52cm, align=center]
        \tikzstyle{opbox}=[seqbox, fill=wfbluefill, line width=0.9pt, draw=wfblue]
        \tikzstyle{scanbox}=[seqbox, fill=wfpurplefill!55]
        \tikzstyle{arrowlink}=[-{Latex[length=1.5mm,width=1.4mm]}, draw=wfline, line width=0.4pt]

        \draw[draw=wfline!30, line width=0.4pt] (0.55,1.35) -- (11.1,1.35);
        \draw[draw=wfline!30, line width=0.4pt] (0.55,0.10) -- (11.1,0.10);

        \node[anchor=east, font=\small\bfseries] at (0.40,1.35) {Active};
        \node[anchor=east, font=\small\bfseries] at (0.40,0.10) {Passive};

        \node[opbox]   (a1) at (1.30,1.35) {5805};
        \node[scanbox] (a2) at (2.75,1.35) {5180};
        \node[opbox]   (a3) at (4.20,1.35) {5805};
        \node[scanbox] (a4) at (5.65,1.35) {5200};
        \node[opbox]   (a5) at (7.10,1.35) {5805};
        \node[scanbox] (a6) at (8.55,1.35) {5240};
        \node[opbox]   (a7) at (10.00,1.35) {5805};
        \node[anchor=west] at (10.70,1.35) {$\cdots$};
        \foreach \i/\j in {a1/a2,a2/a3,a3/a4,a4/a5,a5/a6,a6/a7} {
            \draw[arrowlink] (\i.east) -- (\j.west);
        }
        \draw[arrowlink] (a7.east) -- +(0.45,0);

        \node[scanbox] (p1) at (1.30,0.10) {5180};
        \node[scanbox] (p2) at (2.75,0.10) {5200};
        \node[scanbox] (p3) at (4.20,0.10) {5220};
        \node[scanbox] (p4) at (5.65,0.10) {5240};
        \node[scanbox] (p5) at (7.10,0.10) {5260};
        \node[scanbox] (p6) at (8.55,0.10) {5280};
        \node[scanbox] (p7) at (10.00,0.10) {5300};
        \node[anchor=west] at (10.70,0.10) {$\cdots$};
        \foreach \i/\j in {p1/p2,p2/p3,p3/p4,p4/p5,p5/p6,p6/p7} {
            \draw[arrowlink] (\i.east) -- (\j.west);
        }
        \draw[arrowlink] (p7.east) -- +(0.45,0);

        \node[opbox,   minimum width=0.55cm, minimum height=0.36cm] (lg1) at (1.55,-0.95) {};
        \node[anchor=west, font=\small] at (1.95,-0.95) {operating channel};
        \node[scanbox, minimum width=0.55cm, minimum height=0.36cm] (lg2) at (6.05,-0.95) {};
        \node[anchor=west, font=\small] at (6.45,-0.95) {off-channel scan frequency};
    \end{tikzpicture}
    \caption{Illustrative scan-sequence patterns for representative 5\,GHz files.
    Active scans return to the operating channel (5805\,MHz) between off-channel
    visits; passive scans sweep the 5\,GHz list sequentially.}
    \label{fig:scan_sequence_schematic}
\end{figure}

The dataset includes active and passive spectral scanning. In active scanning, the CM4 receiver remained anchored to an operating channel while briefly visiting other centre frequencies and returning between scan visits (Fig.~\ref{fig:scan_sequence_schematic}). Active files therefore alternate between the operating channel and off-channel discovery frequencies; the operating channel was 5805\,MHz.

In passive scanning, the receiver operated in monitor mode and swept the configured frequencies sequentially without repeated returns to an operating channel. Table~\ref{tab:channel_visit_counts} quantifies the resulting exposure pattern in a fixed validation sample: active files contain roughly 600--700 observations at most off-channel 5\,GHz frequencies but substantially more at 5805\,MHz because of repeated returns, whereas passive observations are distributed across the sequentially visited frequencies. These are empirical observation counts, not measured dwell times, because the raw records contain no explicit timestamps or inter-visit timing.

Figure~\ref{fig:scan_sequence_schematic} illustrates these two sequence patterns using representative 5\,GHz files from the release. The distinction is important because the two modalities differ not only in channel order, but also in operational interpretation: active scans may include scan-related management-frame activity and repeated operating-channel revisits, whereas passive scans provide a receive-only sweep through the configured scan frequencies. For this reason, the benchmark protocols report modality-specific results and include cross-mode transfer experiments rather than assuming that active and passive scans are directly interchangeable.

\begin{table}[!htbp]
    \centering
    \scriptsize
    \caption{Empirical numbers of driver-reported spectral observations per file at each reported centre frequency in the fixed validation sample used to characterise scan-mode channel exposure. Values are mean $\pm$ standard deviation across sampled files, with the median in parentheses. Benign and malicious columns therefore report observation counts per file rather than numbers of files. These counts describe empirical channel exposure and should not be interpreted as measured dwell times.}
    \label{tab:channel_visit_counts}

    \makebox[\textwidth][c]{%
    \begin{minipage}{1.04\textwidth}
        \centering

        \begin{subtable}[t]{0.495\linewidth}
            \centering
            \caption{Active 5\,GHz scans}
            \begin{tabularx}{\linewidth}{@{}lCC@{}}
                \toprule
                \textbf{Freq. (MHz)} &
                \textbf{Benign obs./file} &
                \textbf{Malicious obs./file} \\
                \midrule
                5180 & 678.6 $\pm$ 33.2 (684) & 707.2 $\pm$ 0.5 (707) \\
                5200 & 664.8 $\pm$ 50.1 (673) & 707.1 $\pm$ 0.3 (707) \\
                5240 & 682.3 $\pm$ 48.2 (691) & 706.9 $\pm$ 0.4 (707) \\
                5745 & 687.0 $\pm$ 23.6 (691) & 706.9 $\pm$ 0.3 (707) \\
                5765 & 628.3 $\pm$ 98.9 (705) & 701.3 $\pm$ 21.7 (707) \\
                5785 & 703.6 $\pm$ 12.6 (707) & 706.9 $\pm$ 0.3 (707) \\
                5805 & 3705.2 $\pm$ 1757.0 (4791.5) & 3680.5 $\pm$ 1799.4 (4793) \\
                \bottomrule
            \end{tabularx}
        \end{subtable}
        \hfill
        \begin{subtable}[t]{0.493\linewidth}
            \centering
            \caption{Passive 2.4\,GHz scans}
            \begin{tabularx}{\linewidth}{@{}lCC@{}}
                \toprule
                \textbf{Freq. (MHz)} &
                \textbf{Benign obs./file} &
                \textbf{Malicious obs./file} \\
                \midrule
                2412 & 157.4 $\pm$ 40.3 (157) & 234.7 $\pm$ 0.7 (235) \\
                2417 & 191.1 $\pm$ 28.1 (197) & 216.5 $\pm$ 0.7 (217) \\
                2422 & 193.3 $\pm$ 27.8 (206.5) & 215.4 $\pm$ 2.3 (216) \\
                2427 & 204.9 $\pm$ 16.2 (212) & 213.4 $\pm$ 2.7 (213) \\
                2432 & 195.8 $\pm$ 27.5 (206) & 210.4 $\pm$ 6.0 (212) \\
                2437 & 170.2 $\pm$ 41.8 (171.5) & 212.6 $\pm$ 4.5 (215) \\
                2442 & 184.0 $\pm$ 34.8 (193) & 214.1 $\pm$ 2.5 (214) \\
                2447 & 199.5 $\pm$ 21.1 (210.5) & 215.9 $\pm$ 1.8 (216) \\
                2452 & 198.6 $\pm$ 27.8 (210) & 216.5 $\pm$ 0.6 (217) \\
                2457 & 194.0 $\pm$ 45.3 (213) & 216.5 $\pm$ 0.6 (217) \\
                2462 & 179.4 $\pm$ 45.1 (195) & 213.3 $\pm$ 26.9 (217) \\
                \bottomrule
            \end{tabularx}
        \end{subtable}

        \vspace{0.8em}

        \begin{subtable}[t]{0.95\linewidth}
            \centering
            \caption{Passive 5\,GHz scans}
            \begin{tabularx}{\linewidth}{@{}lCC@{}}
                \toprule
                \textbf{Freq. (MHz)} &
                \textbf{Benign obs./file} &
                \textbf{Malicious obs./file} \\
                \midrule
                5180 & 333.1 $\pm$ 191.2 (235) & 235.0 $\pm$ 0.1 (235) \\
                5200 & 305.3 $\pm$ 206.5 (202) & 217.1 $\pm$ 0.4 (217) \\
                5220 & 308.1 $\pm$ 197.4 (212) & 216.8 $\pm$ 0.4 (217) \\
                5240 & 320.4 $\pm$ 199.7 (216) & 216.7 $\pm$ 0.5 (217) \\
                5260 & 677.9 $\pm$ 27.6 (687) & 688.5 $\pm$ 0.5 (688) \\
                5280 & 657.5 $\pm$ 31.9 (661) & 688.2 $\pm$ 0.4 (688) \\
                5300 & 561.0 $\pm$ 119.1 (588) & 688.0 $\pm$ 0.0 (688) \\
                5320 & 610.6 $\pm$ 127.4 (676) & 688.1 $\pm$ 0.3 (688) \\
                5745 & 327.9 $\pm$ 192.1 (216) & 216.6 $\pm$ 0.5 (217) \\
                5765 & 296.3 $\pm$ 183.3 (217) & 217.0 $\pm$ 0.4 (217) \\
                5785 & 340.3 $\pm$ 205.0 (217) & 216.6 $\pm$ 0.5 (217) \\
                5805 & 338.8 $\pm$ 203.7 (217) & 216.9 $\pm$ 0.4 (217) \\
                5825 & 328.9 $\pm$ 216.0 (217) & 205.6 $\pm$ 48.5 (217) \\
                \bottomrule
            \end{tabularx}
        \end{subtable}

    \end{minipage}%
    }
\end{table}

\subsection{Data collection design}
\label{sec:collection_design}

The data comprise benign (no jamming) and malicious (jamming) conditions. \textit{Benign floor} captures were recorded inside the RF-shielded chamber under both scan modalities, while \textit{benign background} captures were collected in real-world indoor environments. The background location identifiers are scan-mode-specific acquisition labels rather than global physical-site identifiers: in the active subset, \texttt{location1} denotes an open workspace and \texttt{location2} a laboratory environment; in the passive subset, \texttt{location1} and \texttt{location2} denote open-workspace environments and \texttt{location3} the laboratory environment. All \textit{malicious} captures were collected under the controlled RF-chamber setup described above.

Active malicious captures targeted 5805\,MHz using Gaussian-noise and single-tone waveforms at $-40$, $-10$, 0, and $+10$\,dBm. Active files report seven 5\,GHz receiver centre frequencies: 5180, 5200, 5240, 5745, 5765, 5785, and 5805\,MHz. Passive malicious captures used Gaussian-noise jamming at 3, 6, 9, and 12\,dBm, targeting 2412, 2457, 5180, or 5745\,MHz. Passive files are band-specific, covering 11 channels from 2412--2462\,MHz in 2.4\,GHz or 13 channels from 5180--5805\,MHz plus 5825\,MHz in 5\,GHz (Table~\ref{tab:frequency_channels}). Because active and passive scans use different power grids, power settings are treated as modality-specific rather than directly matched across modes. Condition counts are unequal because acquisition duration, time constraints, and collection adjustments varied over the campaign; all valid captures were retained rather than post-hoc equalised. 

\subsection{Dataset labelling and metadata extraction}

Labels were assigned deterministically from the raw folder hierarchy and filenames, with no post-hoc inference. Directory levels encode scan mode, class label, benign subtype and location, or malicious waveform and transmit power; the complete hierarchy is provided in Supplementary Note~\ref{sec:supp_folder_structure}. The manifest-generation script,\footnote{WiFiSpectralJam manifest-generation code: \url{https://www.kaggle.com/code/daniaherzalla/manifest-validation-feature-extraction}.} extracts \texttt{scan\_mode}, \texttt{label}, \texttt{benign\_subtype}, \texttt{location}, \texttt{waveform}, \texttt{power\_dbm}, \texttt{channel\_mhz}, \texttt{band}, and \texttt{collection\_environment}. Structurally inapplicable fields, such as \texttt{waveform} for benign files, are recorded as \texttt{N/A} rather than inferred.

\subsection{Preprocessing and feature aggregation}

Raw CSV files were processed through an automated validation pipeline that verified parsability, confirmed the presence and numeric validity of all eight spectral columns, and checked that each file contained at least one non-empty record. File-level row counts and column dimensionality were recorded in the manifest, and reported in Table~\ref{tab:rowcount_by_scan}.

For the baseline benchmarking layer, a file-level aggregate feature table was derived by computing summary statistics within every raw CSV file. The benchmark feature set and its construction rationale are described in Section~\ref{sec:benchmark_features}. The raw spectral CSV files remain the primary data source; researchers interested in temporal dynamics, sequence models, online detection, or deep learning should work directly from those files.

\subsection{Benchmarking protocol}

Baseline classification establishes reproducible reference results and tests whether the dataset contains learnable jamming-related signal. All train/test splits are constructed at the file level so that observations from one raw recording cannot appear in both partitions. 

The validation suite spans matched-distribution classification, held-out-condition generalisation, RF-chamber floor controls, and cross-scan-mode transfer. Random file-level splits are reported for pooled, active-only, passive-only, passive 2.4\,GHz, and passive 5\,GHz subsets. Held-out experiments withhold jammer power, active waveform, or a scan-mode-specific benign background location. The primary location protocol balances the training classes by down-sampling non-held-out benign files, with an unresampled sensitivity analysis retaining all available training data.

Six classifiers are evaluated throughout: logistic regression (LogReg), random forest (RandomForest), histogram-based gradient boosting (HistGB), XGBoost, LightGBM, and a multilayer perceptron (MLP). Results are reported for all six classifiers. Metrics include accuracy, balanced accuracy, precision, recall, F1-score, ROC-AUC, and average precision (AP), with mean and standard deviation reported over random seeds 1, 7, and 42. F1 uses the fixed decision threshold of 0.5; ROC-AUC and AP are reported alongside it so that ranking quality can be distinguished from threshold-transfer or calibration failures under distribution shift.

The canonical benchmark representation contains 63 spectral-summary features. An ablation repeats the benchmark suite after removing all nine features derived from the driver-reported \texttt{noise} field, yielding a 54-feature \texttt{spectral\_no\_noise} representation. The ablation tests whether the near-fixed active malicious noise floor acts as the sole shortcut behind the strong matched-distribution results. Permutation importance is additionally evaluated for pooled random-split logistic-regression and random-forest models under both feature sets.

\section{Data Records}

\subsection{Repository and access}

WiFiSpectralJam is distributed through the Kaggle dataset record cited in the Data Availability statement~\cite{WiFiSpectralJam_kaggle}. The dataset package contains the raw spectral-scan CSV files together with reusable release artefacts including the file-level metadata manifest, derived aggregate feature tables, and validation outputs. The raw CSV files are the primary research object, while the accompanying artefacts support efficient subset construction, auditing, and reproducible analysis. Public notebooks and source code are distributed separately, as described in the Code Availability section. 

\subsection{Folder structure}
\label{sec:folder_structure}

The Kaggle release has two top-level components: \texttt{rf\_jamming/} contains the raw spectral-scan files organised by scan mode, class, and experimental condition, while \texttt{release\_artifacts/} contains metadata, derived feature tables, and validation outputs. The complete folder tree is provided in Supplementary Note~\ref{sec:supp_folder_structure}.

\subsection{File inventory}

Table~\ref{tab:file_inventory} summarises the release structure. Raw CSV files are stored under \texttt{rf\_jamming/}, reusable processed artefacts under \texttt{release\_artifacts/}, and public notebooks are distributed separately.

\begin{table}[t]
    \centering
    \standardtable
    \caption{Current release structure. Raw spectral-scan records are stored under \texttt{rf\_jamming/}, while reusable processed artefacts are grouped under \texttt{release\_artifacts/}. Public notebooks and code are distributed separately and are described in the Code Availability section.}
    \label{tab:file_inventory}

    \begin{tabularx}{\linewidth}{@{}
        >{\raggedright\arraybackslash}p{0.18\linewidth}
        >{\raggedright\arraybackslash}p{0.37\linewidth}
        L
    @{}}
        \toprule
        \textbf{Release component}
        & \textbf{Path in dataset}
        & \textbf{Role in the dataset package} \\
        \midrule

        Raw active scans
            & \texttt{rf\_jamming/active\_scan/}
            & Primary active-scan CSV records organised by class label and experimental condition. \\

        \midrule
        Raw passive scans
            & \texttt{rf\_jamming/passive\_scan/}
            & Primary passive-scan CSV records organised by class label and experimental condition. \\

        \midrule
        Metadata manifest
            & \texttt{release\_artifacts/metadata/manifest.csv}
            & File-level index containing parsed labels, condition metadata, validation fields, and paths to the raw captures. \\

        \midrule
        Derived features
            & \texttt{release\_artifacts/derived/}\newline
              \texttt{file\_level\_features.csv}\newline
              \texttt{file\_level\_features.parquet}
            & Precomputed file-level spectral summaries for reproducible tabular analysis and benchmark reproduction. \\

        \midrule
        Validation outputs
            & \texttt{release\_artifacts/validation/}
            & Dataset-integrity, composition, row-count, field-range, and related validation summaries generated from the raw release. \\

        \bottomrule
    \end{tabularx}
\end{table}

\subsection{Raw spectral-scan schema}
\label{sec:schema}

Each raw CSV file contains a sequence of driver-reported spectral-scan observations exported by the ath10k spectral-scan pipeline. The QCA9880 performs spectral analysis and exposes FFT-derived summary fields through the driver interface. Each row corresponds to one reported observation at a centre frequency. Table~\ref{tab:raw_schema} reports the schema together with the observed range over the complete release.

\begin{table}[t]
    \centering
    \standardtable
    \caption{Raw spectral-scan fields and observed ranges across the complete release. The ranges report raw stored values before file-level aggregation.}
    \label{tab:raw_schema}
    \begin{tabularx}{\linewidth}{@{}>{\raggedright\arraybackslash}p{0.18\linewidth}L>{\raggedright\arraybackslash}p{0.18\linewidth}>{\centering\arraybackslash}p{0.14\linewidth}@{}}
        \toprule
        \textbf{Field} & \textbf{Description} & \textbf{Unit / representation} & \textbf{Observed range} \\
        \midrule
        \texttt{freq1} & Reported centre frequency for the spectral observation. & MHz & 2412 to 5825 \\
        \texttt{noise} & Driver-estimated noise floor associated with the observation. & dBm & $-114$ to $-95$ \\
        \texttt{max\_magnitude} & Maximum FFT-bin magnitude reported by the spectral-scan interface. & Linear, driver-reported & 1 to 640 \\
        \texttt{total\_gain\_db} & Total receiver-chain gain applied during the spectral measurement. & dB & 0 to 511 \\
        \texttt{base\_pwr\_db} & Base power value reported by the spectral-scan interface. & dB & 371 to 472 \\
        \texttt{rssi} & Received signal-strength indicator on the driver-internal scale. & Driver-reported scale & $-53$ to 109 \\
        \texttt{relpwr\_db} & Relative power summary reported for the observation. & dB & 0 to 63 \\
        \texttt{avgpwr\_db} & Average power over the FFT-derived spectral report. & dB & 0 to 72 \\
        \bottomrule
    \end{tabularx}
\end{table}

The raw \texttt{rssi} field is retained exactly as exported by the collection pipeline. The released derived feature table additionally contains \texttt{rssi\_dbm} summaries produced using the transformation \(\texttt{rssi\_dbm}=\texttt{rssi}-95\), following the documented Atheros RSSI conversion~\cite{wildpackets2002rssi}. Because the mapping is vendor-specific and RSSI is fundamentally a relative indicator, the converted values should be interpreted as an approximate dB-referenced representation rather than an independently calibrated absolute-power measurement. The transformation is a constant offset and therefore preserves ordering; users requiring the original driver scale should use the raw \texttt{rssi} values.

\subsection{Metadata manifest}

The metadata manifest (\texttt{release\_artifacts/metadata/manifest.csv}) provides a file-level index with one record per raw CSV file. It contains the relative file path, parsed scan modality and class label, condition metadata such as benign subtype, location, waveform, transmit power, target channel, band, and collection environment, and validation fields such as row count, feature count, and metadata status. The manifest enables reproducible subset construction without repeatedly traversing the raw folder tree. The complete manifest field list is provided in Supplementary Table~\ref{tab:supp_manifest_fields}.

\subsection{Derived feature and benchmark resources}

The file-level feature table, distributed as CSV and Parquet under \texttt{release\_artifacts/derived/}, provides a precomputed tabular representation for baseline machine learning. Each of its 96{,}090 rows corresponds to one raw file and contains aggregate spectral statistics joined with manifest metadata, for 108 columns in total. The classifier feature subset is defined in Section~\ref{sec:benchmark_features}.

The reported benchmark outputs are produced by the public technical-validation notebook. The notebook reconstructs the train/test splits from the released manifest by applying the documented metadata filters, held-out-condition definitions, and fixed random seeds, and writes detailed per-run and summary result files during execution.

The derived feature table is intended as a reproducible baseline representation for standard tabular machine-learning models. Researchers interested in sequence models, online detection, or within-file spectral dynamics can instead work directly with the raw CSV files, which preserve the order of driver-reported observations within each file. Because the raw records do not include explicit timestamps, however, this ordering should be interpreted as scan order rather than as a uniformly sampled multivariate time series.

\section{Data Overview}
\label{sec:data_overview}

The dataset contains 96{,}090 raw CSV spectral scan files, 522{,}771{,}130 spectral-scan observations, and 14.52\,GB of data. Active scans account for 4{,}358 files and 130{,}711{,}308 observations; passive scans account for 91{,}732 files and 392{,}059{,}822 observations. The class distribution is benign-majority, with 77{,}408 benign files (80.56\,\%) and 18{,}682 malicious files (19.44\,\%).

\begin{figure*}[t]
    \centering
    \begin{tabular}{@{}c@{}}
        \includegraphics[width=0.98\textwidth]
        {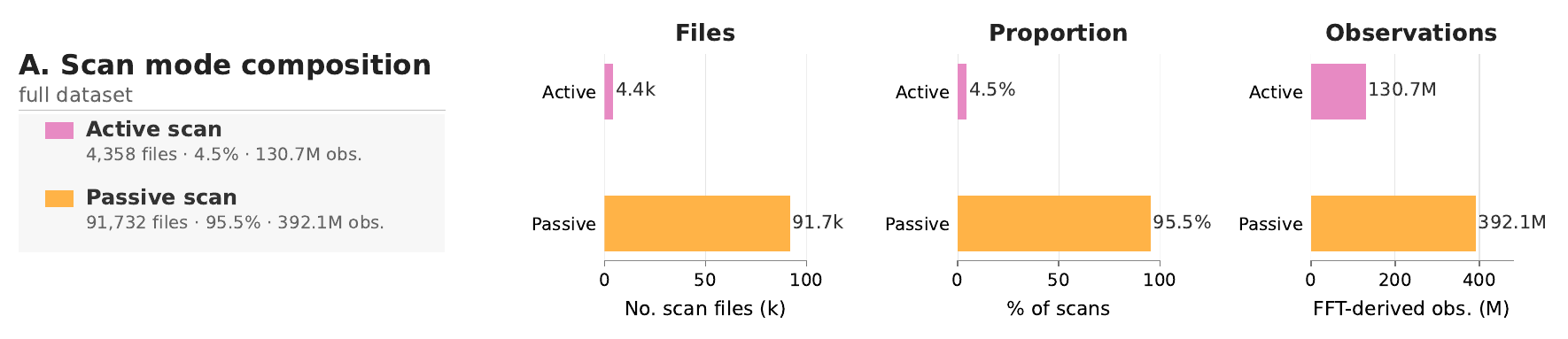}
        \\[-1mm]
        \includegraphics[width=0.98\textwidth]
        {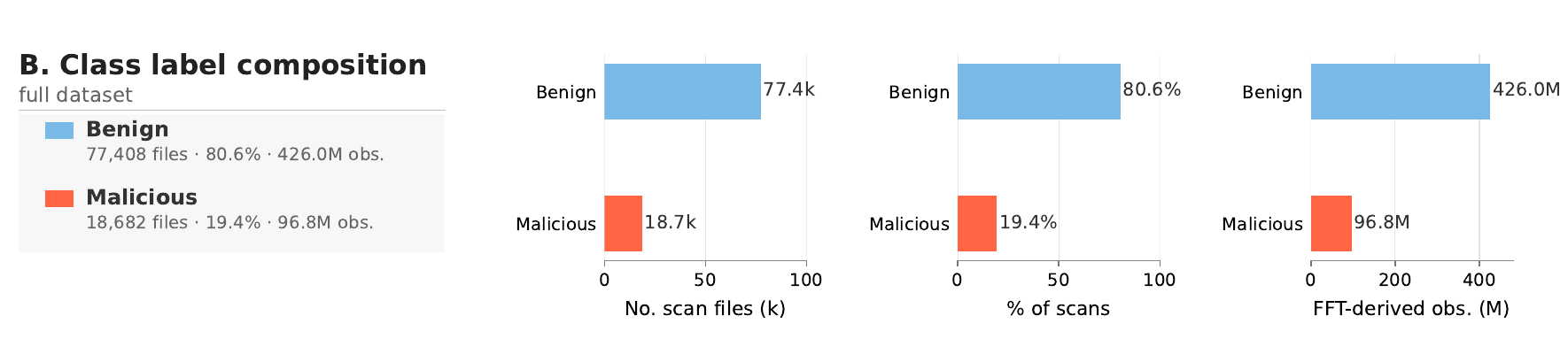}
        \\[-1mm]
        \includegraphics[width=0.98\textwidth]
        {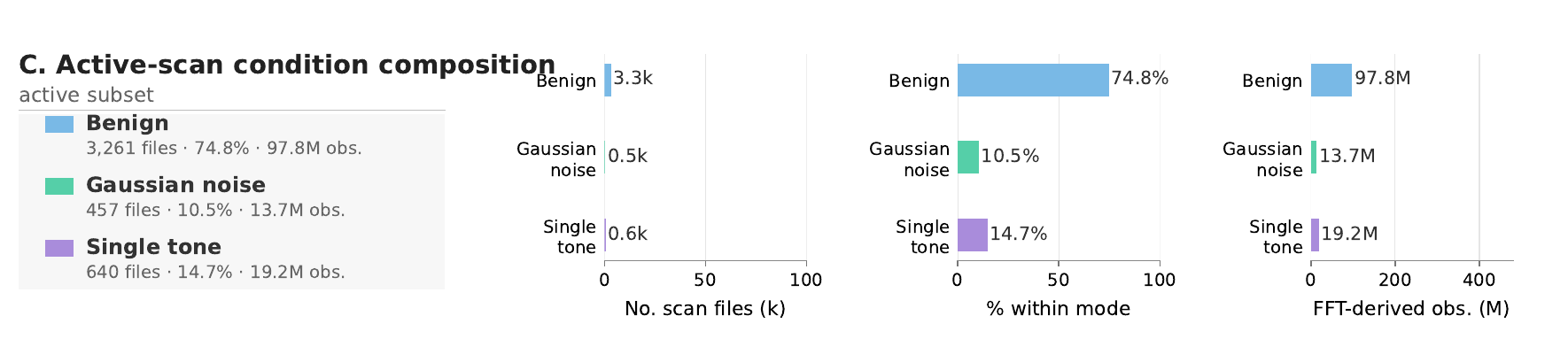}
        \\[-1mm]
        \includegraphics[width=0.98\textwidth]
        {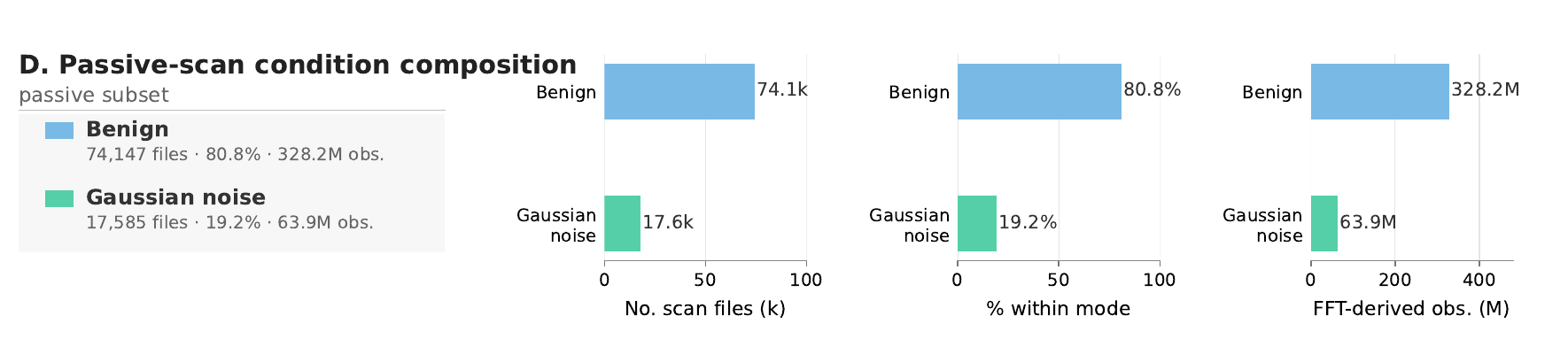}
    \end{tabular}

    \caption{Dataset composition and condition breakdown. Panels summarise
    the WiFiSpectralJam release by scan mode, class label, and
    scan-mode-specific capture condition. Panel A shows the active/passive
    scan-mode split. Panel B shows the benign/malicious class-label split.
    Panel C separates active-scan benign captures from Gaussian-noise and
    single-tone jamming captures. Panel D separates passive-scan benign
    captures from Gaussian-noise jamming captures. Each panel reports file
    count, proportion, and FFT-derived observation count.}
    \label{fig:dataset_composition_breakdown}
\end{figure*}

Figure~\ref{fig:dataset_composition_breakdown} summarises the release by scan mode, class label, and scan-mode-specific jamming condition. Active captures include Gaussian-noise and single-tone jamming, whereas passive malicious captures use Gaussian noise only. Table~\ref{tab:aggregated_dataset_breakdown} provides the corresponding aggregate file, observation, and storage inventory; the full condition-level breakdown is given in Supplementary Table~\ref{tab:supp_condition_breakdown}.

The passive subset is much larger by file count, while passive collection produces shorter per-file channel-sweep records than the longer, near-fixed-length active captures (Table~\ref{tab:rowcount_by_scan}). Similarly, the benign-majority class distribution reflects the greater number of benign background and floor captures relative to the condition-specific malicious captures (Table~\ref{tab:aggregated_dataset_breakdown}). These design characteristics should be considered when constructing train/test splits and selecting evaluation metrics.

\begin{table}[!htbp]
    \centering
    \compacttable
    \caption{Aggregated file, observation, and storage breakdown by dataset label and broad capture category. Rows pool active and passive scans; percentages are relative to the full 96{,}090-file release. The complete condition-level inventory is provided in Supplementary Table~\ref{tab:supp_condition_breakdown}.}\label{tab:aggregated_dataset_breakdown}
    \begin{tabularx}{\linewidth}{@{}>{\hsize=0.70\hsize}L>{\hsize=1.20\hsize}L>{\hsize=0.65\hsize}R>{\hsize=0.65\hsize}R>{\hsize=1.05\hsize}R>{\hsize=0.75\hsize}R@{}}
        \toprule
        \textbf{Label} & \textbf{Category} & \textbf{Files} & \textbf{Percent} & \textbf{Observations} & \textbf{Size (MB)} \\
        \midrule
        \multirow{3}{*}{Benign}
            & Background & 56{,}572 & 58.87\% & 334{,}624{,}472 & 9{,}302.5 \\
            & Floor & 20{,}836 & 21.68\% & 91{,}356{,}985 & 2{,}499.8 \\
            \cmidrule{2-6}
            & \textit{Benign total} & \textbf{77{,}408} & \textbf{80.56\%} & \textbf{425{,}981{,}457} & \textbf{11{,}802.2} \\
        \midrule
        \multirow{3}{*}{Malicious}
            & Gaussian noise & 18{,}042 & 18.78\% & 77{,}589{,}673 & 2{,}154.8 \\
            & Single tone & 640 & 0.67\% & 19{,}200{,}000 & 565.2 \\
            \cmidrule{2-6}
            & \textit{Malicious total} & \textbf{18{,}682} & \textbf{19.44\%} & \textbf{96{,}789{,}673} & \textbf{2{,}720.0} \\
        \midrule
        \textbf{Dataset total} & \textbf{All files} & \textbf{96{,}090} & \textbf{100.00\%} & \textbf{522{,}771{,}130} & \textbf{14{,}522.2} \\
        \bottomrule
    \end{tabularx}
\end{table}

Table~\ref{tab:frequency_channels} summarises the jammed centre frequencies used in malicious captures together with the receiver centre frequencies reported in the released raw files for each scan mode and band. Passive files are band-specific, so the passive 2.4\,GHz and 5\,GHz receiver-frequency sets are listed separately.

\begin{table}[H]
    \centering
    \compacttable
    \caption{Jammed and receiver-reported centre frequencies by scan mode and band. Frequencies are given in MHz and correspond to the \texttt{freq1} values used in the released raw files.}\label{tab:frequency_channels}
    \begin{tabularx}{\linewidth}{@{}>{\hsize=0.43\hsize}L>{\hsize=0.34\hsize}L>{\hsize=0.58\hsize}L>{\hsize=2.65\hsize}L@{}}
        \toprule
        \textbf{Scan mode} & \textbf{Band} & \textbf{Jamm. freq.} & \textbf{Receiver scan centre frequencies (MHz)} \\
        \midrule
        Active & 5\,GHz & 5805 & 5180, 5200, 5240, 5745, 5765, 5785, 5805 \\
        \midrule
        \multirow{2}{*}{Passive}
         & 2.4\,GHz & 2412, 2457 & 2412, 2417, 2422, 2427, 2432, 2437, 2442, 2447, 2452, 2457, 2462 \\
         & 5\,GHz   & 5180, 5745 & 5180, 5200, 5220, 5240, 5260, 5280, 5300, 5320, 5745, 5765, 5785, 5805, 5825 \\
        \bottomrule
    \end{tabularx}
\end{table}

\FloatBarrier
\subsection{Malicious capture conditions}
\label{sec:malicious_capture_conditions}

Because waveform, transmit power, scan mode, and jammed frequency jointly define the malicious collection design, their structure is summarised separately from the aggregate dataset inventory. Table~\ref{tab:waveform_power_distribution} reports the malicious-file distribution by scan mode, waveform, and jammed frequency, while listing file counts under the actual transmit-power values used in each scan mode. Active malicious captures target 5805\,MHz and include both Gaussian-noise and single-tone jamming at $-40$, $-10$, 0, and $+10$\,dBm. Passive malicious captures use Gaussian-noise jamming only, with transmit powers of $+3$, $+6$, $+9$, and $+12$\,dBm split across the four passive jammed frequencies.

\begin{table}[H]
    \centering
    \compacttable
    \caption{Malicious-file distribution by scan mode, waveform, jammed frequency, and transmit power. Table entries are file counts. Frequencies are reported in MHz and transmit powers are reported in dBm. Active and passive captures are shown separately because they use different transmit-power settings. The complete condition-level inventory, including observation counts and storage sizes, is provided in Supplementary Table~\ref{tab:supp_condition_breakdown}.}
    \label{tab:waveform_power_distribution}

    \begin{subtable}[t]{0.49\linewidth}
        \centering
        \caption{Active scans}
        \label{tab:active_power_distribution}
        \footnotesize
        \begin{tabular*}{\linewidth}{@{\extracolsep{\fill}}l c r r r r r@{}}
            \toprule
            \textbf{Waveform} & \textbf{Freq.}
                & \textbf{$-40$} & \textbf{$-10$}
                & \textbf{$0$} & \textbf{$+10$}
                & \textbf{Total files} \\
            \midrule
            Gaussian noise & 5805 & 101 & 102 & 153 & 101 & 457 \\
            Single tone    & 5805 & 101 & 205 & 104 & 230 & 640 \\
            \midrule
            \textbf{All files}
                &  & \textbf{202} & \textbf{307} & \textbf{257} & \textbf{331} & \textbf{1{,}097} \\
            \bottomrule
        \end{tabular*}
    \end{subtable}
    \hfill
    \begin{subtable}[t]{0.49\linewidth}
        \centering
        \caption{Passive scans}
        \label{tab:passive_power_distribution}
        \scriptsize
        \begin{tabular*}{\linewidth}{@{\extracolsep{\fill}}l c r r r r r@{}}
            \toprule
            \textbf{Waveform} & \textbf{Freq.}
                & \textbf{$+3$} & \textbf{$+6$}
                & \textbf{$+9$} & \textbf{$+12$}
                & \textbf{Total files} \\
            \midrule
            \multirow{4}{*}{Gaussian noise}
                & 2412 & 1{,}004 & 1{,}016 & 1{,}128 & 1{,}013 & 4{,}161 \\
                & 2457 & 1{,}006 & 1{,}005 & 1{,}006 & 1{,}012 & 4{,}029 \\
                & 5180 & 1{,}002 & 1{,}019 & 1{,}109 & 1{,}010 & 4{,}140 \\
                & 5745 & 1{,}076 & 1{,}171 & 1{,}985 & 1{,}023 & 5{,}255 \\
            \midrule
            \textbf{All files}
                &  & \textbf{4{,}088} & \textbf{4{,}211} & \textbf{5{,}228} & \textbf{4{,}058} & \textbf{17{,}585} \\
            \bottomrule
        \end{tabular*}
    \end{subtable}
\end{table}

This design supports active-scan waveform comparisons within a fixed 5\,GHz jamming setting, where both Gaussian-noise and single-tone captures are available. Passive validation results should instead be interpreted as Gaussian-noise power and frequency-response diagnostics across the passive jammed-frequency set, rather than waveform-generalisation experiments.

\section{Technical Validation}

Technical validation is organised around six validation axes: file integrity, metadata consistency, acquisition-composition checks, spectral-field behaviour, active/passive scan-mode differences, and baseline learnability under matched-distribution, held-out-condition, and cross-scan-mode evaluation. All validation results reported here are generated at the file level unless explicitly stated otherwise. The classification benchmarks use the conservative 63-feature spectral-summary representation described in Section~\ref{sec:benchmark_features} and keep raw CSV files intact across train/test partitions to avoid row-level leakage. 

\subsection{Data Integrity and Composition}

The validation pipeline confirms that all 96{,}090 raw CSV files contain the expected eight spectral fields (\texttt{freq1}, \texttt{noise}, \texttt{max\_magnitude}, \texttt{total\_gain\_db}, \texttt{base\_pwr\_db}, \texttt{rssi}, \texttt{relpwr\_db}, and \texttt{avgpwr\_db}) and at least one non-empty row. The generated schema-issue and metadata-warning outputs contained no failed records. Row counts, file sizes, header status, and parse status were propagated into the manifest to support file-level auditing. Metadata validation also distinguishes missing values from fields that are structurally non-applicable. For example, \texttt{waveform} and \texttt{power\_dbm} apply to malicious files but not to benign files. Recording such fields as non-applicable prevents their absence from being mistaken for metadata-extraction errors.

Table~\ref{tab:rowcount_by_scan} validates the file-length structure of the release by summarising the number of spectral-scan observations per raw CSV file. Active-scan recordings are nearly fixed in length at approximately 30{,}000 rows per file, whereas passive-scan recordings are shorter and more variable.
\begin{table}[!htbp]
\centering
\compacttable
\caption{File-length validation by scan mode, label, and file group. Statistics report the number of spectral-scan observations per raw CSV file, confirming the near-fixed active recording length and the shorter, more variable passive channel-sweep files.}
\label{tab:rowcount_by_scan}
\begin{tabularx}{\linewidth}{@{}>{\hsize=0.48\hsize}L>{\hsize=0.52\hsize}L>{\hsize=0.76\hsize}L>{\hsize=0.48\hsize}C>{\hsize=0.78\hsize}C>{\hsize=0.62\hsize}C>{\hsize=0.58\hsize}C>{\hsize=0.58\hsize}R@{}}
\toprule
\textbf{Scan mode} & \textbf{Label} & \textbf{File group} & \textbf{Files} & \textbf{Mean $\pm$ SD obs.} & \textbf{Median obs.} & \textbf{Min obs.} & \textbf{Max obs.} \\
\midrule
\multirow{4}{*}{Active}
& \multirow{2}{*}{Benign} & Background & 2{,}857 & 29{,}990 $\pm$ 134 & 30{,}000 & 26{,}331 & 30{,}000 \\
&                         & Floor & 404 & 30{,}000 $\pm$ 0 & 30{,}000 & 30{,}000 & 30{,}000 \\
\cmidrule(lr){2-8}
& \multirow{2}{*}{Malicious} & Gaussian noise & 457 & 30{,}000 $\pm$ 0 & 30{,}000 & 30{,}000 & 30{,}000 \\
&                            & Single tone & 640 & 30{,}000 $\pm$ 0 & 30{,}000 & 30{,}000 & 30{,}000 \\
\midrule
\multirow{3}{*}{Passive}
& \multirow{2}{*}{Benign} & Background & 53{,}715 & 4{,}635 $\pm$ 2{,}287 & 4{,}341 & 1{,}184 & 17{,}040 \\
&                         & Floor & 20{,}432 & 3{,}878 $\pm$ 1{,}131 & 4{,}721 & 2{,}162 & 4{,}732 \\
\cmidrule(lr){2-8}
& Malicious & Gaussian noise & 17{,}585 & 3{,}633 $\pm$ 1{,}167 & 4{,}720 & 2{,}343 & 4{,}729 \\
\bottomrule
\end{tabularx}
\end{table}

File-length variation is largest for passive benign backgrounds and reflects the passive collection process rather than a data-quality failure. Differences in channel-visit structure affect per-channel exposure and the number of observations accumulated per file, which matters for sequence models and batching strategies. Because the raw records contain no explicit timestamps, such analyses should preserve observation and scan-frequency order without assuming fixed inter-sample timing.

Dataset composition was checked against the intended collection design at the level of scan mode, label, benign subtype, location, waveform, transmit power, band, and target channel. Active-scan malicious captures are confined to jamming the 5805\,MHz target channel, whereas passive-scan malicious captures cover the four intended target channels: 2412, 2457, 5180, and 5745\,MHz. Each passive malicious file corresponds to one target channel and one transmit-power setting. The condition counts in Supplementary Table~\ref{tab:supp_condition_breakdown} are therefore the result of deterministic parsing of the dataset hierarchy and validation against expected condition fields.

The active scan malicious subset is comparatively small at the per-condition level, with 101--230 files per waveform-power combination. The passive scan malicious subset is larger, with 4{,}058--5{,}228 files per transmit-power condition. This difference affects the statistical stability of fine-grained held-out evaluations and motivates reporting active, passive, and pooled benchmark slices separately.

\subsection{Signal-level validation}
\label{sec:signal_level_validation}

\begin{figure}[!htbp]
    \centering
    \captionsetup[subfigure]{font=scriptsize,skip=1pt}
    \begin{subfigure}[t]{0.50\linewidth}
        \centering
        \includegraphics[width=\linewidth]{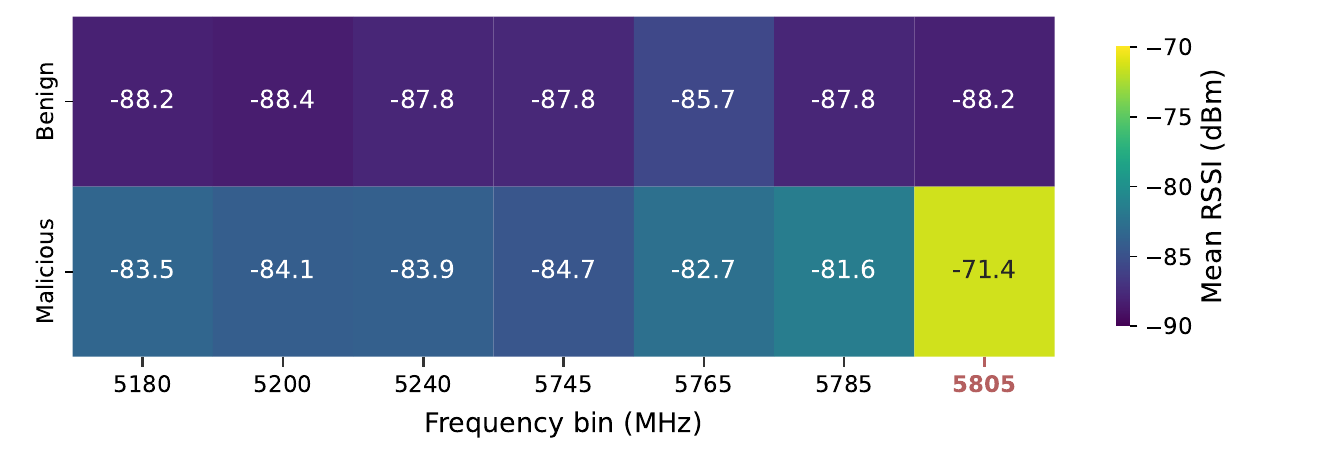}
        \caption{Active, RSSI.}
    \end{subfigure}\hfill
    \begin{subfigure}[t]{0.50\linewidth}
        \centering
        \includegraphics[width=\linewidth]{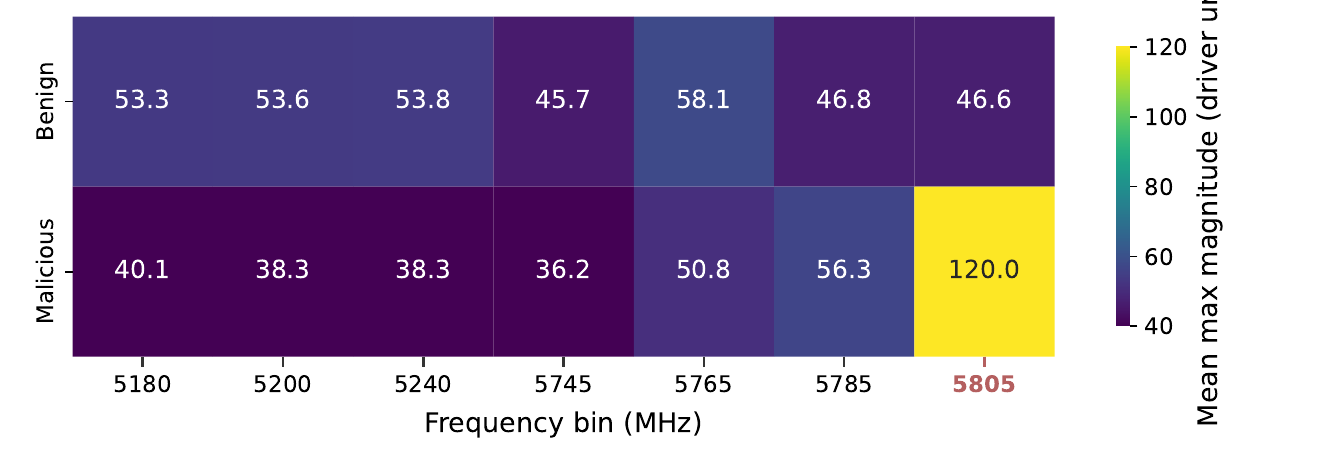}
        \caption{Active, maximum magnitude.}
    \end{subfigure}

    \vspace{0.20em}
    \begin{subfigure}[t]{0.50\linewidth}
        \centering
        \includegraphics[width=\linewidth]{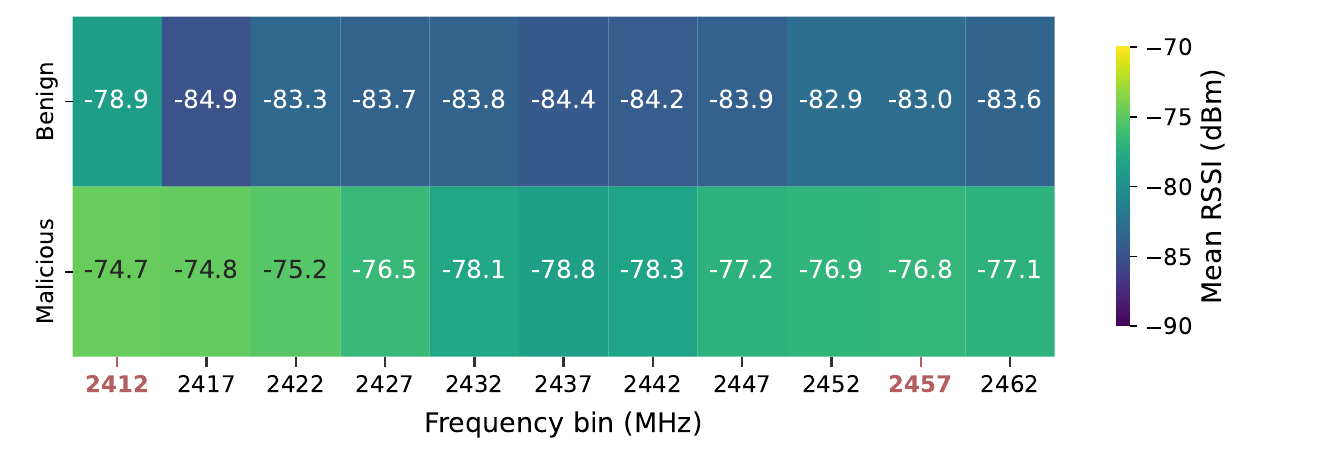}
        \caption{Passive 2.4\,GHz, RSSI.}
    \end{subfigure}\hfill
    \begin{subfigure}[t]{0.50\linewidth}
        \centering
        \includegraphics[width=\linewidth]{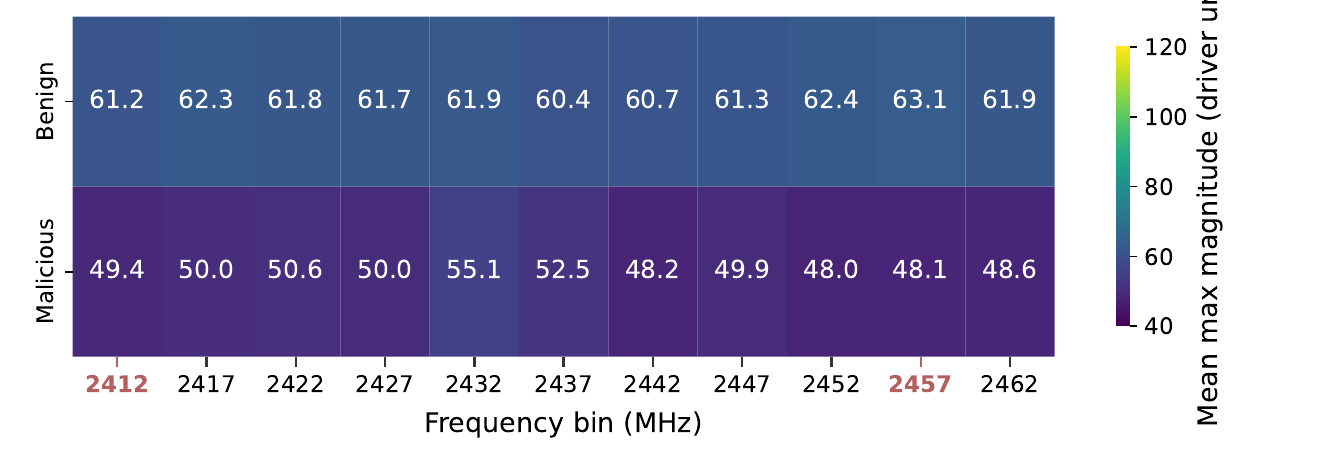}
        \caption{Passive 2.4\,GHz, maximum magnitude.}
    \end{subfigure}

    \vspace{0.20em}
    \begin{subfigure}[t]{0.50\linewidth}
        \centering
        \includegraphics[width=\linewidth]{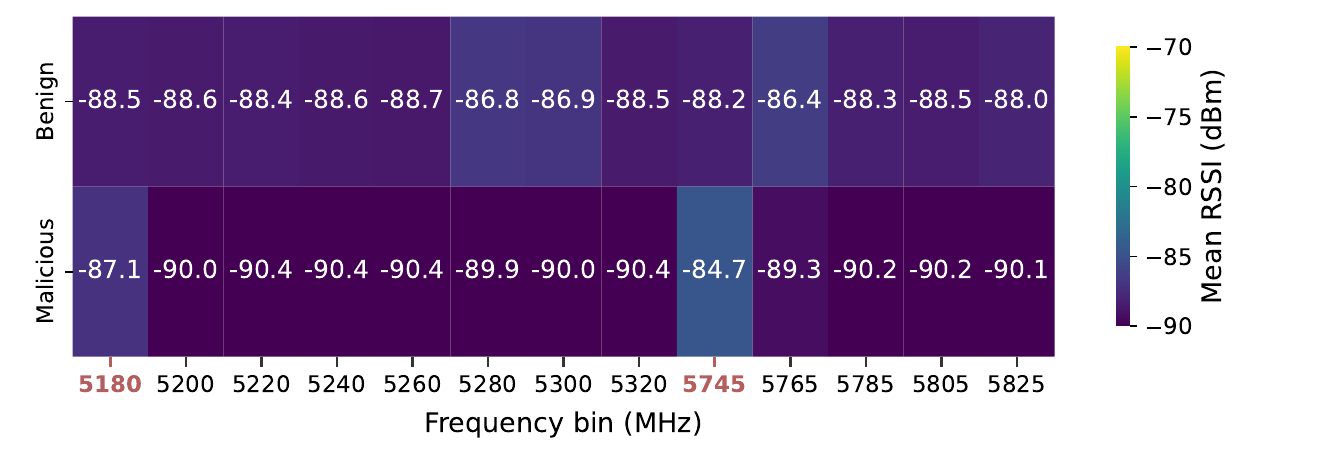}
        \caption{Passive 5\,GHz, RSSI.}
    \end{subfigure}\hfill
    \begin{subfigure}[t]{0.50\linewidth}
        \centering
        \includegraphics[width=\linewidth]{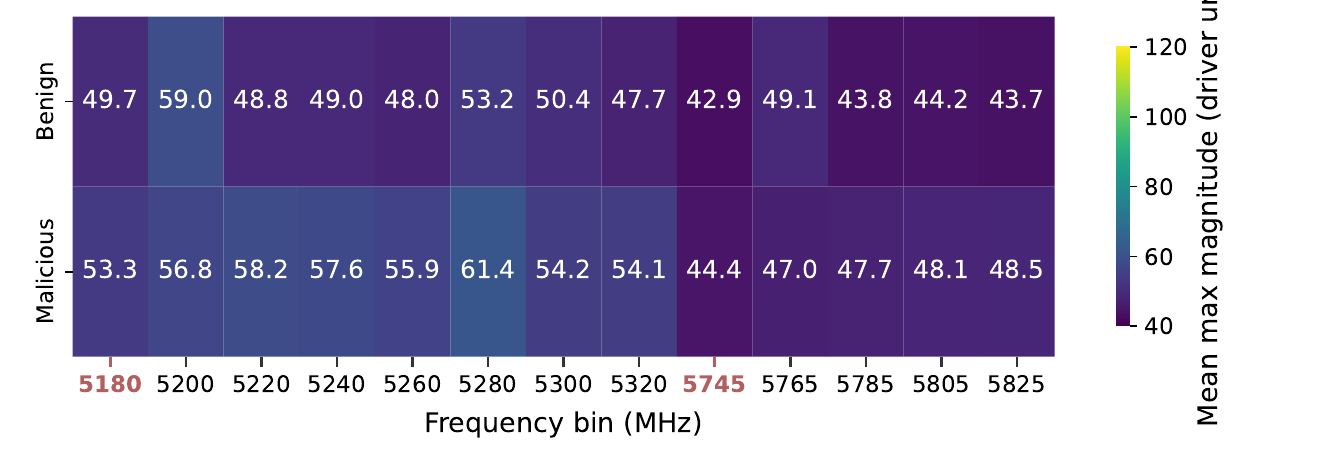}
        \caption{Passive 5\,GHz, maximum magnitude.}
    \end{subfigure}
    \caption{Frequency-bin validation across scan modes and bands. Cells report class-wise file-level means by reported \texttt{freq1} centre frequency; within each sampled file, observations are first averaged within each rounded frequency bin before averaging across files. Active malicious captures target 5805\,MHz. Passive malicious captures include Gaussian-noise jamming at 2412 and 2457\,MHz in the 2.4\,GHz band and at 5180 and 5745\,MHz in the 5\,GHz band.}
    \label{fig:frequency_bin_validation}
    \label{fig:active_frequency_bin_validation}
    \label{fig:passive_2p4ghz_frequency_bin_validation}
    \label{fig:passive_5ghz_frequency_bin_validation}
\end{figure}

Signal-level validation focuses on frequency-bin summaries and stratified spectral-response plots. Each raw file contains observations at multiple reported centre frequencies; therefore, the heatmaps in Figure~\ref{fig:frequency_bin_validation} summarise spectral fields by the reported \texttt{freq1} value and class label. Heatmap cells are computed from file-level means: for each sampled file, observations are first averaged within each \texttt{freq1} bin, and the resulting file-frequency means are then averaged across sampled files within each class. This prevents longer files from contributing more heavily than shorter files to the displayed class-wise frequency-bin summaries.

The main heatmap panels use two matched metrics: RSSI and maximum magnitude. RSSI provides a receiver-strength view of class- and frequency-dependent changes, while maximum magnitude captures peak-like spectral behaviour. In this setting, RSSI and related power summaries should be interpreted as aggregate received RF energy at the NIC, not as source-attributed useful-link quality. A jammer can therefore increase the reported RSSI by raising the total received energy, even though the added energy is interference rather than decodable communication. This behaviour is expected because the spectral-scan interface does not distinguish legitimate Wi-Fi energy from jammer energy when reporting aggregate channel measurements. Passive heatmaps are shown separately for 2.4\,GHz and 5\,GHz. Supplementary Figure~\ref{fig:supp_frequency_bin_validation_additional} provides the corresponding average-power and noise-floor frequency-bin heatmaps, while power- and waveform-stratified supplementary summaries are reported in the Supplementary Information for the spectral fields that add distinct visual information.

Table~\ref{tab:file_level_metric_statistics} summarises the same signal fields at the file level for each scan-mode and class pair. Each entry is computed from the released file-level feature table: observations are first aggregated within each raw CSV file, and the table then reports the mean and standard deviation of those per-file means across files. This file-level aggregation matches the benchmark design and avoids allowing longer recordings to dominate the class summaries. Table~\ref{tab:file_level_metric_delta} reports the corresponding malicious-minus-benign shift within each scan mode, while Supplementary Table~\ref{tab:supp_file_level_metric_statistics_detailed} provides the full distributional version of these summaries, including medians, interquartile ranges, and observed extrema.

Active-scan malicious captures produce the largest file-level shifts, with higher RSSI, maximum magnitude, base power, and relative power and lower total gain than active-scan benign captures. Passive-scan shifts are weaker and less uniformly signed, indicating greater overlap between benign and malicious files and stronger dependence on band, target frequency, and ambient occupancy.

As shown in Supplementary Table~\ref{tab:supp_file_level_metric_statistics_detailed}, the driver-reported noise-floor field is comparatively stable across benign and malicious conditions. This is most pronounced in the active subset, where malicious files report an almost fixed noise-floor value, while jamming-related differences appear more clearly in RSSI, average power, base power, maximum magnitude, relative power, and total gain. In the passive subset, the noise-floor summaries show slightly greater variation, but the benign and malicious ranges remain strongly overlapping. This behaviour is consistent with interpreting the noise floor as a calibrated driver-reported baseline rather than a direct instantaneous measure of jammer power. Accordingly, the noise-floor field should be treated as contextual calibration metadata rather than as a standalone jamming indicator. This observation motivates the noise-feature ablation and permutation-importance analyses in Section~\ref{sec:benchmark_features}, which test whether the strong benchmark performance depends disproportionately on noise-derived features.

\begin{table}[!htbp]
    \centering
    \footnotesize
    \setlength{\tabcolsep}{5pt}
    \renewcommand{\arraystretch}{1.15}
    \caption{File-level spectral metric statistics by scan mode and class label. Values are mean $\pm$ standard deviation across files, where each file contributes its own mean value for the corresponding spectral field. Units follow the raw spectral-field definitions in Table~\ref{tab:raw_schema}; RSSI is reported using the derived dBm conversion.}
    \label{tab:file_level_metric_statistics}
    \resizebox{\linewidth}{!}{%
    \begin{tabular}{@{}llrrrrrrrr@{}}
        \toprule
        \textbf{Scan mode} & \textbf{Label} & \textbf{Files} &
        \makecell{\textbf{RSSI}\\\textbf{(dBm)}} &
        \makecell{\textbf{Max mag.}\\\textbf{(linear)}} &
        \makecell{\textbf{Base pwr.}\\\textbf{(dB)}} &
        \makecell{\textbf{Total gain}\\\textbf{(dB)}} &
        \makecell{\textbf{Avg. pwr.}\\\textbf{(dB)}} &
        \makecell{\textbf{Rel. pwr.}\\\textbf{(dB)}} &
        \makecell{\textbf{Noise}\\\textbf{(dBm)}} \\
        \midrule
        Active  & Benign    & 3{,}261  & $-88.06 \pm 2.06$  & $47.31 \pm 2.94$   & $381.88 \pm 4.24$  & $81.42 \pm 6.61$   & $54.18 \pm 2.49$ & $9.49 \pm 0.13$  & $-104.61 \pm 2.53$ \\
          & Malicious & 1{,}097  & $-73.05 \pm 10.50$ & $108.72 \pm 65.31$ & $394.05 \pm 11.36$ & $66.45 \pm 11.36$ & $54.91 \pm 3.60$ & $11.95 \pm 2.26$ & $-102.00 \pm 0.00$ \\
        \midrule
        Passive & Benign    & 74{,}147 & $-86.86 \pm 3.95$  & $52.57 \pm 9.09$   & $380.66 \pm 4.32$  & $80.16 \pm 7.10$   & $56.39 \pm 1.88$ & $11.16 \pm 2.21$ & $-102.05 \pm 3.48$ \\
         & Malicious & 17{,}585 & $-83.71 \pm 6.77$  & $52.16 \pm 4.16$   & $385.46 \pm 9.08$  & $75.84 \pm 11.59$ & $55.96 \pm 1.97$ & $12.20 \pm 3.55$ & $-102.68 \pm 2.75$ \\
        \bottomrule
    \end{tabular}%
    }
\end{table}

\begin{table}[!htbp]
    \centering
    \scriptsize
    \setlength{\tabcolsep}{4pt}
    \renewcommand{\arraystretch}{1.08}
    \caption{Jamming-minus-non-jamming shifts in file-level spectral metrics. Each entry is the malicious mean minus the benign mean from Table~\ref{tab:file_level_metric_statistics} within the same scan mode. Positive RSSI values indicate that the received energy became less negative under jamming.}
    \label{tab:file_level_metric_delta}
    \begin{tabularx}{\linewidth}{@{}>{\hsize=0.95\hsize}L*{7}{>{\hsize=1.00\hsize}R}@{}}
        \toprule
        \textbf{Scan mode} &
        \makecell[r]{\textbf{$\Delta$RSSI}\\\textbf{(dBm)}} &
        \makecell[r]{\textbf{$\Delta$Max mag.}\\\textbf{(linear)}} &
        \makecell[r]{\textbf{$\Delta$Base pwr.}\\\textbf{(dB)}} &
        \makecell[r]{\textbf{$\Delta$Total gain}\\\textbf{(dB)}} &
        \makecell[r]{\textbf{$\Delta$Avg. pwr.}\\\textbf{(dB)}} &
        \makecell[r]{\textbf{$\Delta$Rel. pwr.}\\\textbf{(dB)}} &
        \makecell[r]{\textbf{$\Delta$Noise}\\\textbf{(dBm)}} \\
        \midrule
        Active  & $+15.02$ & $+61.41$ & $+12.17$ & $-14.97$ & $+0.73$ & $+2.47$ & $+2.61$ \\
        Passive & $+3.15$  & $-0.41$  & $+4.80$  & $-4.32$  & $-0.43$ & $+1.04$ & $-0.64$ \\
        \bottomrule
    \end{tabularx}
\end{table}

\begin{table}[!htbp]
    \centering
    \scriptsize
    \setlength{\tabcolsep}{2.4pt}
    \renewcommand{\arraystretch}{1.02}
    \caption{Band-stratified passive file-level spectral summaries. Values are means of per-file means. Separating 2.4 and 5\,GHz reveals that several malicious-minus-benign feature shifts reverse direction across bands and are obscured by pooled passive statistics.}
    \label{tab:passive_band_statistics}
    \begin{tabularx}{\linewidth}{@{}LLRRRRRRR@{}}
        \toprule
        \textbf{Band} &
        \textbf{Label} &
        \textbf{Files} &
        \makecell[r]{\textbf{RSSI}\\\textbf{(dBm)}} &
        \makecell[r]{\textbf{Max mag.}\\\textbf{(linear)}} &
        \makecell[r]{\textbf{Base pwr.}\\\textbf{(dB)}} &
        \makecell[r]{\textbf{Total gain}\\\textbf{(dB)}} &
        \makecell[r]{\textbf{Rel. pwr.}\\\textbf{(dB)}} &
        \makecell[r]{\textbf{Noise}\\\textbf{(dBm)}} \\
        \midrule
        2.4\,GHz & Benign & 21,336 & -83.82 & 61.66 & 385.63 & 71.13 & 12.80 & -97.42 \\
         & Malicious & 8,190 & -76.81 & 49.85 & 394.84 & 63.72 & 15.62 & -99.94 \\
        \midrule
        5\,GHz & Benign & 52,811 & -88.08 & 48.90 & 378.66 & 83.81 & 10.49 & -103.91 \\
         & Malicious & 9,395 & -89.73 & 54.17 & 377.28 & 86.41 & 9.21 & -105.07 \\
        \bottomrule
    \end{tabularx}
\end{table}

The band-stratified summaries reveal qualitatively different passive regimes: several malicious--benign shifts, including RSSI, base power, relative power, and total gain, reverse direction between 2.4 and 5\,GHz. Passive benchmarks should therefore be reported separately by band rather than treating passive scans as homogeneous.

Beyond the frequency-bin heatmaps, we also provide per-condition spectral-response summaries that stratify the scanned centre frequencies by jammer transmit power and waveform. In these summaries a value at a non-target centre frequency represents the feature observed at that scanned frequency while the jamming condition was active, not an additional jammed channel. Stratifying by transmit power isolates waveform effects from power effects: active scan single-tone jamming produces a localised response concentrated near the 5805\,MHz target (most visible in maximum magnitude), whereas Gaussian-noise jamming produces a broader, less sharply concentrated profile (most visible in aggregate RSSI). For passive scans, which contain Gaussian-noise jamming only, the corresponding power sweeps show that low-power passive jamming can sit closer to benign passive variability than low-power active scan jamming, motivating the feature-space diagnostics and held-out-condition analysis below. The complete supplementary spectral-response set is provided for the metrics most relevant to the raw spectral fields: RSSI, base power, maximum magnitude, and noise floor. These figures include active scan waveform comparisons at matched transmit powers (Figs.~\ref{fig:supp_active_waveform_stratified_rssi}--\ref{fig:supp_active_waveform_stratified_noise}), active waveform-conditioned power sweeps (Figs.~\ref{fig:supp_active_power_sweep_rssi}--\ref{fig:supp_active_power_sweep_noise}), and passive Gaussian-noise power sweeps (Figs.~\ref{fig:supp_passive_power_sweep_rssi}--\ref{fig:supp_passive_power_sweep_noise}). The file-level distributional consequences of these effects are summarised in the supplementary active/passive feature-distribution diagnostics in Figs.~\ref{fig:supp_active_weak_power_combined_additional} and~\ref{fig:supp_passive_weak_power_feature_distributions}.

\begin{figure}[!htbp]
    \centering
    \captionsetup[subfigure]{font=scriptsize,skip=2pt}

    \begin{subfigure}[t]{0.24\linewidth}
        \centering
        \includegraphics[width=\linewidth,trim=8pt 8pt 8pt 8pt,clip]{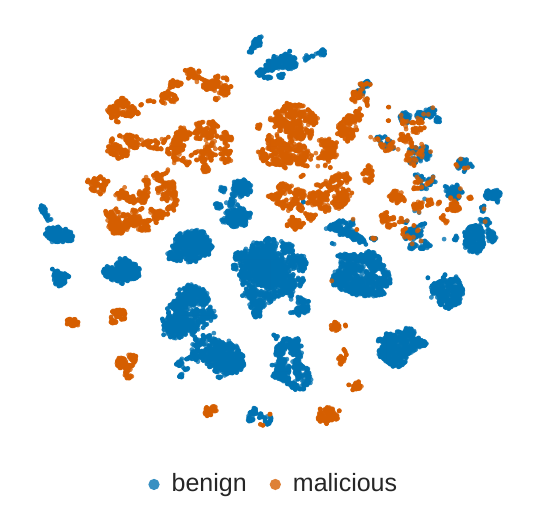}
        \par\vspace{1.2em}
        \caption{Class label.}
    \end{subfigure}\hfill
    \begin{subfigure}[t]{0.24\linewidth}
        \centering
        \includegraphics[width=\linewidth,trim=8pt 8pt 8pt 8pt,clip]{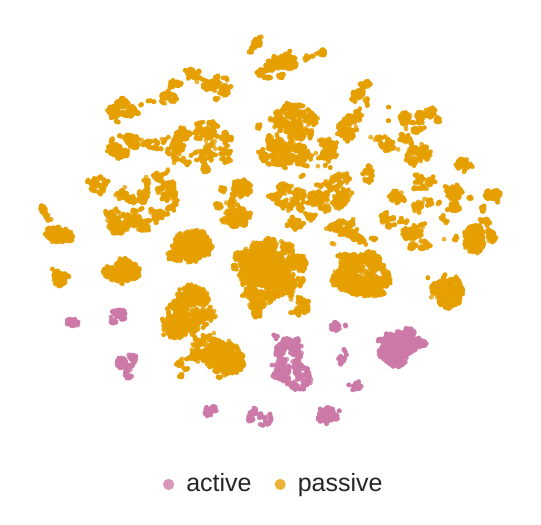}
        \par\vspace{1.2em}
        \caption{Scan mode.}
    \end{subfigure}\hfill
    \begin{subfigure}[t]{0.24\linewidth}
        \centering
        \includegraphics[width=\linewidth,trim=8pt 8pt 8pt 8pt,clip]{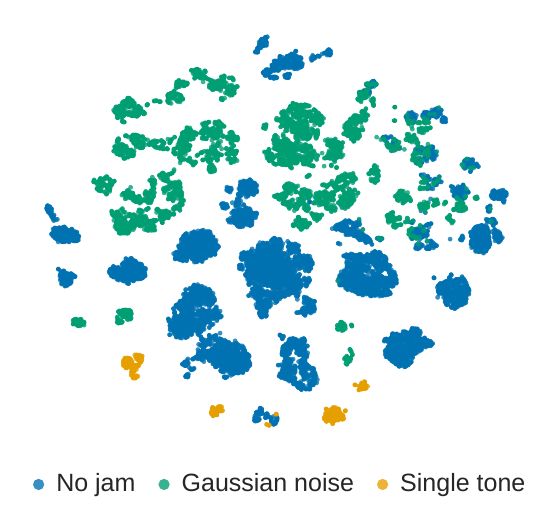}
        \par\vspace{1.2em}
        \caption{Waveform type.}
    \end{subfigure}\hfill
    \begin{subfigure}[t]{0.24\linewidth}
        \centering
        \raisebox{-0.7em}{%
            \includegraphics[width=0.98\linewidth,trim=8pt 8pt 8pt 8pt,clip]{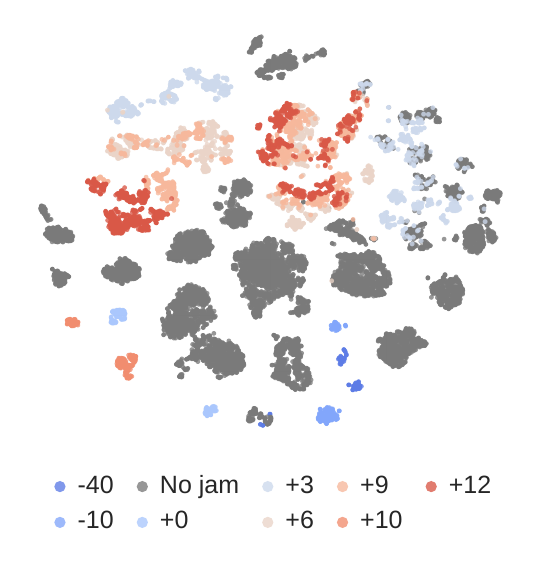}
        }
        \par\vspace{0.54em}
        \caption{Transmit power.}
    \end{subfigure}

    \caption{t-SNE embeddings of file-level aggregate spectral features, coloured by (a) class label, (b) scan mode, (c) waveform type, and (d) transmit power. Each point corresponds to one scan file after file-level aggregation, imputation, standardisation, and PCA pre-reduction; a fixed-seed stratified sample capped at 20{,}000 files is used. The embedding is a qualitative diagnostic of feature-space structure and is not used for model scoring.}
    \label{fig:tsne_feature_space}
\end{figure}

Figure~\ref{fig:tsne_feature_space} provides a complementary t-SNE view of the aggregate feature space, coloured by class, scan mode, waveform, and transmit power. Each point represents one file after aggregation, imputation, standardisation, and PCA pre-reduction, using a fixed-seed stratified sample capped at 20{,}000 files. The embeddings show class- and condition-related local structure and distinct active/passive neighbourhoods, but are interpreted qualitatively because t-SNE is sensitive to sampling and hyperparameters.

\subsection{File-level spectral-summary feature representation and benchmark setup}
\label{sec:benchmark_features}

For all reported tabular benchmarks, each raw file is converted into a file-level spectral-summary vector. The canonical representation contains 63 features derived from seven measurement quantities: \texttt{avgpwr\_db}, \texttt{base\_pwr\_db}, \texttt{max\_magnitude}, \texttt{noise}, \texttt{relpwr\_db}, \texttt{rssi\_dbm}, and \texttt{total\_gain\_db}. For each quantity, nine summaries are retained: mean, standard deviation, minimum, first quartile, median, third quartile, maximum, skewness, and kurtosis. Centre-frequency summaries, unique-count features, raw unconverted RSSI summaries, row counts, file size, target-channel metadata, jammer power, and validation-only quantities are excluded from classifier inputs.

Because the active malicious \texttt{noise} field is nearly fixed at $-102$\,dBm, the complete benchmark suite is also repeated after removing every \texttt{noise\_*} feature. This complementary representation contains 54 features and is used as an ablation rather than as a replacement for the canonical 63-feature release benchmark. All six classifiers are reported for each experiment.

\subsection{Baseline and held-out classification}
\label{sec:baseline_classification}

\subsubsection{Matched-distribution benign-versus-malicious file classification}

Table~\ref{tab:baseline_all_models} reports F1 for all six classifiers under matched-distribution random file-level splits. Performance is saturated or near saturated for the tree and boosting models across all slices, while logistic regression and the MLP perform weaker in the larger pooled and passive 5\,GHz settings. All classifiers reach F1~=~1.000 on the active-only split. Passive 2.4\,GHz is similarly saturated, and passive 5\,GHz remains highly separable when the relevant conditions are represented in training. These results demonstrate strong within-distribution separability but should not be interpreted as evidence of generalisation to unseen jammer conditions or acquisition domains.

\begin{table}[!htbp]
    \centering
    \scriptsize
    \caption{Matched-distribution random file-level F1-score for all six classifiers using the canonical 63-feature spectral-summary representation. Values are mean $\pm$ standard deviation across seeds 1, 7, and 42. ROC-AUC and AP are retained in the released all-model result files; both are approximately saturated for the high-performing tree and boosting models.}
    \label{tab:baseline_all_models}
    \begin{tabularx}{\linewidth}{@{}LCCCCC@{}}
        \toprule
        \textbf{Model} & \textbf{Active 5 GHz} & \textbf{Passive 2.4 GHz} & \textbf{Passive 5 GHz} & \textbf{Passive pooled} & \textbf{All data} \\
        \midrule
        LogReg & 1.0000 $\pm$ 0.0000 & 1.0000 $\pm$ 0.0000 & 0.9651 $\pm$ 0.0031 & 0.9796 $\pm$ 0.0024 & 0.9721 $\pm$ 0.0032 \\
        RandomForest & 1.0000 $\pm$ 0.0000 & 1.0000 $\pm$ 0.0000 & 0.9996 $\pm$ 0.0004 & 0.9994 $\pm$ 0.0001 & 0.9996 $\pm$ 0.0002 \\
        HistGB & 1.0000 $\pm$ 0.0000 & 0.9994 $\pm$ 0.0005 & 0.9996 $\pm$ 0.0002 & 0.9995 $\pm$ 0.0004 & 0.9997 $\pm$ 0.0002 \\
        XGBoost & 1.0000 $\pm$ 0.0000 & 0.9994 $\pm$ 0.0005 & 0.9999 $\pm$ 0.0002 & 0.9996 $\pm$ 0.0003 & 0.9998 $\pm$ 0.0003 \\
        LightGBM & 1.0000 $\pm$ 0.0000 & 0.9993 $\pm$ 0.0012 & 0.9995 $\pm$ 0.0000 & 0.9997 $\pm$ 0.0002 & 0.9998 $\pm$ 0.0003 \\
        MLP & 1.0000 $\pm$ 0.0000 & 1.0000 $\pm$ 0.0000 & 0.9948 $\pm$ 0.0014 & 0.9933 $\pm$ 0.0004 & 0.9948 $\pm$ 0.0041 \\
        \bottomrule
    \end{tabularx}
\end{table}

\subsubsection{Held-out jammer transmit-power and active-waveform generalisation}

\begin{figure}[!htbp]
    \centering
    \includegraphics[width=\linewidth]{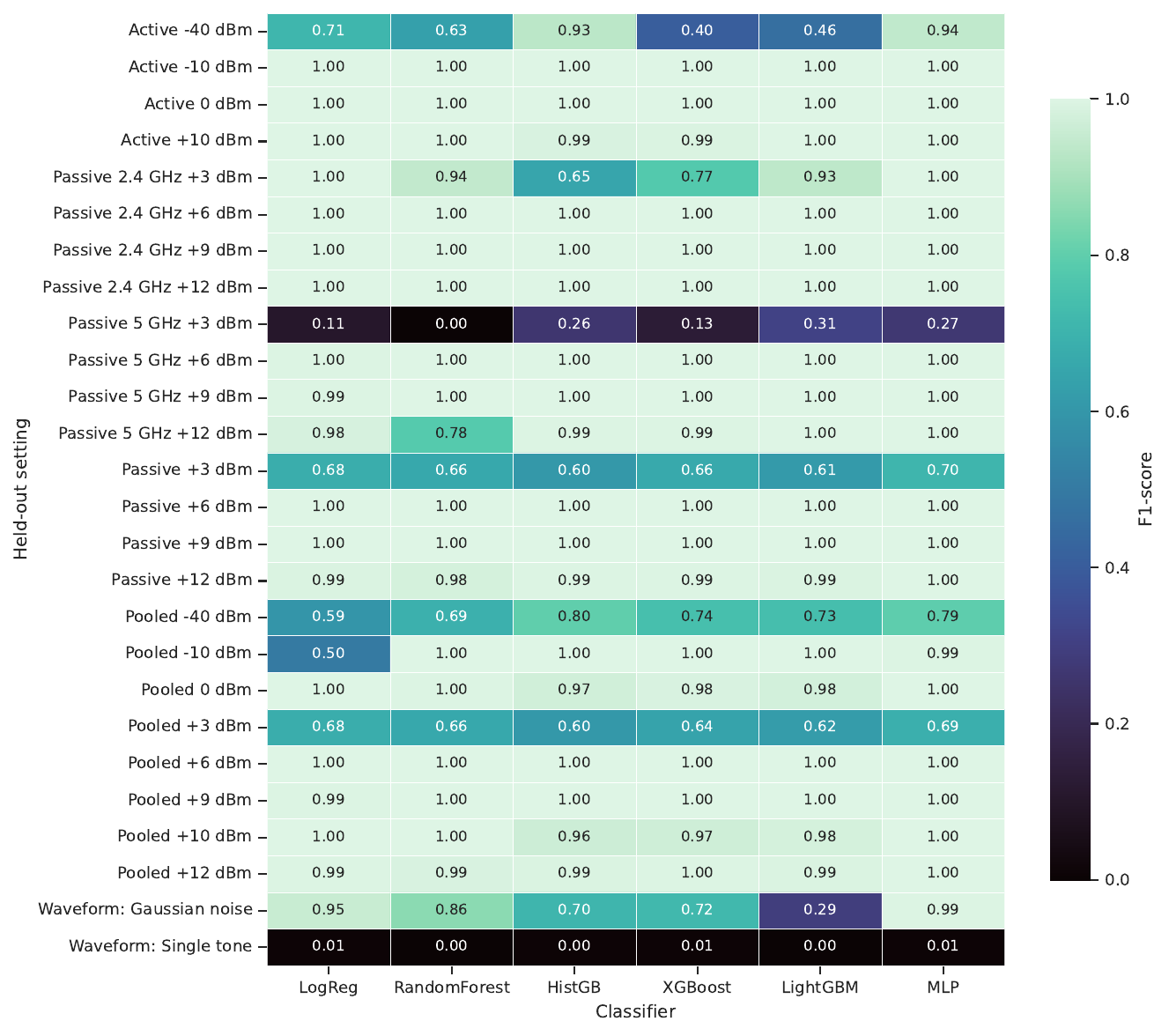}
    \caption{Per-model held-out-condition F1-score heatmap using the canonical 63-feature representation. Rows are held-out power or waveform settings and columns are classifiers.}
    \label{fig:heldout_all_model_heatmap}
\end{figure}

Held-out-condition experiments expose substantially more structure than the matched-distribution splits. Figure~\ref{fig:heldout_all_model_heatmap} reports F1 for every classifier and held-out condition, while Table~\ref{tab:heldout_difficult_all_models} gives F1, ROC-AUC and AP for three representative difficult shifts. Moderate and high held-out powers generally transfer well, but active $-40$\,dBm is classifier-dependent and low-power passive transfer is strongly band-dependent.

Passive $+3$\,dBm is strongly band-dependent. The held-out 2.4\,GHz test set reaches F1~=~1.000 for multiple classifiers, whereas the similarly sized 5\,GHz test set yields poor fixed-threshold F1 across all six models. This is consistent with the band-stratified feature shifts in Table~\ref{tab:passive_band_statistics}, several of which reverse direction across bands. High ROC-AUC for several 5\,GHz models despite low F1 further indicates a threshold-transfer problem rather than complete loss of ranking information. The pooled passive result therefore combines an easy 2.4\,GHz regime with a much harder 5\,GHz regime.

Held-out waveform transfer is also asymmetric. Gaussian-noise holdout remains strong, whereas single-tone F1 is near zero for every classifier at the default threshold. Yet RandomForest reaches ROC-AUC/AP~=~1.000 and XGBoost/LightGBM remain near one, showing that the failure is largely one of score-scale transfer rather than intrinsic class indistinguishability.

\begin{table}[!htbp]
    \centering
    \scriptsize
    \setlength{\tabcolsep}{4.0pt}
    \renewcommand{\arraystretch}{0.96}
    \caption{All-model results for representative difficult held-out conditions using the canonical 63-feature representation. Values are mean $\pm$ standard deviation across three seeds. ROC-AUC and AP are threshold-free and should be interpreted alongside the fixed-threshold F1.}
    \label{tab:heldout_difficult_all_models}

    \begin{tabularx}{0.95\textwidth}{@{}>{\raggedright\arraybackslash}X
                                      >{\raggedright\arraybackslash}X
                                      >{\centering\arraybackslash}X
                                      >{\centering\arraybackslash}X
                                      >{\centering\arraybackslash}X@{}}
        \toprule
        \textbf{Held-out setting} & \textbf{Model} & \textbf{F1} & \textbf{ROC-AUC} & \textbf{AP} \\
        \midrule
        Active $-40$ dBm & LogReg & 0.711 $\pm$ 0.002 & 1.000 $\pm$ 0.000 & 1.000 $\pm$ 0.000 \\
         & RandomForest & 0.632 $\pm$ 0.056 & 0.999 $\pm$ 0.001 & 0.999 $\pm$ 0.002 \\
         & HistGB & 0.932 $\pm$ 0.002 & 0.941 $\pm$ 0.037 & 0.956 $\pm$ 0.017 \\
         & XGBoost & 0.404 $\pm$ 0.046 & 0.993 $\pm$ 0.003 & 0.992 $\pm$ 0.004 \\
         & LightGBM & 0.458 $\pm$ 0.000 & 0.992 $\pm$ 0.001 & 0.992 $\pm$ 0.002 \\
         & MLP & 0.939 $\pm$ 0.037 & 1.000 $\pm$ 0.000 & 1.000 $\pm$ 0.000 \\
        \midrule
        Passive 5 GHz $+3$ dBm & LogReg & 0.105 $\pm$ 0.002 & 0.969 $\pm$ 0.004 & 0.961 $\pm$ 0.008 \\
         & RandomForest & 0.000 $\pm$ 0.000 & 0.973 $\pm$ 0.007 & 0.973 $\pm$ 0.007 \\
         & HistGB & 0.259 $\pm$ 0.020 & 0.730 $\pm$ 0.024 & 0.738 $\pm$ 0.013 \\
         & XGBoost & 0.129 $\pm$ 0.028 & 0.958 $\pm$ 0.008 & 0.950 $\pm$ 0.009 \\
         & LightGBM & 0.311 $\pm$ 0.140 & 0.744 $\pm$ 0.068 & 0.788 $\pm$ 0.029 \\
         & MLP & 0.266 $\pm$ 0.145 & 0.954 $\pm$ 0.027 & 0.959 $\pm$ 0.023 \\
        \midrule
        Active single tone & LogReg & 0.006 $\pm$ 0.000 & 0.017 $\pm$ 0.000 & 0.319 $\pm$ 0.000 \\
         & RandomForest & 0.000 $\pm$ 0.000 & 1.000 $\pm$ 0.000 & 1.000 $\pm$ 0.000 \\
         & HistGB & 0.004 $\pm$ 0.004 & 0.896 $\pm$ 0.087 & 0.866 $\pm$ 0.080 \\
         & XGBoost & 0.006 $\pm$ 0.000 & 0.992 $\pm$ 0.001 & 0.979 $\pm$ 0.008 \\
         & LightGBM & 0.002 $\pm$ 0.004 & 0.996 $\pm$ 0.002 & 0.991 $\pm$ 0.010 \\
         & MLP & 0.006 $\pm$ 0.000 & 0.159 $\pm$ 0.002 & 0.454 $\pm$ 0.002 \\
        \bottomrule
    \end{tabularx}
\end{table}

\subsubsection{Noise-feature ablation and shortcut audit}

Removing all nine \texttt{noise\_*} summaries leaves matched-distribution performance and the RF-chamber floor control saturated, showing that the strong benchmark is not explained solely by the driver-reported noise floor. Noise summaries nevertheless contribute under selected shifts: pooled $-40$\,dBm and several Passive$\rightarrow$Active models degrade after removal, while Active$\rightarrow$Passive changes slightly. Thus, noise features are unnecessary for matched-condition separability but can improve robustness in specific transfer settings.

\begin{table}[!htbp]
    \centering
    \scriptsize
    \caption{Noise-feature ablation for all six classifiers. Each entry reports F1 as 63-feature / 54-feature-no-noise. Matched-distribution performance remains near saturated without noise summaries, while the transfer effect is model- and direction-dependent.}
    \label{tab:noise_ablation_all_models}
    \begin{tabularx}{\linewidth}{@{}LC>{\centering\arraybackslash}p{2.8cm}CCC@{}}
        \toprule
        \textbf{Model} & \textbf{Active matched} & \textbf{Passive 5 GHz matched} & \textbf{All-data matched} & \textbf{A$\rightarrow$P} & \textbf{P$\rightarrow$A} \\
        \midrule
        LogReg & 1.000/1.000 & 0.965/0.957 & 0.972/0.967 & 0.426/0.439 & 0.000/0.002 \\
        RandomForest & 1.000/1.000 & 1.000/0.999 & 1.000/1.000 & 0.349/0.411 & 0.037/0.039 \\
        HistGB & 1.000/1.000 & 1.000/1.000 & 1.000/1.000 & 0.364/0.364 & 0.262/0.091 \\
        XGBoost & 1.000/1.000 & 1.000/1.000 & 1.000/1.000 & 0.439/0.438 & 0.197/0.004 \\
        LightGBM & 1.000/1.000 & 0.999/1.000 & 1.000/1.000 & 0.424/0.424 & 0.564/0.133 \\
        MLP & 1.000/1.000 & 0.995/0.990 & 0.995/0.991 & 0.415/0.417 & 0.000/0.002 \\
        \bottomrule
    \end{tabularx}
\end{table}

\subsubsection{Permutation-importance audit}

Permutation importance supports the same conclusion. In pooled logistic regression, \texttt{noise\_max} is the strongest 63-feature predictor, but after noise removal importance shifts to gain, base-power, relative-power, and RSSI summaries while matched performance remains saturated. Random forest relies most strongly on \texttt{relpwr\_db\_std} in both representations. Noise therefore contributes signal without being the sole basis for classification.

\begin{table}[!htbp]
    \centering
    \scriptsize
    \caption{Highest permutation-importance features for pooled random-split models. Importance is the mean decrease in average precision under feature permutation, reported as mean $\pm$ standard deviation over repeated permutations on the fixed importance subset.}
    \label{tab:permutation_importance}
    \begin{tabularx}{\linewidth}{@{}LLLR@{}}
        \toprule
        \textbf{Feature set} & \textbf{Model} & \textbf{Feature} & \textbf{Importance} \\
        \midrule
        63 features & LogReg & \texttt{noise\_max} & 0.2890 $\pm$ 0.0092 \\
         &  & \texttt{relpwr\_db\_q75} & 0.2033 $\pm$ 0.0057 \\
         &  & \texttt{rssi\_dbm\_skew} & 0.1581 $\pm$ 0.0060 \\
         &  & \texttt{base\_pwr\_db\_mean} & 0.1513 $\pm$ 0.0061 \\
         &  & \texttt{rssi\_dbm\_std} & 0.1243 $\pm$ 0.0070 \\
        \cmidrule(lr){2-4}
         & RandomForest & \texttt{relpwr\_db\_std} & 0.0010 $\pm$ 0.0001 \\
         &  & \texttt{rssi\_dbm\_min} & 0.0003 $\pm$ 0.0002 \\
         &  & \texttt{rssi\_dbm\_kurtosis} & 0.0001 $\pm$ 0.0000 \\
         &  & \texttt{noise\_q25} & 0.0000 $\pm$ 0.0000 \\
         &  & \texttt{relpwr\_db\_skew} & 0.0000 $\pm$ 0.0000 \\
        \midrule
        54 no-noise & LogReg & \texttt{total\_gain\_db\_mean} & 0.4409 $\pm$ 0.0083 \\
         &  & \texttt{base\_pwr\_db\_mean} & 0.3420 $\pm$ 0.0101 \\
         &  & \texttt{total\_gain\_db\_min} & 0.3299 $\pm$ 0.0086 \\
         &  & \texttt{relpwr\_db\_mean} & 0.3284 $\pm$ 0.0121 \\
         &  & \texttt{rssi\_dbm\_std} & 0.2289 $\pm$ 0.0106 \\
        \cmidrule(lr){2-4}
         & RandomForest & \texttt{relpwr\_db\_std} & 0.0202 $\pm$ 0.0024 \\
         &  & \texttt{rssi\_dbm\_min} & 0.0003 $\pm$ 0.0002 \\
         &  & \texttt{relpwr\_db\_skew} & 0.0001 $\pm$ 0.0001 \\
         &  & \texttt{rssi\_dbm\_kurtosis} & 0.0001 $\pm$ 0.0000 \\
         &  & \texttt{relpwr\_db\_mean} & 0.0001 $\pm$ 0.0000 \\
        \bottomrule
    \end{tabularx}
\end{table}

\subsubsection{Scan-mode-specific robustness to unseen benign background locations}

Background-location evaluation is performed within scan mode. Active held-out backgrounds and passive \texttt{location2}/\texttt{location3} remain strong, while passive \texttt{location1} is consistently hardest (best fixed-threshold F1 $\approx0.873$, ROC-AUC $\approx0.977$). The same ordering persists without training rebalancing. Because passive \texttt{location3} contains substantially more observations per file than locations 1 and 2, residual sample-count sensitivity in higher-order summaries cannot be excluded.

\begin{figure}[!htbp]
    \centering
    \includegraphics[width=\linewidth]{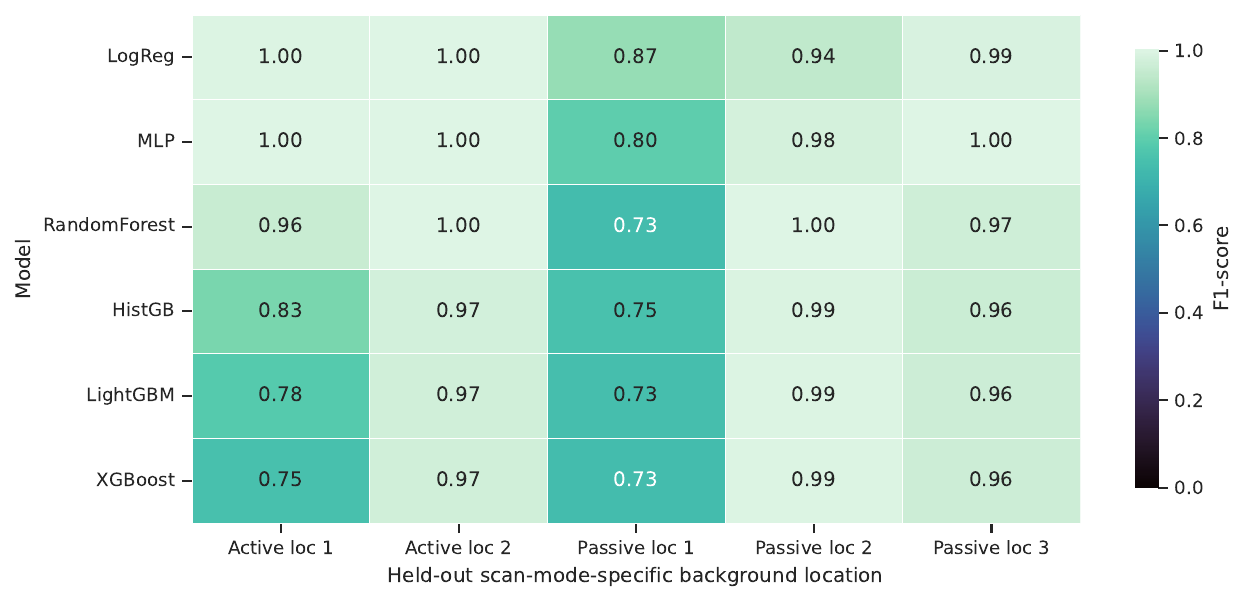}
    \caption{Per-model F1-score heatmap for scan-mode-specific held-out background-location evaluation. Columns are held-out benign background locations and rows are classifiers. The primary protocol uses class-balanced training and balanced held-out test sets.}
    \label{fig:background_location_withinmode_heatmap}
\end{figure}

\subsubsection{RF-chamber floor-control validation against chamber-domain shortcuts}

The RF-chamber floor control compares benign floor captures with chamber-collected malicious captures. Performance is saturated or near saturated across active, passive, and pooled scopes and remains so after noise removal, reducing the likelihood that matched results arise only from a coarse real-world-versus-chamber domain difference. The control does not establish transfer to unseen chamber geometries, jammer placements, or real-world jamming conditions.

\begin{figure}[!htbp]
    \centering
    \begin{subfigure}[t]{0.497\linewidth}
        \centering
        \includegraphics[width=\linewidth]{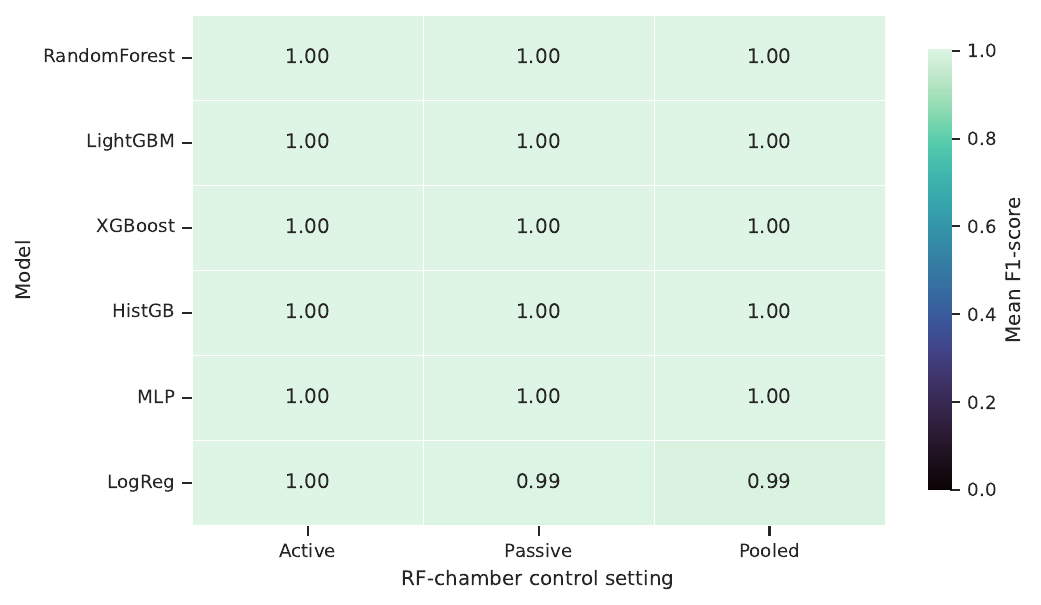}
        \caption{F1-score.}
    \end{subfigure}
    \hfill
    \begin{subfigure}[t]{0.483\linewidth}
        \centering
        \includegraphics[width=\linewidth]{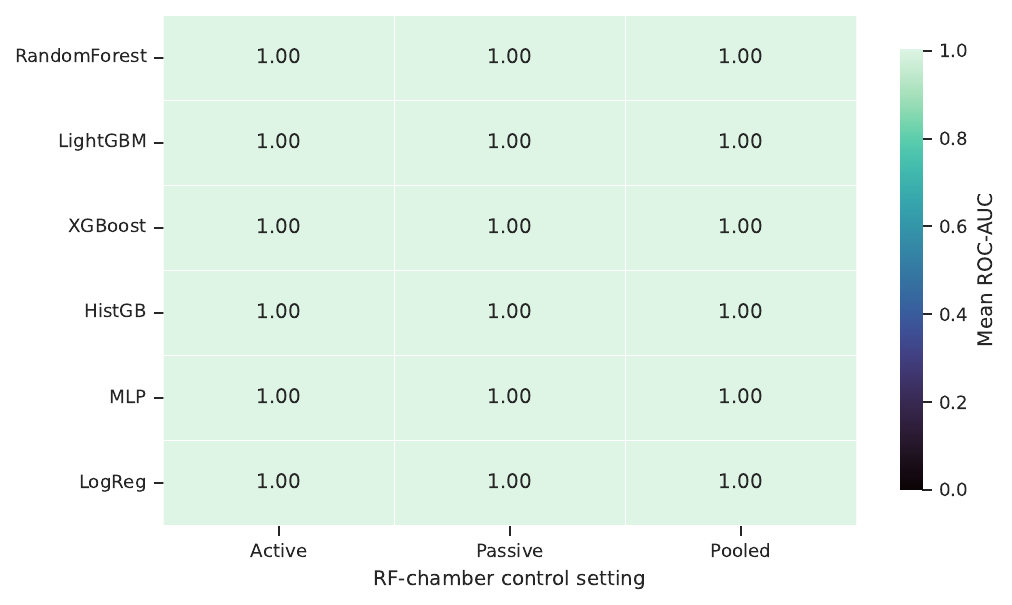}
        \caption{ROC-AUC.}
    \end{subfigure}
    \caption{Per-model RF-chamber floor-control results. The task uses only chamber-domain files and reports all six classifiers rather than selecting a test-set winner.}
    \label{fig:chamber_control_all_models}
\end{figure}

\subsubsection{Held-out single-tone score-threshold diagnostic}

Held-out single tone shows the clearest divergence between thresholded and ranking metrics. Several models retain high ranking performance despite near-zero fixed-threshold F1; RandomForest is the most extreme example, with ROC-AUC/AP~=~1.000 but F1~=~0.000. Fixed-threshold F1 here therefore reflects threshold transfer as much as ranking separability.

\subsection{Cross-scan-mode transfer}
\label{sec:cross_scan_transfer}

Cross-scan-mode transfer is substantially harder than matched classification. Active$\rightarrow$Passive F1 peaks around 0.44, while Passive$\rightarrow$Active reaches about 0.56 for LightGBM; ROC-AUC and AP also fall, confirming a genuine acquisition-domain shift. The MLP fails particularly strongly under Passive$\rightarrow$Active transfer, with a below-1\% predicted-positive rate and F1~=~0, unlike the high-AUC/low-F1 calibration pattern seen for single tone.

\begin{table}[!htbp]
    \centering
    \scriptsize
    \caption{Cross-scan-mode transfer results for all six classifiers using the canonical 63-feature representation. Values are mean $\pm$ standard deviation across three seeds.}
    \label{tab:cross_scan_all_models}
    \begin{tabularx}{\linewidth}{@{}LLCCC@{}}
        \toprule
        \textbf{Direction} & \textbf{Model} & \textbf{F1} & \textbf{ROC-AUC} & \textbf{AP} \\
        \midrule
        Active $\rightarrow$ Passive & LogReg & 0.426 $\pm$ 0.000 & 0.750 $\pm$ 0.000 & 0.311 $\pm$ 0.000 \\
         & RandomForest & 0.349 $\pm$ 0.040 & 0.632 $\pm$ 0.011 & 0.379 $\pm$ 0.019 \\
         & HistGB & 0.364 $\pm$ 0.000 & 0.706 $\pm$ 0.000 & 0.459 $\pm$ 0.000 \\
         & XGBoost & 0.439 $\pm$ 0.001 & 0.703 $\pm$ 0.002 & 0.444 $\pm$ 0.007 \\
         & LightGBM & 0.424 $\pm$ 0.000 & 0.745 $\pm$ 0.000 & 0.470 $\pm$ 0.000 \\
         & MLP & 0.415 $\pm$ 0.010 & 0.746 $\pm$ 0.031 & 0.367 $\pm$ 0.079 \\
        \midrule
        Passive $\rightarrow$ Active & LogReg & 0.000 $\pm$ 0.000 & 0.708 $\pm$ 0.000 & 0.388 $\pm$ 0.000 \\
         & RandomForest & 0.037 $\pm$ 0.032 & 0.861 $\pm$ 0.022 & 0.701 $\pm$ 0.052 \\
         & HistGB & 0.262 $\pm$ 0.070 & 0.793 $\pm$ 0.056 & 0.572 $\pm$ 0.114 \\
         & XGBoost & 0.197 $\pm$ 0.009 & 0.909 $\pm$ 0.008 & 0.702 $\pm$ 0.011 \\
         & LightGBM & 0.564 $\pm$ 0.000 & 0.897 $\pm$ 0.000 & 0.777 $\pm$ 0.000 \\
         & MLP & 0.000 $\pm$ 0.000 & 0.055 $\pm$ 0.013 & 0.140 $\pm$ 0.001 \\
        \bottomrule
    \end{tabularx}
\end{table}

The validation results should be interpreted within the collection design. Active malicious conditions contain comparatively few files per waveform--power combination, and residual sensitivity of higher-order file-level summaries to variable passive file length cannot be excluded even though row count and file size are not model features. Broader dataset-level limitations, including hardware specificity, sequential channel scanning, absence of explicit timestamps, and the controlled-chamber scope of malicious captures, are summarised in the Usage Notes section.

\section{Usage Notes}

\subsection{Loading and exploring the data}

The metadata manifest (\texttt{release\_artifacts/metadata/manifest.csv}) is the recommended entry point for selecting subsets by scan mode, label, band, channel, location, power, or waveform without loading raw files. Filtered raw files can then be loaded from \texttt{rf\_jamming/} using the manifest paths. For ready-to-use tabular analysis, the CSV and Parquet files under \texttt{release\_artifacts/derived/} provide the 96{,}090-row aggregate feature table used by the reported benchmarks. The public Kaggle notebooks demonstrate release preparation, benchmark reproduction, and figure generation.

\subsection{Recommended evaluation protocols}

File-level splits are required for valid generalisation evaluation. Row-level random splitting places highly correlated rows from the same recording session into both partitions simultaneously. Because observations within a single file are drawn from an identical experimental condition (same label, same location, same jammer power), a row-level split produces training and test sets that are effectively identically distributed, yielding performance estimates that do not reflect any meaningful generalisation. All splits must ensure that every observation from a given file is assigned exclusively to one partition.

The accompanying technical-validation notebook reconstructs the reported evaluation protocols from the released manifest and fixed random seeds and writes the corresponding per-run and summary result files during execution. Researchers who define custom splits should verify that no file contributes observations to more than one partition and should explicitly account for scan modality, frequency band, and held-out experimental conditions where relevant. Exact file-level assignments can be regenerated from the notebook using the same seed and split-construction logic.

\subsection{Active versus passive scan interpretation}

Active and passive scans should be treated as distinct acquisition domains rather than interchangeable samples. This distinction is reflected in the cross-scan-mode benchmark setting, where direct transfer between acquisition regimes is substantially more challenging than matched scan-mode evaluation (see Table~\ref{tab:cross_scan_all_models}). Models trained on active scans should therefore be interpreted as calibrated to the operating-channel-anchored active-scan context described in Section~\ref{sec:active_passive_scanning}, while models trained on passive scans should be interpreted as calibrated to the sequential spectrum-monitoring context.

Within active scans, waveform should be treated as a condition variable. Gaussian-noise and single-tone jamming produce different feature responses at the same nominal power, and the held-out waveform results show that ranking separability can persist even when the default decision threshold fails. Stronger cross-mode or cross-waveform transfer may be possible with scan-aware preprocessing, calibration, representation learning, or domain adaptation, but results should explicitly report scan modality, waveform condition, and preprocessing choices.

\subsection{Signal representation and scope}

The released files contain FFT-derived, driver-reported spectral-scan measurements. Each observation contains the eight fields described in Table~\ref{tab:raw_schema}, and the records preserve observation order but not explicit timestamps or measured inter-sample intervals. The dataset should therefore be interpreted as ordered multivariate spectral-scan sequences rather than uniformly timestamped multivariate time series. The release does not preserve phase, complex samples, precise timing, or the complete FFT-bin vector for each spectral report. It is consequently well suited to commodity-NIC spectrum monitoring and spectral-summary learning, but not to tasks requiring raw baseband access such as phase-sensitive channel estimation, custom demodulation, or waveform-level I/Q modelling.

\subsection{Recommended use cases and limitations}

The dataset supports RF jamming detection, interference classification, spectral anomaly detection, spectrum monitoring, and distribution-shift studies. Tasks may use binary benign/malicious labels or condition metadata such as benign subtype, location, jammer power, scan mode, and waveform. Benign chamber-floor or real-world backgrounds can also serve as reference distributions for one-class, semi-supervised, or out-of-distribution studies. Because the raw records contain no explicit timestamps, event-level temporal localisation should not be assumed.

Several limitations should be considered. The collection is tied to a specific CM4/QCA9880 receiver platform, Linux driver stack, HackRF One jammer, antenna configuration, and fixed chamber geometry, so models may require adaptation for other hardware or deployment environments. Channel observations are sequential rather than simultaneous, and the raw records preserve scan order without explicit timestamps; they should therefore not be interpreted as instantaneous wideband snapshots or uniformly sampled time series. This is especially relevant for mobile or rapidly changing environments, where the sensed state may evolve between channel visits. Passive file lengths are also more variable than active file lengths, and although row count and file size are not classifier features, higher-order file-level summaries can retain some sensitivity to the number of observations used to estimate them. Finally, all malicious captures were collected in the RF-shielded chamber and did not include a separate legitimate Wi-Fi traffic stream. The malicious labels therefore characterise controlled jammer-present spectral conditions rather than packet-level link degradation, and the release does not represent mobile jammers, multiple simultaneous jammers, reactive attacks, or uncontrolled real-world malicious deployments.

\subsection{Hardware replication notes and possible extensions}

Researchers wishing to replicate or extend the collection pipeline should account for the hardware and driver dependencies of the CM4/ath10k setup used here. For direct replication, a CM4 remains closest to the original sensing platform; however, if CM4 availability is constrained, a Compute Module~5 (CM5) or Raspberry Pi~5 host could be used as a practical replacement, provided that the selected Wi-Fi adapter and kernel stack expose the required spectral-scan interface. The same collection methodology may also be extended to newer Qualcomm Wi-Fi driver families where comparable spectral-scan functionality is exposed by the selected chipset, firmware, and kernel stack. In particular, the \texttt{ath11k} driver is documented for Qualcomm IEEE~802.11ax devices and exposes optional spectral-scan support through \texttt{CONFIG\_ATH11K\_SPECTRAL}~\cite{linux_ath11k,linux_ath11k_spectral}, while \texttt{ath12k} is documented as the Linux driver for Qualcomm Wi-Fi~7/IEEE~802.11be devices~\cite{linux_ath12k}. Future datasets collected on such platforms could extend the present release toward Wi-Fi~6/6E and Wi-Fi~7 deployments, including 6\,GHz operation where supported, and would enable studies of cross-hardware generalisation, wider-band interference observability, and continuity of spectral-scan analytics across Wi-Fi driver generations. These extensions should be validated empirically because spectral-scan availability, reported fields, FFT configuration, and timing behaviour may differ across chipsets, firmware versions, and kernel implementations.

\section{Data Availability}

WiFiSpectralJam is publicly available on Kaggle~\cite{WiFiSpectralJam_kaggle}. The dataset record contains the raw scans, manifest, derived feature tables, and validation summaries; notebooks and source code are provided in the Code Availability section.

\section{Code Availability}

The ath10k spectral-scan collection utility is available online.\footnote{Spectral-scan collection utility: \url{https://github.com/daniaherzalla/ath-spectral-scan}.} The WiFiSpectralJam preprocessing and analysis code is available on Kaggle.\footnote{Preprocessing code: \url{https://www.kaggle.com/code/daniaherzalla/manifest-validation-feature-extraction}.} The executable technical-validation and benchmark-reproduction notebook is also available on Kaggle,\footnote{Technical-validation: \url{https://www.kaggle.com/code/daniaherzalla/technical-validation-benchmark-reproduction}.} reproducing the validation checks, split logic, and benchmark outputs.

\section{Author Contributions}

D.H. and G.S. configured the CM4/QCA9880 spectral-sensing platform, HackRF One jammer, RF-shielded chamber setup, and acquisition tools. D.H. collected the dataset; developed the metadata, validation, and feature-extraction pipelines; led the design and implementation of the experiments; analysed the results; generated the figures and tables; prepared the dataset release; and wrote the manuscript. M.A. contributed to the design of technical-validation experiments. D.H., M.A., and G.S. contributed to interpretation of the results. D.H., G.S., W.T.L., and M.A. contributed to manuscript revision and editing. All authors approved the final manuscript.

\section{Competing Interests}

The authors declare no competing interests.


\clearpage
\FloatBarrier
\section*{Supplementary Information}
\label{sec:supplementary_information}
\setcounter{figure}{0}
\renewcommand{\thefigure}{S\arabic{figure}}
\setcounter{table}{0}
\renewcommand{\thetable}{S\arabic{table}}
\setcounter{subsection}{0}
\renewcommand{\thesubsection}{S\arabic{subsection}}

The Supplementary Information is organised into dataset documentation, extended signal-level validation, and extended benchmark diagnostics. It provides the repository folder structure, related-dataset survey, condition-level inventory, manifest field list, detailed spectral summaries, and additional diagnostics generated from the same validation pipeline used for the main manuscript.

\subsection{Repository Folder Structure}
\label{sec:supp_folder_structure}

The released Kaggle dataset follows the folder structure below. Raw captures are stored under \texttt{rf\_jamming/} and preserve the scan-mode and experimental-condition hierarchy, while reusable processed outputs are grouped under \texttt{release\_artifacts/}.

\begingroup
\renewcommand*\DTstyle{\ttfamily}
\setlength{\DTbaselineskip}{12pt}
\par\vspace{1.1em}

\dirtree{%
.1 dataset\_root/.
.2 rf\_jamming/.
.3 active\_scan/.
.4 benign/.
.5 background/.
.6 location1/.
.6 location2/.
.5 floor/.
.4 malicious/.
.5 gaussian\_noise/.
.6 0dbm/.
.6 10dbm/.
.6 neg10dbm/.
.6 neg40dbm/.
.5 singletone/.
.6 0dbm/.
.6 10dbm/.
.6 neg10dbm/.
.6 neg40dbm/.
.3 passive\_scan/.
.4 benign/.
.5 background/.
.6 location1/.
.6 location2/.
.6 location3/.
.5 floor/.
.4 malicious/.
.5 gaussian\_noise/.
.6 12dbm/.
.6 3dbm/.
.6 6dbm/.
.6 9dbm/.
.2 release\_artifacts/.
.3 metadata/.
.4 manifest.csv.
.3 derived/.
.4 file\_level\_features.csv.
.4 file\_level\_features.parquet.
.3 validation/.
}
\endgroup

\begin{landscape}
\begin{table}[p]
\centering
\fontsize{7.0}{7.8}\selectfont
\setlength{\tabcolsep}{1.8pt}
\setlength{\extrarowheight}{0.7pt}
\renewcommand{\arraystretch}{1.10}

\begin{minipage}{\dimexpr\paperheight-2in\relax}
\caption{Characteristics of selected adversarial-jamming datasets and experimental benchmarks relevant to the present work. Entries are separated into non-Wi-Fi or cross-technology jamming benchmarks and Wi-Fi-specific jamming studies. The table reports only attributes supported by the cited source; dashes indicate information not reported or not established from the cited publication.}
\label{tab:related_datasets_grouped}

\resizebox{\dimexpr\paperheight-2in\relax}{!}{%
\begin{tabular}{@{}
>{\centering\arraybackslash}p{0.90cm}
>{\raggedright\arraybackslash}p{2.15cm}
>{\centering\arraybackslash}p{0.70cm}
>{\raggedright\arraybackslash}p{1.55cm}
>{\raggedright\arraybackslash}p{1.25cm}
>{\raggedright\arraybackslash}p{2.35cm}
>{\raggedright\arraybackslash}p{1.15cm}
>{\raggedright\arraybackslash}p{1.15cm}
>{\raggedright\arraybackslash}p{4.15cm}
>{\raggedright\arraybackslash}p{4.35cm}
>{\centering\arraybackslash}p{0.85cm}
>{\centering\arraybackslash}p{0.90cm}
@{}}
\toprule
\textbf{Category} &
\textbf{Dataset / study} &
\textbf{Year} &
\textbf{Scope} &
\textbf{Signal Rep.} &
\textbf{Capture Hardware} &
\textbf{Freq. Band} &
\textbf{Size} &
\textbf{Observables} &
\textbf{Interference / Scenario} &
\textbf{Format} &
\textbf{Access} \\
\midrule

\multirow{4}{*}{\rotatebox{90}{\makecell[c]{Other / cross-technology}}}
& Pu\~nal \textit{et al.}\ / CRAWDAD~\cite{punal2012crawdad,punal2014crawdad}
& 2012--14
& 802.11p VANET jamming
& Packet/link traces
& Commodity 802.11p vehicular hardware
& 5.9\,GHz
& $\sim$1.3\,GB
& Packet-level traces including timing/sequence information and received-signal measurements
& Experimental vehicular jamming traces spanning multiple jammer behaviours
& PCAP
& Gated \\

& Hussain \textit{et al.}~\cite{hussain2022edgeaijamming}
& 2022
& IoT/Wi-Fi jamming detection
& Scalar RSS
& Raspberry Pi + software-defined radio
& Wi-Fi
& --
& Received-signal-strength measurements
& Normal, constant-jamming, and periodic-jamming conditions
& --
& Open \\

& Li \textit{et al.}~\cite{li2023uavjammingdetection}
& 2022
& UAV/OFDM jamming detection and classification
& PHY features; spectrograms
& COEX Clover UAV platform + HackRF One / GNU Radio
& 2.4\,GHz
& 23{,}565 samples
& Eight OFDM/PHY-derived features and spectrogram representations
& Barrage, single-tone, successive-pulse, and protocol-aware jamming
& --
& Open \\

& Alhazbi \textit{et al.}~\cite{alhazbi2023indoorjammingdataset}
& 2023
& Indoor physical-layer jamming
& Raw I/Q
& USRP X310 ($\times$7); BPSK link
& --
& --
& Raw I/Q samples under jammer-present and jammer-absent conditions
& Tone and Gaussian-noise jamming with variable transmitter--receiver geometry and jammer power
& MAT
& Open \\

\midrule

\multirow{4}{*}{\rotatebox{90}{\makecell[c]{Wi-Fi jamming }}}
& Davaslioglu \textit{et al.}\ (DeepWiFi)~\cite{davaslioglu2019deepwifi}
& 2021
& Wi-Fi jamming detection
& Simulated I/Q
& MATLAB WLAN Toolbox; 802.11ac TGac channel models
& 5.25\,GHz
& 12{,}000 samples
& Synthetic 802.11ac VHT I/Q frames across six TGac channel models; idle/Wi-Fi/jammer classification
& Probabilistic, sensing-based, and adaptive jamming under simulated channel fading
& MAT
& Not released \\

& Ali \textit{et al.}~\cite{ali2024rfjammingdataset,ali2022rfjammingdataport}
& 2022--24
& Wi-Fi jamming characterisation
& Commodity-NIC spectral scan
& CM4 / QCA9880-class Wi-Fi hardware + HackRF One / JamRF
& 5\,GHz
& 6{,}750 scans
& Frequency, noise, maximum magnitude, total gain, base power, RSSI/received-power, relative power, and average power
& Controlled Wi-Fi jamming conditions across multiple jammer behaviours and bandwidth configurations; limited benign reference
& CSV
& Gated \\

& JamShield~\cite{panitsas2025jamshield}
& 2024--25
& Wi-Fi jamming detection
& Cross-layer KPIs
& ASUS RT-AX88U Pro + USRP X310
& 2.4\,GHz; AP on ch.\,6
& 39{,}861 samples
& PHY/MAC/network KPIs; 40 collected features with 20 selected for modelling
& Constant, random, and reactive jamming using multiple jammer waveforms and channel/gain settings
& CSV
& Open \\

\cmidrule(lr){2-12}

& \textbf{WiFiSpectralJam (ours)}
& \textbf{2023-24}
& \textbf{Wi-Fi jamming detection}
& \textbf{Commodity-NIC spectral scan}
& \textbf{CM4 / QCA9880 + HackRF One}
& \textbf{2.4/5\,GHz}
& \textbf{14.52\,GB; 96{,}090 scans}
& \textbf{Frequency, noise, max magnitude, gain, base power, RSSI, relative power, average power}
& \textbf{Continuous Gaussian-noise and single-tone jamming; multiple powers and target channels; waveform variation in active subset}
& \textbf{CSV}
& \textbf{Open} \\

\bottomrule
\end{tabular}%
}
\end{minipage}
\end{table}
\end{landscape}

\subsection{Condition-level inventory}

Supplementary Table~\ref{tab:supp_condition_breakdown} provides the complete condition-level inventory referenced from the Data Records and Technical Validation sections.

\begin{table}[!p]
    \centering
    \compacttable
    \caption{Condition-level file, observation, and storage breakdown for benign and malicious subsets. Percentages are relative to the full 96{,}090-file dataset. Background locations correspond to real-world indoor captures; floor and all malicious captures were collected inside the RF-shielded chamber. Active malicious captures target 5805\,MHz; passive malicious captures use Gaussian-noise jamming across the passive target-channel set. Power values are in dBm.}\label{tab:supp_condition_breakdown}
    \begin{tabularx}{\linewidth}{@{}>{\hsize=0.42\hsize}L>{\hsize=0.52\hsize}L>{\hsize=0.72\hsize}L>{\hsize=1.10\hsize}L>{\hsize=0.46\hsize}C>{\hsize=0.46\hsize}C>{\hsize=0.88\hsize}C>{\hsize=0.56\hsize}R@{}}
        \toprule
        \textbf{Label} & \textbf{Scan mode} & \textbf{Interference} & \textbf{Category} & \textbf{Files} & \textbf{Percent} & \textbf{Observations} & \textbf{Size (MB)} \\
        \midrule
        \multirow{8}{*}{Benign}
         & \multirow{3}{*}{Active}
         & \multirow{2}{*}{Background} & location1 (high-activity) & 1{,}333 & 1.39\% & 39{,}990{,}000 & 1{,}097.9 \\
         &                               &                             & location2 (low-activity) & 1{,}524 & 1.59\% & 45{,}691{,}308 & 1{,}266.4 \\
         &                               & Floor                       & RF chamber & 404 & 0.42\% & 12{,}120{,}000 & 337.8 \\
        \cmidrule(lr){2-8}
         & \multirow{4}{*}{Passive}
         & \multirow{3}{*}{Background} & location1 (high-activity) & 18{,}188 & 18.93\% & 67{,}371{,}597 & 1{,}851.9 \\
         &                               &                             & location2 (high-activity) & 17{,}792 & 18.52\% & 65{,}368{,}570 & 1{,}806.3 \\
         &                               &                             & location3 (low-activity) & 17{,}735 & 18.46\% & 116{,}202{,}997 & 3{,}280.0 \\
         &                               & Floor                       & RF chamber & 20{,}432 & 21.26\% & 79{,}236{,}985 & 2{,}162.0 \\
        \cmidrule(lr){2-8}
         & \multicolumn{3}{l}{\textit{Benign total}} & \textit{77{,}408} & \textit{80.56\%} & \textit{425{,}981{,}457} & \textit{11{,}802.2} \\
        \midrule
        \multirow{13}{*}{Malicious}
         & \multirow{8}{*}{Active}
         & \multirow{4}{*}{Gaussian noise} & $-40$ dBm & 101 & 0.11\% & 3{,}030{,}000 & 84.6 \\
         &                                  &                & $-10$ dBm & 102 & 0.11\% & 3{,}060{,}000 & 88.0 \\
         &                                  &                & $0$ dBm & 153 & 0.16\% & 4{,}590{,}000 & 132.4 \\
         &                                  &                & $+10$ dBm & 101 & 0.11\% & 3{,}030{,}000 & 86.9 \\
        \cmidrule(lr){3-8}
         &                                  & \multirow{4}{*}{Single tone} & $-40$ dBm & 101 & 0.11\% & 3{,}030{,}000 & 86.2 \\
         &                                  &                & $-10$ dBm & 205 & 0.21\% & 6{,}150{,}000 & 180.8 \\
         &                                  &                & $0$ dBm & 104 & 0.11\% & 3{,}120{,}000 & 92.7 \\
         &                                  &                & $+10$ dBm & 230 & 0.24\% & 6{,}900{,}000 & 205.5 \\
        \cmidrule(lr){2-8}
         & \multirow{4}{*}{Passive}
         & \multirow{4}{*}{Gaussian noise} & $+3$ dBm & 4{,}088 & 4.25\% & 14{,}605{,}665 & 403.1 \\
         &                                  &                & $+6$ dBm & 4{,}211 & 4.38\% & 15{,}156{,}003 & 418.5 \\
         &                                  &                & $+9$ dBm & 5{,}228 & 5.44\% & 19{,}698{,}681 & 543.2 \\
         &                                  &                & $+12$ dBm & 4{,}058 & 4.22\% & 14{,}419{,}324 & 398.1 \\
        \cmidrule(lr){2-8}
         & \multicolumn{3}{l}{\textit{Malicious total}} & \textit{18{,}682} & \textit{19.44\%} & \textit{96{,}789{,}673} & \textit{2{,}720.0} \\
        \midrule
        \multicolumn{4}{@{}l}{\textbf{Dataset total}} & \textbf{96{,}090} & \textbf{100.00\%} & \textbf{522{,}771{,}130} & \textbf{14{,}522.2} \\
        \bottomrule
    \end{tabularx}
\end{table}

\subsection{Manifest field list}

Supplementary Table~\ref{tab:supp_manifest_fields} lists the main fields included in \texttt{release\_artifacts/metadata/manifest.csv}. Structurally inapplicable fields are recorded as \texttt{N/A} rather than inferred.

\begin{table}[!htbp]
    \centering
    \compacttable
    \caption{Metadata manifest fields. The manifest contains one row per raw spectral-scan file and provides the file paths, parsed labels, condition metadata, and validation fields used to construct dataset subsets and reproduce the reported splits.}\label{tab:supp_manifest_fields}
    \begin{tabularx}{\linewidth}{@{}>{\hsize=0.55\hsize}L>{\hsize=1.45\hsize}L@{}}
        \toprule
        \textbf{Column} & \textbf{Description} \\
        \midrule
        \texttt{file\_path} & Relative path to the raw CSV file. \\
        \texttt{scan\_mode} & Scan modality: active or passive. \\
        \texttt{label} & Class label: benign or malicious. \\
        \texttt{benign\_subtype} & Benign subtype, either floor or background, for benign files. \\
        \texttt{location} & Location identifier for real-world background files. \\
        \texttt{waveform} & Jamming waveform for malicious files, either Gaussian noise or single tone. \\
        \texttt{power\_dbm} & Jamming transmit power for malicious files. \\
        \texttt{channel\_mhz} & Jammed centre frequency for malicious files. \\
        \texttt{band} & Wi-Fi band associated with the file or target channel: 2.4\,GHz or 5\,GHz. \\
        \texttt{collection\_environment} & Collection environment, either RF chamber or real-world. \\
        \texttt{num\_rows} & Number of spectral-scan observations in the raw CSV file. \\
        \texttt{num\_features} & Number of raw spectral/FFT columns in the file. \\
        \texttt{metadata\_status} & Metadata completeness status used by the validation pipeline. \\
        \bottomrule
    \end{tabularx}
\end{table}

\subsection{Detailed file-level spectral metric statistics}

Supplementary Table~\ref{tab:supp_file_level_metric_statistics_detailed} expands the compact metric summaries in Tables~\ref{tab:file_level_metric_statistics} and~\ref{tab:file_level_metric_delta}. For each raw CSV file, the relevant spectral field is first reduced to a per-file mean; the table then summarises those per-file means within each scan-mode and class group. The median, interquartile range (IQR), and extrema are included to show the degree of overlap and skew that is hidden by mean--standard-deviation summaries alone.

\begin{table}[p]
    \centering
    \scriptsize
    \setlength{\tabcolsep}{4pt}
    \renewcommand{\arraystretch}{1.08}
    \caption{Detailed file-level spectral metric statistics by scan mode, metric, and class label. Each row summarises the distribution across files of the corresponding per-file mean spectral field. RSSI is reported using the derived dBm conversion. IQR denotes the interquartile range. The table shows that power- and magnitude-related fields exhibit larger shifts under jamming than the driver-reported noise-floor field, which remains comparatively stable and should be interpreted as a calibration-related baseline rather than a standalone jamming indicator.}
    \label{tab:supp_file_level_metric_statistics_detailed}

    \begin{tabularx}{0.97\textwidth}{
        @{}
        >{\hsize=0.42\hsize\centering\arraybackslash}X
        >{\hsize=1.99\hsize\raggedright\arraybackslash}X
        >{\hsize=0.85\hsize\raggedright\arraybackslash}X
        >{\hsize=0.95\hsize\centering\arraybackslash}X
        >{\hsize=0.95\hsize\centering\arraybackslash}X
        >{\hsize=0.95\hsize\centering\arraybackslash}X
        >{\hsize=0.95\hsize\centering\arraybackslash}X
        >{\hsize=0.95\hsize\centering\arraybackslash}X
        >{\hsize=0.95\hsize\centering\arraybackslash}X
        >{\hsize=0.95\hsize\centering\arraybackslash}X
        @{}
    }
        \toprule
        \textbf{Scan mode} & \textbf{Metric} & \textbf{Label} &
        \textbf{Files} & \textbf{Mean} & \textbf{Std.} &
        \textbf{Median} & \textbf{IQR} & \textbf{Min.} & \textbf{Max.} \\
        \midrule

        \multirow{14}{*}{\rotatebox[origin=c]{90}{Active}}
        & \multirow{2}{*}{RSSI (dBm)} & Benign
        & 3,261 & -88.06 & 2.06 & -86.94 & 4.30 & -91.07 & -79.66 \\
        & & Malicious
        & 1,097 & -73.05 & 10.50 & -72.26 & 21.13 & -85.80 & -52.00 \\
        \cmidrule(lr){2-10}

        & \multirow{2}{*}{Noise floor (dBm)} & Benign
        & 3,261 & -104.61 & 2.53 & -102.93 & 5.14 & -108.43 & -102.00 \\
        & & Malicious
        & 1,097 & -102.00 & 0.00 & -102.00 & 0.00 & -102.00 & -102.00 \\
        \cmidrule(lr){2-10}

        & \multirow{2}{*}{Average power (dB)} & Benign
        & 3,261 & 54.18 & 2.49 & 55.19 & 4.99 & 48.20 & 57.53 \\
        & & Malicious
        & 1,097 & 54.91 & 3.60 & 55.78 & 6.93 & 47.77 & 58.56 \\
        \cmidrule(lr){2-10}

        & \multirow{2}{*}{Base power (dB)} & Benign
        & 3,261 & 381.88 & 4.24 & 380.69 & 8.10 & 376.73 & 394.35 \\
        & & Malicious
        & 1,097 & 394.05 & 11.36 & 395.87 & 23.48 & 378.10 & 413.89 \\
        \cmidrule(lr){2-10}

        & \multirow{2}{*}{Max magnitude (linear)} & Benign
        & 3,261 & 47.31 & 2.94 & 47.21 & 4.06 & 38.34 & 67.85 \\
        & & Malicious
        & 1,097 & 108.72 & 65.31 & 76.57 & 138.01 & 37.03 & 182.07 \\
        \cmidrule(lr){2-10}

        & \multirow{2}{*}{Relative power (dB)} & Benign
        & 3,261 & 9.49 & 0.13 & 9.44 & 0.18 & 9.12 & 10.56 \\
        & & Malicious
        & 1,097 & 11.95 & 2.26 & 11.55 & 4.34 & 9.03 & 15.27 \\
        \cmidrule(lr){2-10}

        & \multirow{2}{*}{Total gain (dB)} & Benign
        & 3,261 & 81.42 & 6.61 & 80.80 & 13.52 & 66.18 & 89.73 \\
        & & Malicious
        & 1,097 & 66.45 & 11.36 & 64.63 & 23.47 & 46.60 & 82.40 \\

        \midrule

        \multirow{14}{*}{\rotatebox[origin=c]{90}{Passive}}
        & \multirow{2}{*}{RSSI (dBm)} & Benign
        & 74,147 & -86.86 & 3.95 & -87.88 & 3.97 & -91.53 & -55.75 \\
        & & Malicious
        & 17,585 & -83.71 & 6.77 & -89.27 & 13.04 & -91.05 & -69.54 \\
        \cmidrule(lr){2-10}

        & \multirow{2}{*}{Noise floor (dBm)} & Benign
        & 74,147 & -102.05 & 3.48 & -102.44 & 3.34 & -106.03 & -95.00 \\
        & & Malicious
        & 17,585 & -102.68 & 2.75 & -104.46 & 5.11 & -105.88 & -95.67 \\
        \cmidrule(lr){2-10}

        & \multirow{2}{*}{Average power (dB)} & Benign
        & 74,147 & 56.39 & 1.88 & 57.18 & 2.52 & 47.68 & 58.40 \\
        & & Malicious
        & 17,585 & 55.96 & 1.97 & 56.13 & 3.17 & 47.92 & 58.34 \\
        \cmidrule(lr){2-10}

        & \multirow{2}{*}{Base power (dB)} & Benign
        & 74,147 & 380.66 & 4.32 & 380.06 & 6.16 & 374.71 & 411.63 \\
        & & Malicious
        & 17,585 & 385.46 & 9.08 & 379.41 & 18.03 & 375.04 & 401.74 \\
        \cmidrule(lr){2-10}

        & \multirow{2}{*}{Max magnitude (linear)} & Benign
        & 74,147 & 52.57 & 9.09 & 50.70 & 9.87 & 35.31 & 119.58 \\
        & & Malicious
        & 17,585 & 52.16 & 4.16 & 52.03 & 5.62 & 39.96 & 64.15 \\
        \cmidrule(lr){2-10}

        & \multirow{2}{*}{Relative power (dB)} & Benign
        & 74,147 & 11.16 & 2.21 & 10.76 & 3.80 & 8.55 & 25.33 \\
        & & Malicious
        & 17,585 & 12.20 & 3.55 & 9.36 & 5.77 & 8.93 & 19.75 \\
        \cmidrule(lr){2-10}

        & \multirow{2}{*}{Total gain (dB)} & Benign
        & 74,147 & 80.16 & 7.10 & 80.86 & 10.09 & 45.17 & 88.37 \\
        & & Malicious
        & 17,585 & 75.84 & 11.59 & 83.78 & 22.97 & 57.96 & 88.13 \\

        \bottomrule
    \end{tabularx}
\end{table}

The detailed table reinforces the interpretation in the main text. In the active subset, the malicious RSSI distribution is shifted upward by about 15~dB relative to benign files, and the maximum-magnitude distribution is both larger and much more variable, indicating a strong jammer-induced spectral signature. Active total gain shifts downward, which is consistent with stronger received energy causing the receiver chain to operate at lower gain. In the passive subset, the malicious and benign distributions overlap more strongly: the RSSI mean increases, but the malicious median remains close to the lower end of the benign range, and several metrics show small or opposite-signed mean shifts. This supports treating passive detection as the more heterogeneous setting and motivates the separate passive-only and held-out-condition evaluations reported in the benchmark section.

\subsection{Additional frequency-bin heatmaps}

\begin{figure}[!htbp]
    \centering
    \captionsetup[subfigure]{font=scriptsize,skip=1pt}
    \begin{subfigure}[t]{0.49\linewidth}
        \centering
        \includegraphics[width=\linewidth]{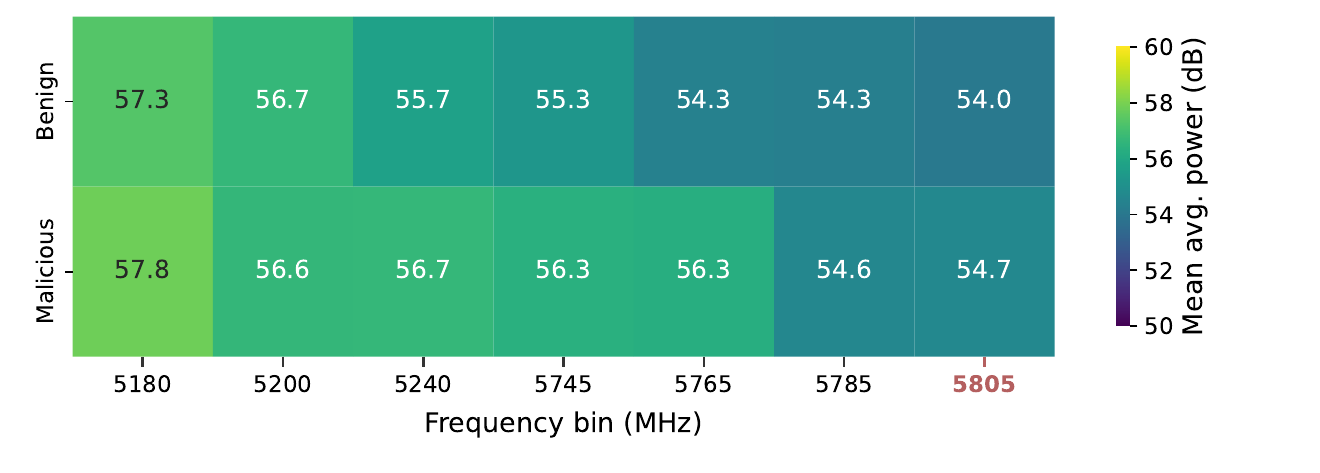}
        \caption{Active, average power.}
    \end{subfigure}\hfill
    \begin{subfigure}[t]{0.49\linewidth}
        \centering
        \includegraphics[width=\linewidth]{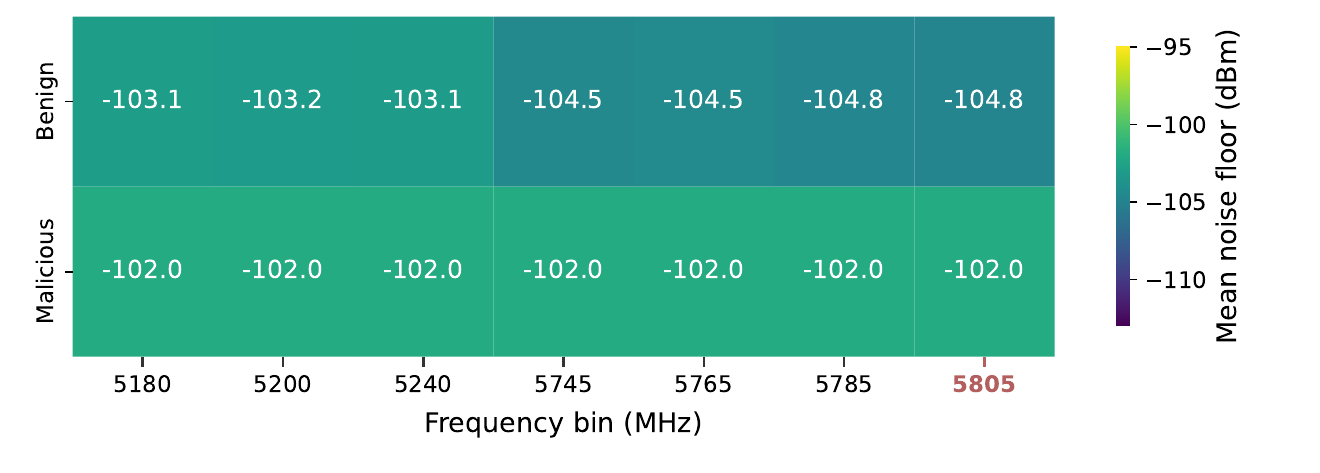}
        \caption{Active, noise floor.}
    \end{subfigure}

    \vspace{0.20em}
    \begin{subfigure}[t]{0.50\linewidth}
        \centering
        \includegraphics[width=\linewidth]{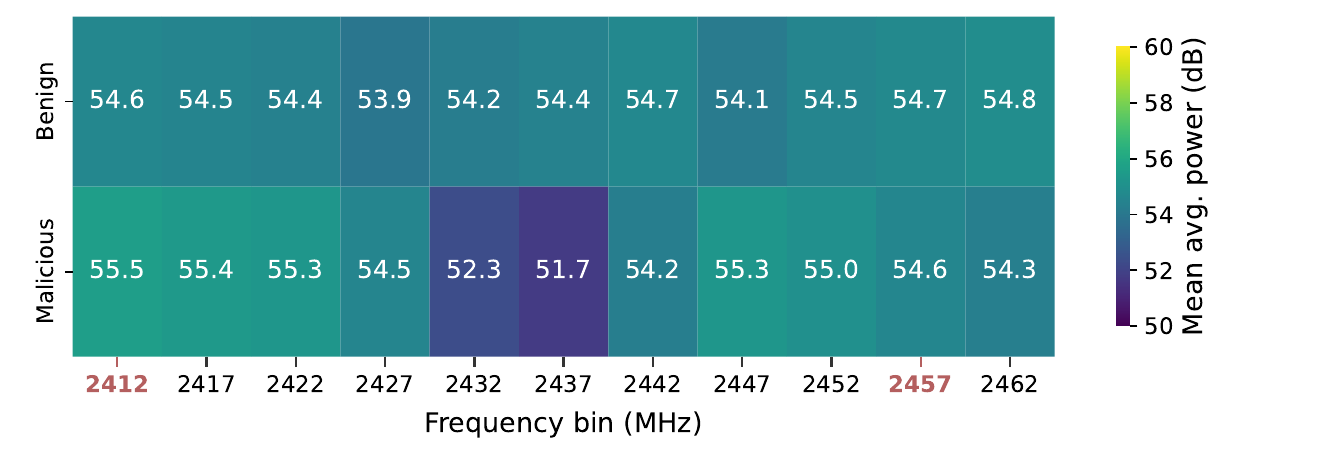}
        \caption{Passive 2.4\,GHz, average power.}
    \end{subfigure}\hfill
    \begin{subfigure}[t]{0.50\linewidth}
        \centering
        \includegraphics[width=\linewidth]{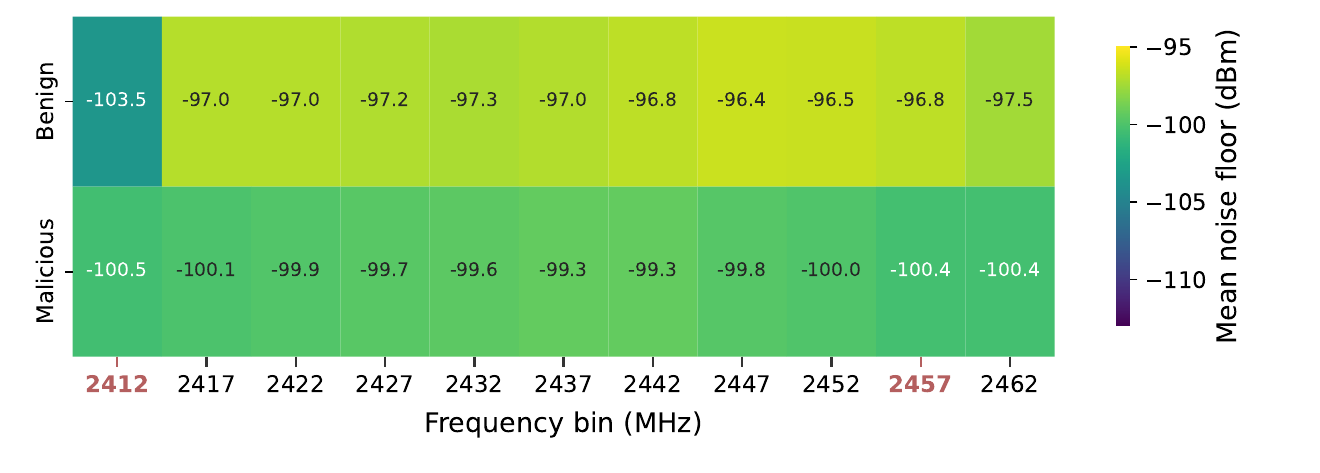}
        \caption{Passive 2.4\,GHz, noise floor.}
    \end{subfigure}

    \vspace{0.20em}
    \begin{subfigure}[t]{0.50\linewidth}
        \centering
        \includegraphics[width=\linewidth]{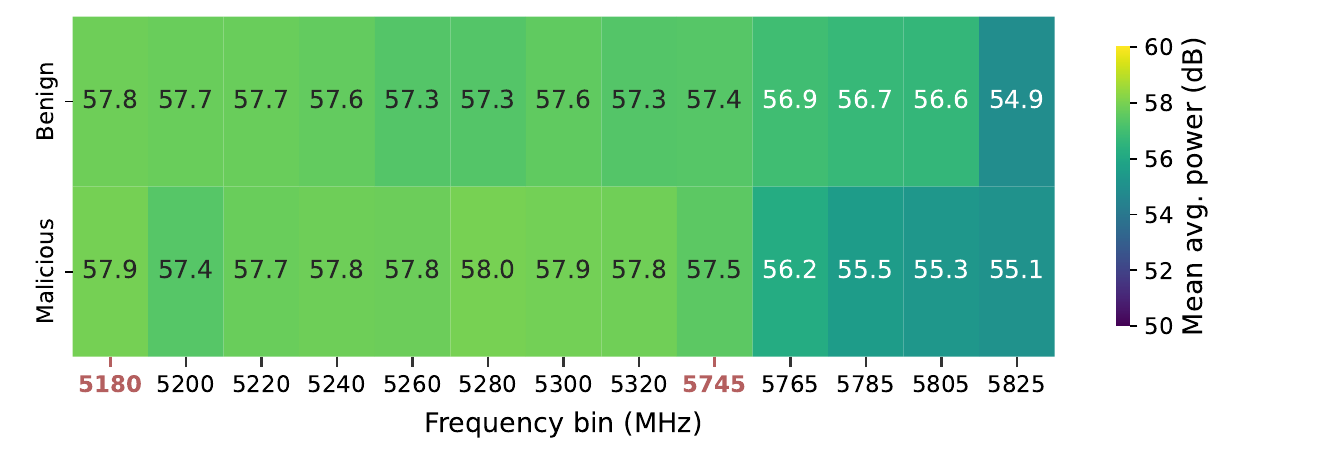}
        \caption{Passive 5\,GHz, average power.}
    \end{subfigure}\hfill
    \begin{subfigure}[t]{0.50\linewidth}
        \centering
        \includegraphics[width=\linewidth]{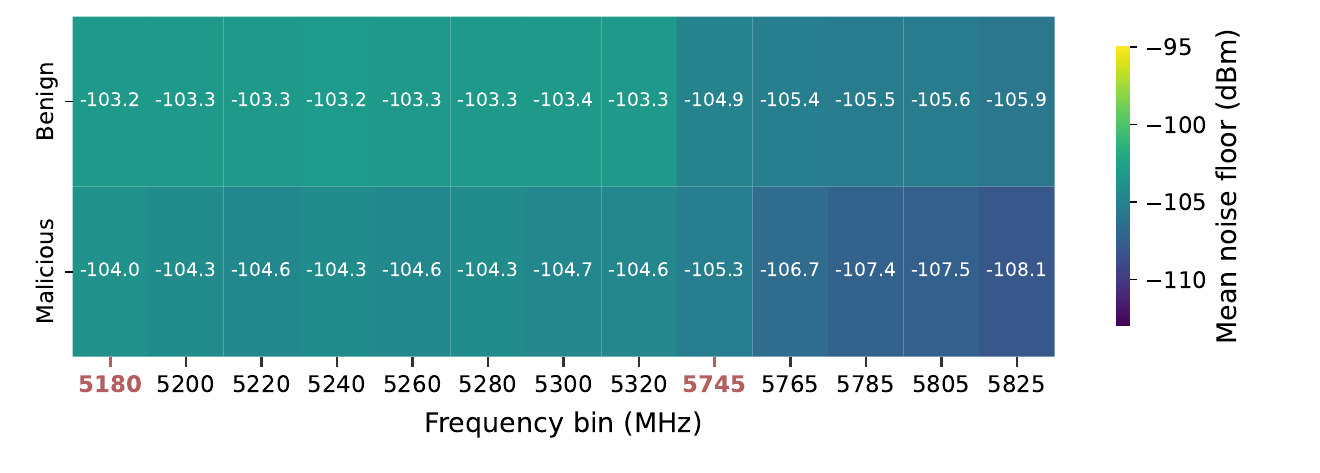}
        \caption{Passive 5\,GHz, noise floor.}
    \end{subfigure}
    \caption{Supplementary frequency-bin validation heatmaps for average power and noise floor. Cells report class-wise file-level means by reported \texttt{freq1} centre frequency. Active-scan malicious captures target 5805\,MHz. Passive-scan malicious captures include Gaussian-noise jamming at 2412 and 2457\,MHz in the 2.4\,GHz band and at 5180 and 5745\,MHz in the 5\,GHz band.}
    \label{fig:supp_frequency_bin_validation_additional}
\end{figure}

Supplementary Figure~\ref{fig:supp_frequency_bin_validation_additional} extends the main frequency-bin validation heatmaps by reporting average power and noise floor across the same scan-mode and band partitions. As in the main heatmaps, each cell is computed from file-level frequency-bin means: observations are first averaged within each rounded \texttt{freq1} bin inside each sampled file, and those file-level values are then averaged across files within each class label. These panels are provided as supplementary diagnostics because they complement the RSSI and maximum-magnitude views in Fig.~\ref{fig:frequency_bin_validation} without changing the main interpretation.

\subsection{Benchmark Classifier Configuration}
\label{sec:supp_classifier_configuration}

Table~\ref{tab:supp_classifier_configuration} reports the preprocessing and hyperparameter configurations used for the six baseline classifiers. Parameters not listed in the table retain the corresponding library defaults. All experiments used random seeds 1, 7, and 42, and binary predictions were obtained using a decision threshold of 0.5.

\begin{table}[!htbp]
    \centering
    \footnotesize
    \setlength{\tabcolsep}{4pt}
    \renewcommand{\arraystretch}{1.15}
    \caption{Preprocessing and hyperparameter configurations used for the baseline classifiers. Median imputation was applied to all models. Standardisation was applied only to logistic regression and the multilayer perceptron. Parameters not listed retain their corresponding library defaults.}
    \label{tab:supp_classifier_configuration}

    \begin{tabularx}{\linewidth}{@{}p{0.12\linewidth}p{0.28\linewidth}X@{}}
        \toprule
        \textbf{Classifier}
        & \textbf{Preprocessing}
        & \textbf{Configuration} \\
        \midrule

        LogReg
        & Median imputation; standardisation
        & Maximum iterations: 3{,}000; seed-specific \texttt{random\_state}. \\

        RandomForest
        & Median imputation
        & 300 trees; \texttt{class\_weight=balanced\_subsample}; unrestricted tree depth; seed-specific \texttt{random\_state}. \\

        HistGB
        & Median imputation
        & Scikit-learn default histogram-gradient-boosting configuration; seed-specific \texttt{random\_state}. \\

        XGBoost
        & Median imputation
        & 300 trees; maximum depth 6; learning rate 0.05; row subsampling 0.9; column subsampling 0.9; binary logistic objective; log-loss evaluation metric; histogram tree method. \\

        LightGBM
        & Median imputation
        & 300 trees; learning rate 0.05; balanced class weighting; seed-specific \texttt{random\_state}. \\

        MLP
        & Median imputation; standardisation
        & Hidden layers of 256, 128, and 64 units with ReLU activation; batch normalisation after the first two hidden layers; dropout rates of 0.30, 0.25, and 0.20; Adam optimiser with learning rate $10^{-3}$; binary cross-entropy loss; maximum 25 epochs; batch size 256; validation fraction 0.15; early-stopping patience of four epochs with restoration of the best weights. \\

        \bottomrule
    \end{tabularx}
\end{table}

\subsection{Stratified spectral summaries}

The main text summarises the power- and waveform-stratified spectral response in prose and in compact feature-distribution diagnostics. The supplementary figures below provide the full spectral-response validation set for RSSI, base power, maximum magnitude, and noise floor. These views are intentionally retained because different spectral fields respond differently to jamming: some show clear changes at jammed frequencies, while others show weak or compressed responses that are also informative for dataset reuse. In all panels, the x-axis denotes scanned centre frequency; dashed vertical lines indicate jammed frequencies. Values at non-target centre frequencies correspond to measurements collected at those scanned frequencies while the relevant jamming condition was active, not to additional jammed channels.

\begin{figure}[!htbp]
    \centering
    \begin{subfigure}[t]{0.49\linewidth}\centering\includegraphics[width=\linewidth]{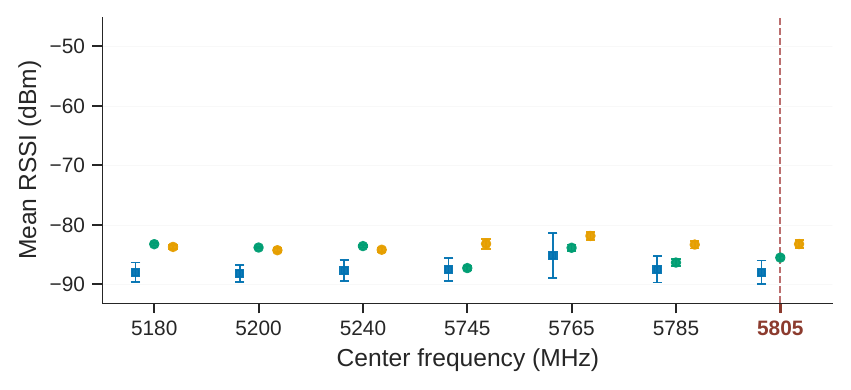}\caption{$-40$\,dBm.}\end{subfigure}\hfill
    \begin{subfigure}[t]{0.49\linewidth}\centering\includegraphics[width=\linewidth]{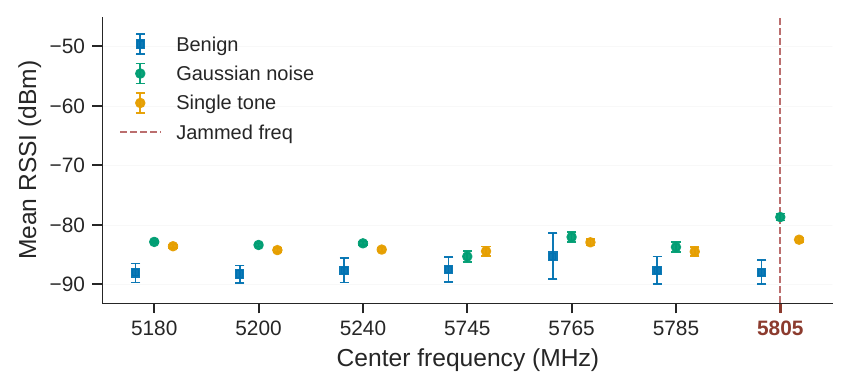}\caption{$-10$\,dBm.}\end{subfigure}

    \vspace{0.45em}
    \begin{subfigure}[t]{0.49\linewidth}\centering\includegraphics[width=\linewidth]{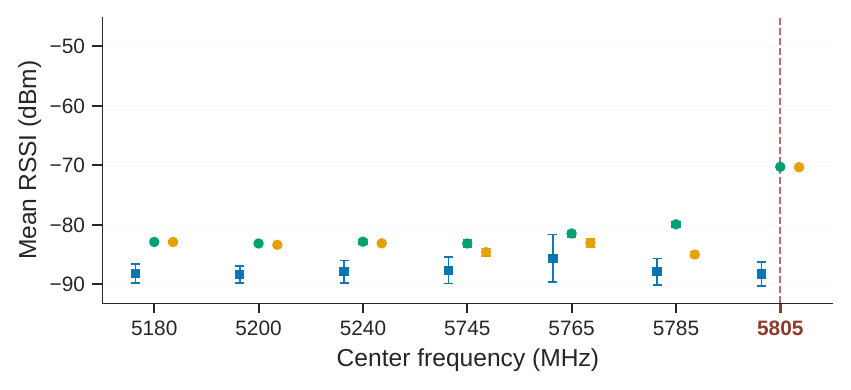}\caption{$0$\,dBm.}\end{subfigure}\hfill
    \begin{subfigure}[t]{0.49\linewidth}\centering\includegraphics[width=\linewidth]{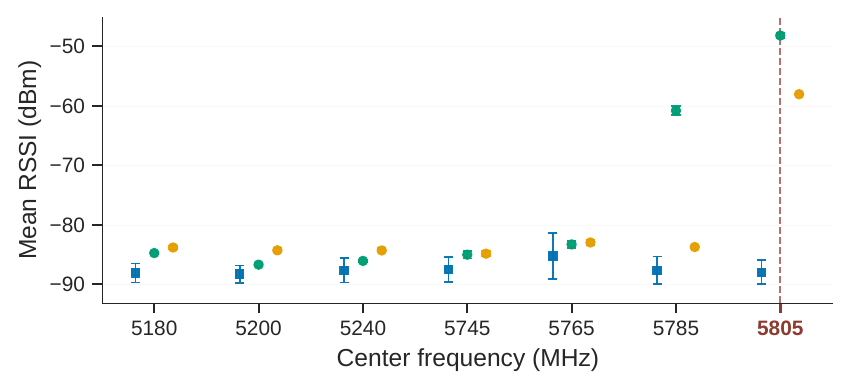}\caption{$+10$\,dBm.}\end{subfigure}
    \caption{Active-scan waveform comparison using RSSI, stratified by transmit power.}
    \label{fig:supp_active_waveform_stratified_rssi}
\end{figure}

\begin{figure}[!htbp]
    \centering
    \begin{subfigure}[t]{0.49\linewidth}\centering\includegraphics[width=\linewidth]{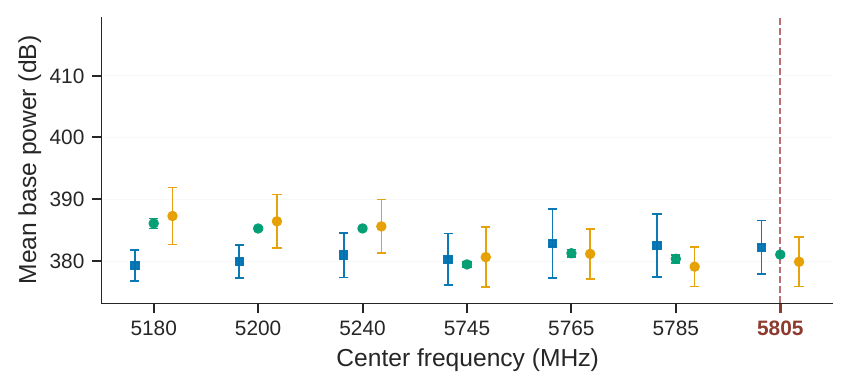}\caption{$-40$\,dBm.}\end{subfigure}\hfill
    \begin{subfigure}[t]{0.49\linewidth}\centering\includegraphics[width=\linewidth]{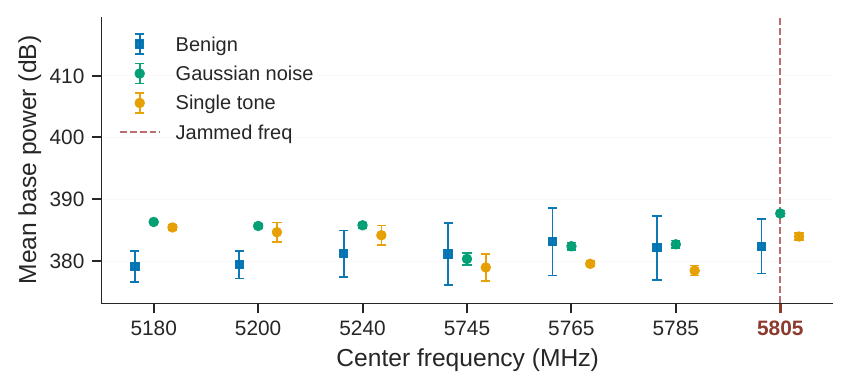}\caption{$-10$\,dBm.}\end{subfigure}

    \vspace{0.45em}
    \begin{subfigure}[t]{0.49\linewidth}\centering\includegraphics[width=\linewidth]{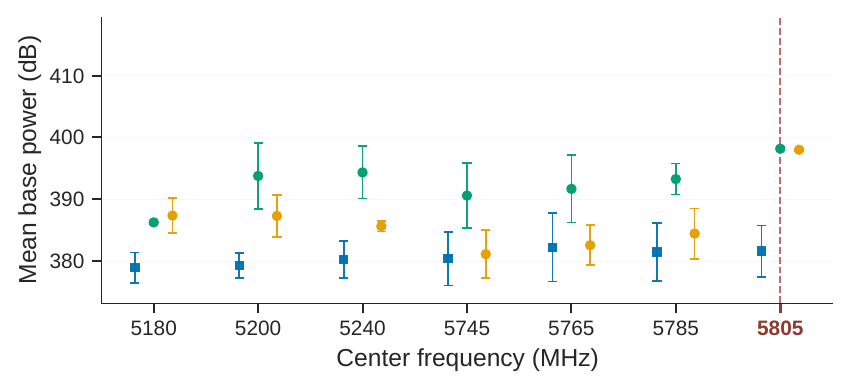}\caption{$0$\,dBm.}\end{subfigure}\hfill
    \begin{subfigure}[t]{0.49\linewidth}\centering\includegraphics[width=\linewidth]{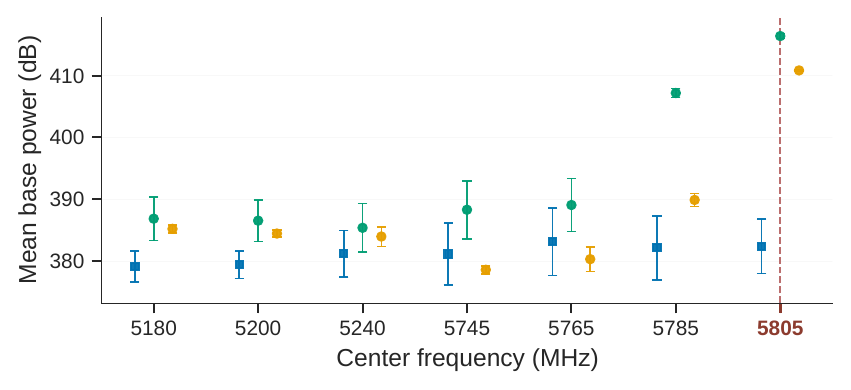}\caption{$+10$\,dBm.}\end{subfigure}
    \caption{Active-scan waveform comparison using base power, stratified by transmit power.}
    \label{fig:supp_active_waveform_stratified_base_power}
\end{figure}

\begin{figure}[!htbp]
    \centering
    \begin{subfigure}[t]{0.49\linewidth}\centering\includegraphics[width=\linewidth]{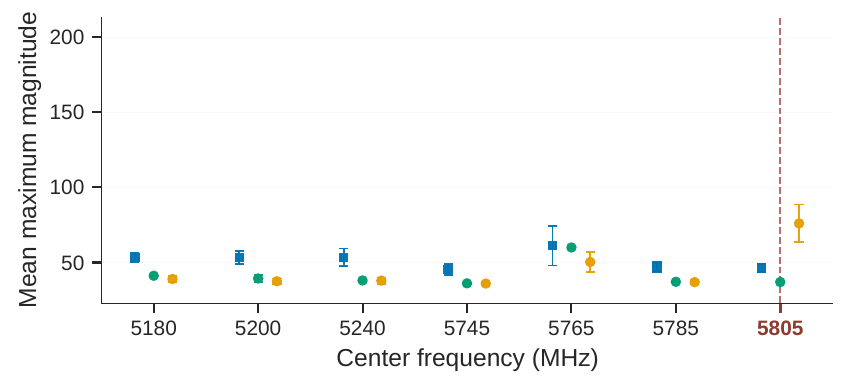}\caption{$-40$\,dBm.}\end{subfigure}\hfill
    \begin{subfigure}[t]{0.49\linewidth}\centering\includegraphics[width=\linewidth]{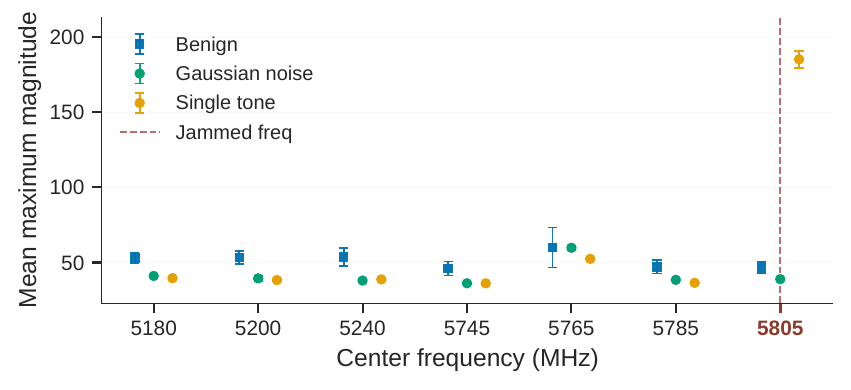}\caption{$-10$\,dBm.}\end{subfigure}

    \vspace{0.45em}
    \begin{subfigure}[t]{0.49\linewidth}\centering\includegraphics[width=\linewidth]{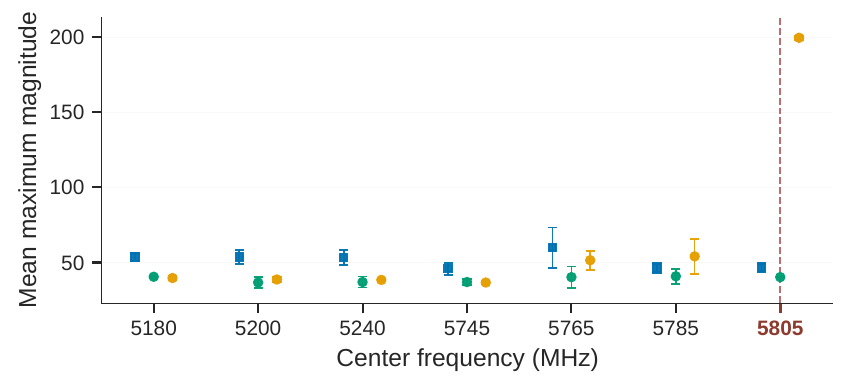}\caption{$0$\,dBm.}\end{subfigure}\hfill
    \begin{subfigure}[t]{0.49\linewidth}\centering\includegraphics[width=\linewidth]{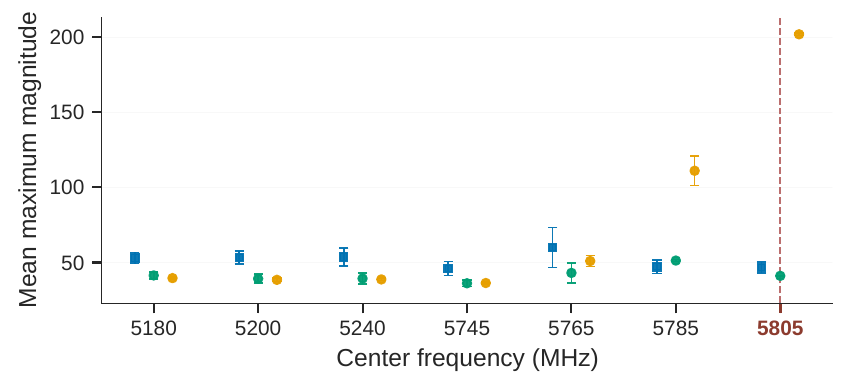}\caption{$+10$\,dBm.}\end{subfigure}
    \caption{Active-scan waveform comparison using maximum magnitude, stratified by transmit power.}
    \label{fig:supp_active_waveform_stratified_max_magnitude}
\end{figure}

\begin{figure}[!htbp]
    \centering
    \begin{subfigure}[t]{0.49\linewidth}\centering\includegraphics[width=\linewidth]{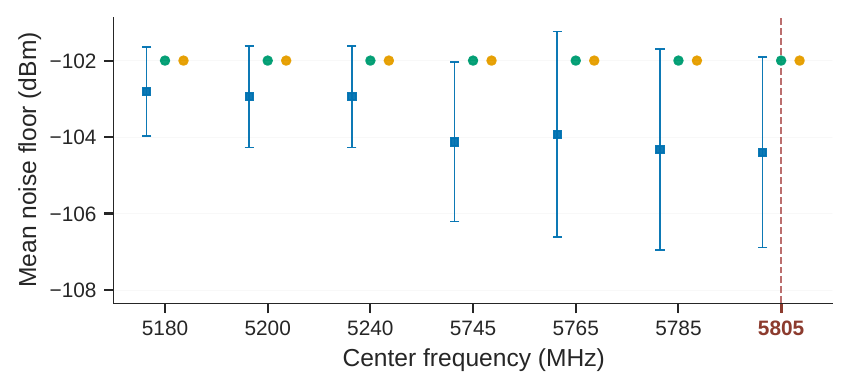}\caption{$-40$\,dBm.}\end{subfigure}\hfill
    \begin{subfigure}[t]{0.49\linewidth}\centering\includegraphics[width=\linewidth]{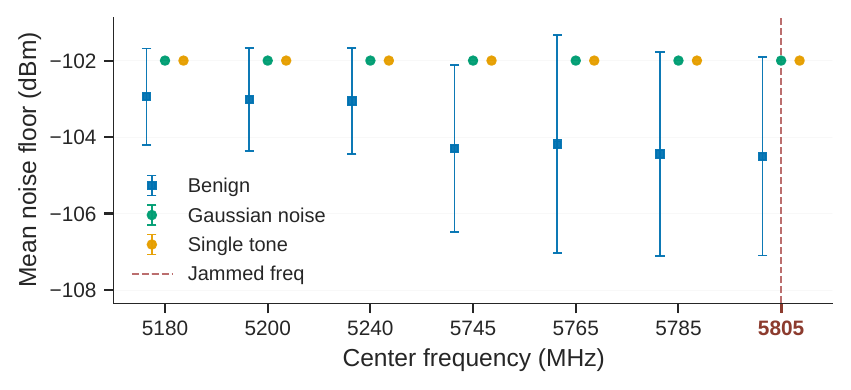}\caption{$-10$\,dBm.}\end{subfigure}

    \vspace{0.45em}
    \begin{subfigure}[t]{0.49\linewidth}\centering\includegraphics[width=\linewidth]{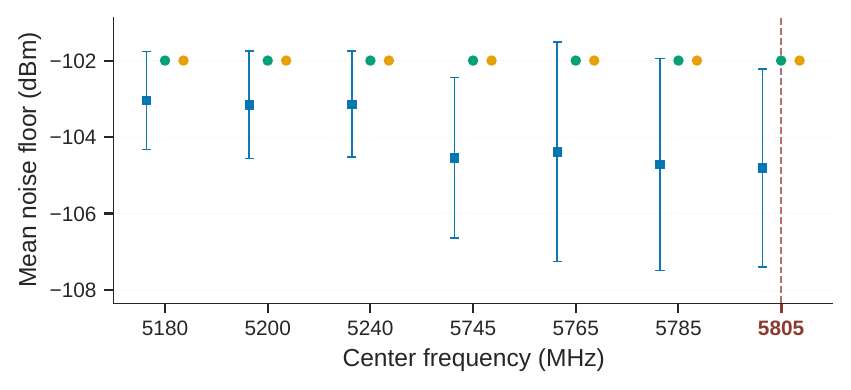}\caption{$0$\,dBm.}\end{subfigure}\hfill
    \begin{subfigure}[t]{0.49\linewidth}\centering\includegraphics[width=\linewidth]{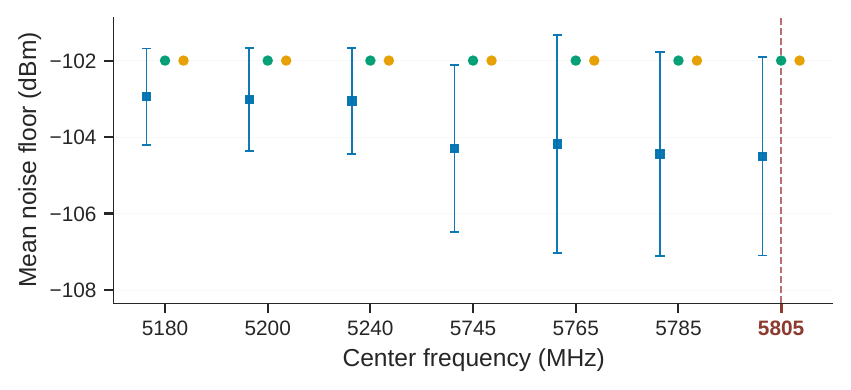}\caption{$+10$\,dBm.}\end{subfigure}
    \caption{Active-scan waveform comparison using noise floor, stratified by transmit power.}
    \label{fig:supp_active_waveform_stratified_noise}
\end{figure}

\paragraph{Active-scan waveform comparison stratified by transmit power.}
Figures~\ref{fig:supp_active_waveform_stratified_rssi}--\ref{fig:supp_active_waveform_stratified_noise} compare Gaussian-noise and single-tone active jamming at matched transmit powers. This family is the most direct way to inspect waveform effects without conflating them with jammer transmit power.

\begin{figure}[!htbp]
    \centering
    \begin{subfigure}[t]{0.49\linewidth}\centering\includegraphics[width=\linewidth]{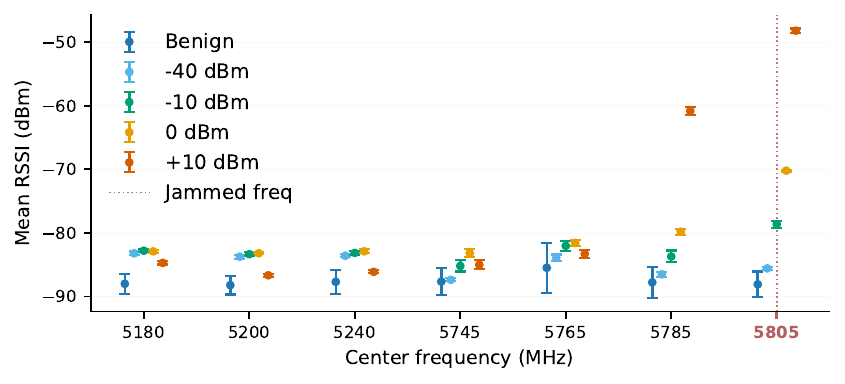}\caption{Gaussian noise.}\end{subfigure}\hfill
    \begin{subfigure}[t]{0.49\linewidth}\centering\includegraphics[width=\linewidth]{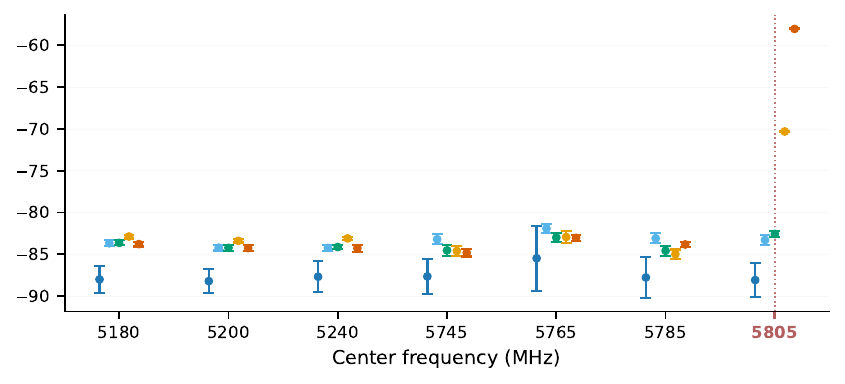}\caption{Single tone.}\end{subfigure}
    \caption{Active-scan waveform-stratified transmit-power sweeps using RSSI.}
    \label{fig:supp_active_power_sweep_rssi}
\end{figure}

\begin{figure}[!htbp]
    \centering
    \begin{subfigure}[t]{0.49\linewidth}\centering\includegraphics[width=\linewidth]{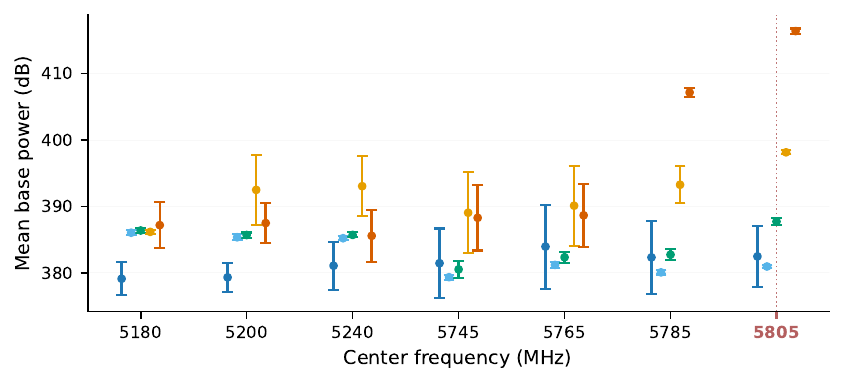}\caption{Gaussian noise.}\end{subfigure}\hfill
    \begin{subfigure}[t]{0.49\linewidth}\centering\includegraphics[width=\linewidth]{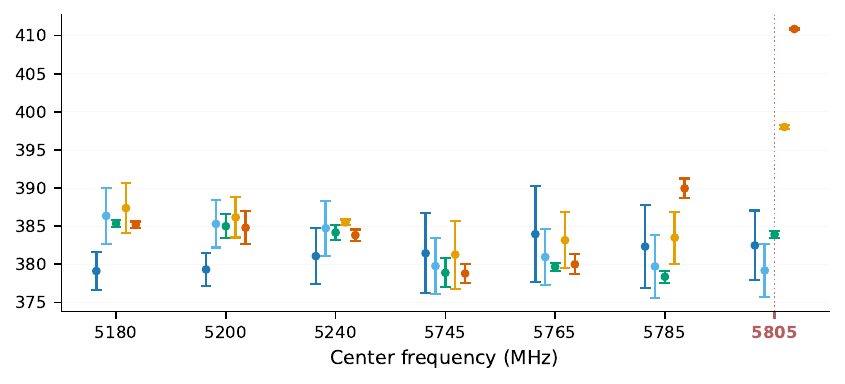}\caption{Single tone.}\end{subfigure}
    \caption{Active-scan waveform-stratified transmit-power sweeps using base power.}
    \label{fig:supp_active_power_sweep_base_power}
\end{figure}

\begin{figure}[!htbp]
    \centering
    \begin{subfigure}[t]{0.49\linewidth}\centering\includegraphics[width=\linewidth]{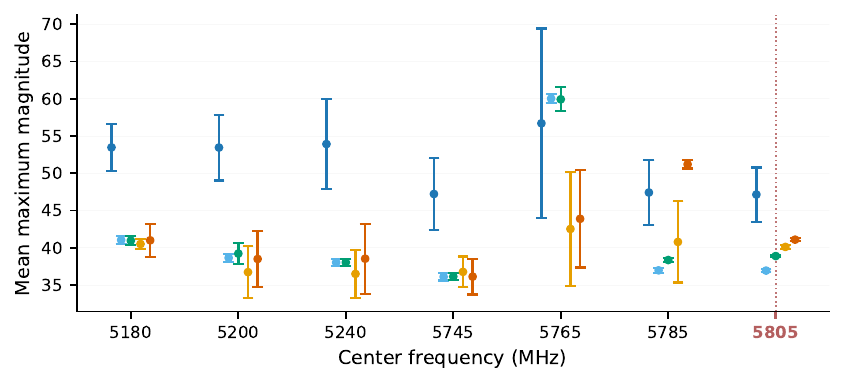}\caption{Gaussian noise.}\end{subfigure}\hfill
    \begin{subfigure}[t]{0.49\linewidth}\centering\includegraphics[width=\linewidth]{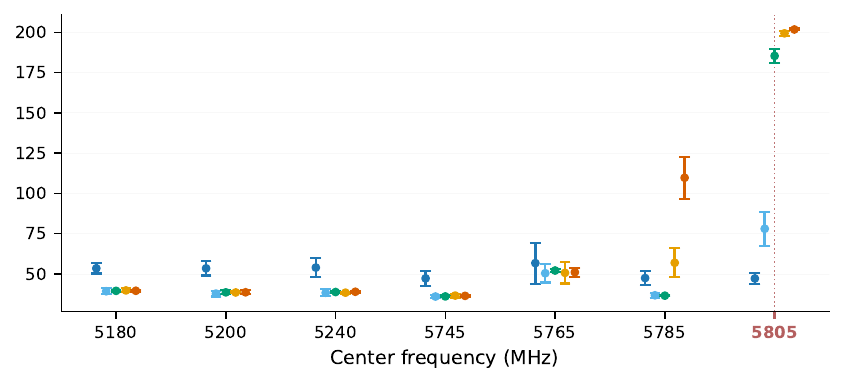}\caption{Single tone.}\end{subfigure}
    \caption{Active-scan waveform-stratified transmit-power sweeps using maximum magnitude.}
    \label{fig:supp_active_power_sweep_max_magnitude}
\end{figure}

\begin{figure}[!htbp]
    \centering
    \begin{subfigure}[t]{0.49\linewidth}\centering\includegraphics[width=\linewidth]{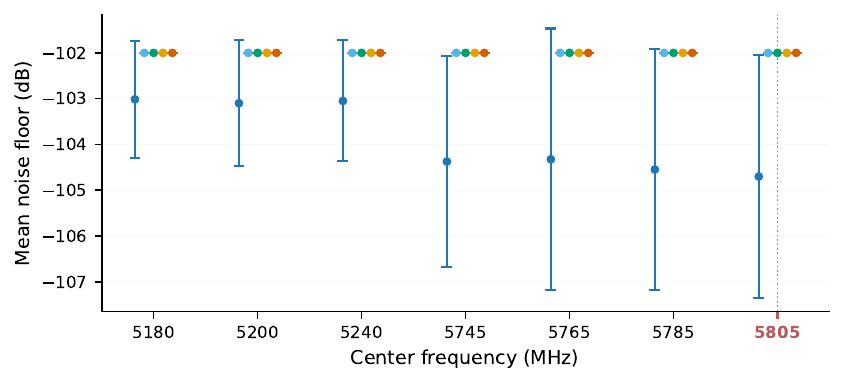}\caption{Gaussian noise.}\end{subfigure}\hfill
    \begin{subfigure}[t]{0.49\linewidth}\centering\includegraphics[width=\linewidth]{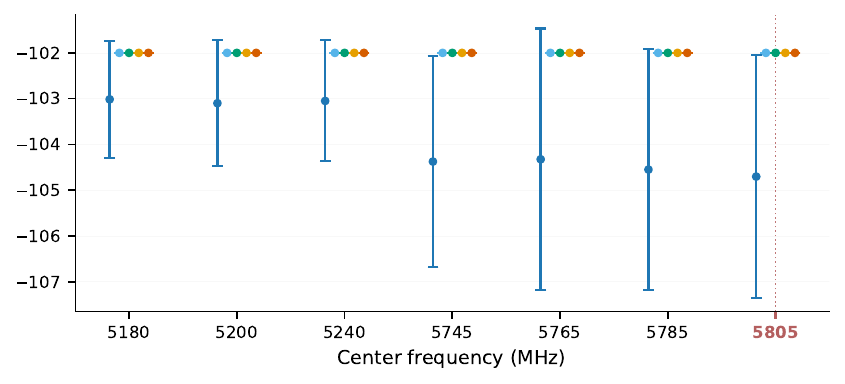}\caption{Single tone.}\end{subfigure}
    \caption{Active-scan waveform-stratified transmit-power sweeps using noise floor.}
    \label{fig:supp_active_power_sweep_noise}
\end{figure}

\paragraph{Active-scan waveform-stratified transmit-power sweeps.}
Figures~\ref{fig:supp_active_power_sweep_rssi}--\ref{fig:supp_active_power_sweep_noise} show the complementary active view: each panel holds waveform fixed and varies transmit power. These should be interpreted as waveform-conditioned power responses rather than pooled active malicious trends.

\begin{figure}[!htbp]
    \centering
    \begin{subfigure}[t]{0.49\linewidth}\centering\includegraphics[width=\linewidth]{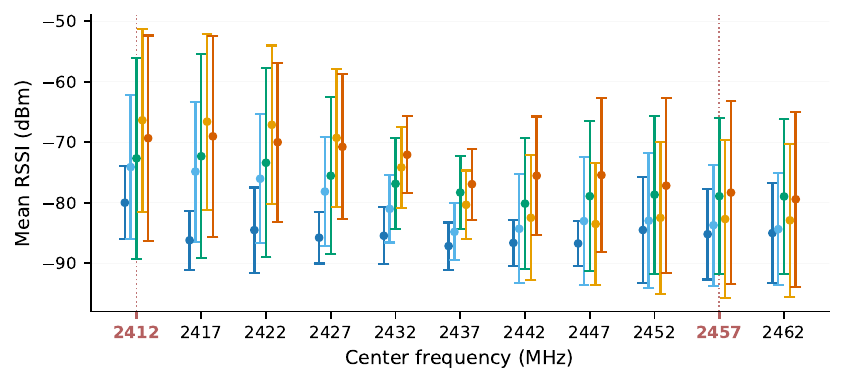}\caption{2.4\,GHz.}\end{subfigure}\hfill
    \begin{subfigure}[t]{0.49\linewidth}\centering\includegraphics[width=\linewidth]{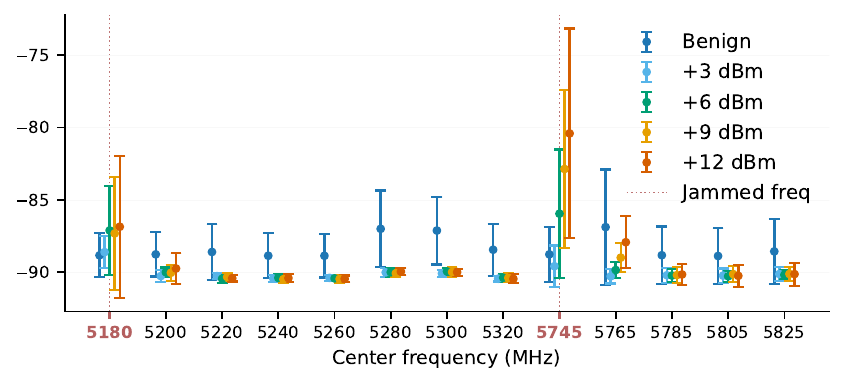}\caption{5\,GHz.}\end{subfigure}
    \caption{Passive-scan Gaussian-noise transmit-power sweeps using RSSI.}
    \label{fig:supp_passive_power_sweep_rssi}
\end{figure}

\begin{figure}[!htbp]
    \centering
    \begin{subfigure}[t]{0.49\linewidth}\centering\includegraphics[width=\linewidth]{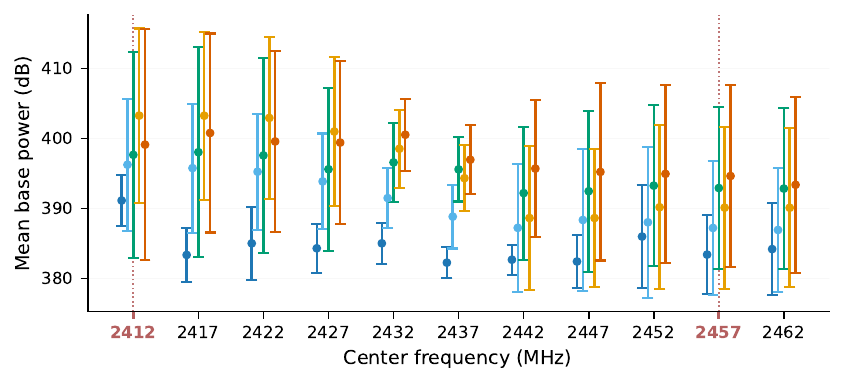}\caption{2.4\,GHz.}\end{subfigure}\hfill
    \begin{subfigure}[t]{0.49\linewidth}\centering\includegraphics[width=\linewidth]{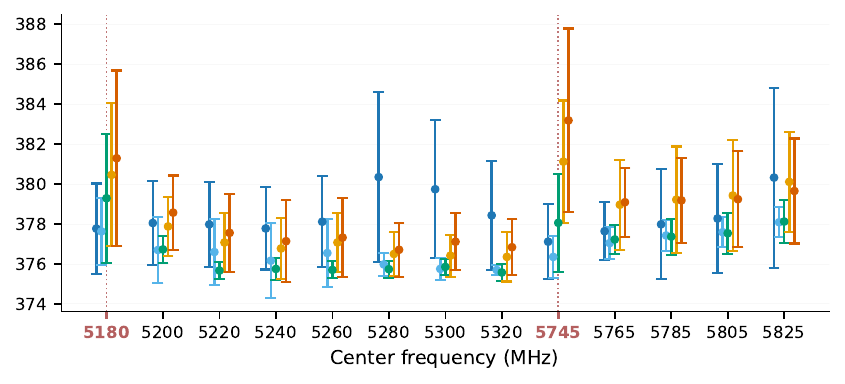}\caption{5\,GHz.}\end{subfigure}
    \caption{Passive-scan Gaussian-noise transmit-power sweeps using base power.}
    \label{fig:supp_passive_power_sweep_base_power}
\end{figure}

\begin{figure}[!htbp]
    \centering
    \begin{subfigure}[t]{0.49\linewidth}\centering\includegraphics[width=\linewidth]{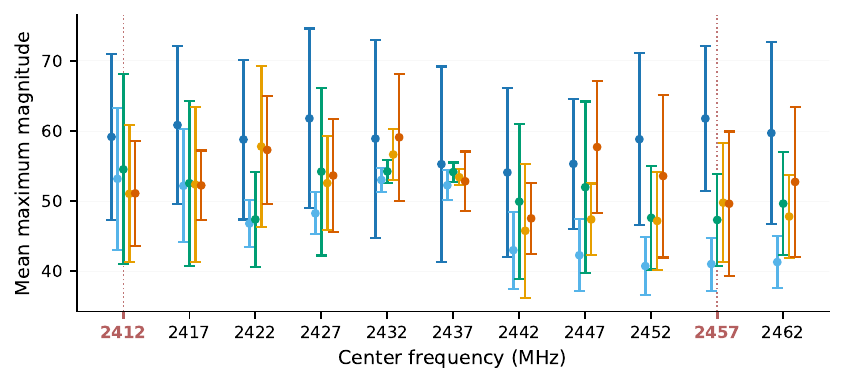}\caption{2.4\,GHz.}\end{subfigure}\hfill
    \begin{subfigure}[t]{0.49\linewidth}\centering\includegraphics[width=\linewidth]{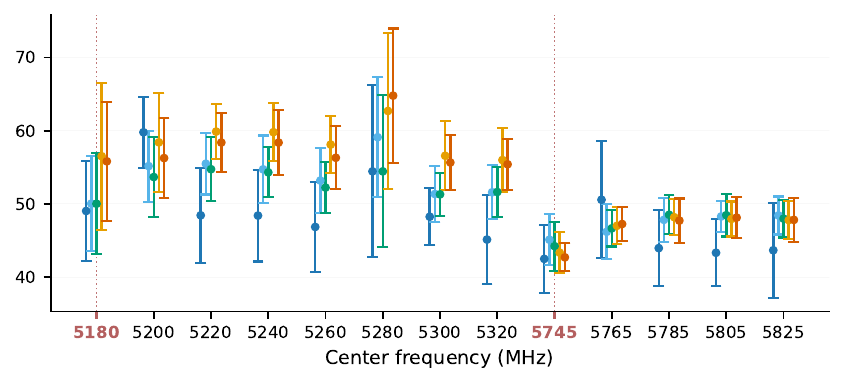}\caption{5\,GHz.}\end{subfigure}
    \caption{Passive-scan Gaussian-noise transmit-power sweeps using maximum magnitude.}
    \label{fig:supp_passive_power_sweep_max_magnitude}
\end{figure}

\begin{figure}[!htbp]
    \centering
    \begin{subfigure}[t]{0.49\linewidth}\centering\includegraphics[width=\linewidth]{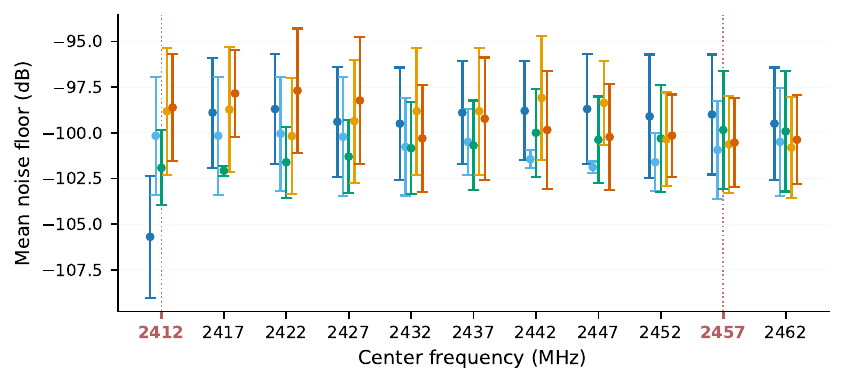}\caption{2.4\,GHz.}\end{subfigure}\hfill
    \begin{subfigure}[t]{0.49\linewidth}\centering\includegraphics[width=\linewidth]{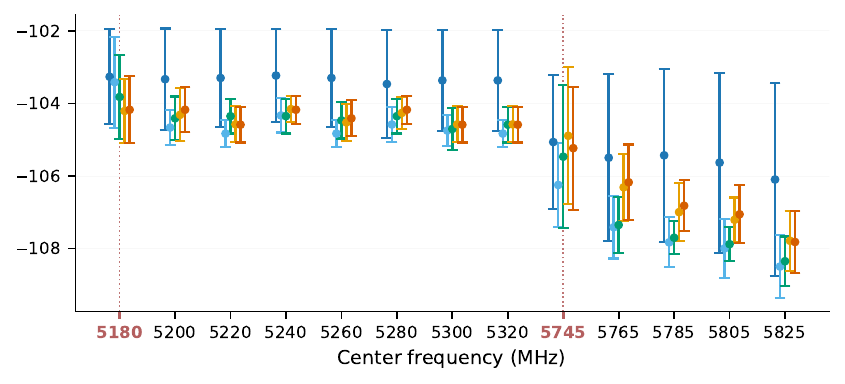}\caption{5\,GHz.}\end{subfigure}
    \caption{Passive-scan Gaussian-noise transmit-power sweeps using noise floor.}
    \label{fig:supp_passive_power_sweep_noise}
\end{figure}

\paragraph{Passive-scan Gaussian-noise transmit-power sweeps.}
Figures~\ref{fig:supp_passive_power_sweep_rssi}--\ref{fig:supp_passive_power_sweep_noise} show the passive power response for both 2.4\,GHz and 5\,GHz. Unlike the active subset, passive malicious captures contain Gaussian-noise jamming only, so these panels provide the cleanest power-conditioned view of the passive collection.

\FloatBarrier

\begin{figure}[!p]
    \centering
    \begin{subfigure}[t]{0.49\linewidth}\centering\includegraphics[width=\linewidth]{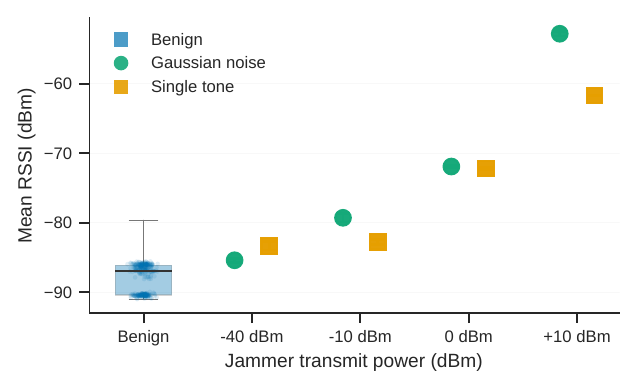}\caption{RSSI.}\end{subfigure}\hfill
    \begin{subfigure}[t]{0.49\linewidth}\centering\includegraphics[width=\linewidth]{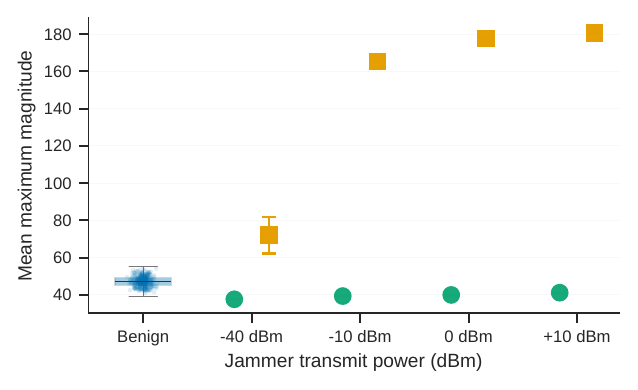}\caption{Maximum magnitude.}\end{subfigure}

    \vspace{0.5em}
    \begin{subfigure}[t]{0.49\linewidth}\centering\includegraphics[width=\linewidth]{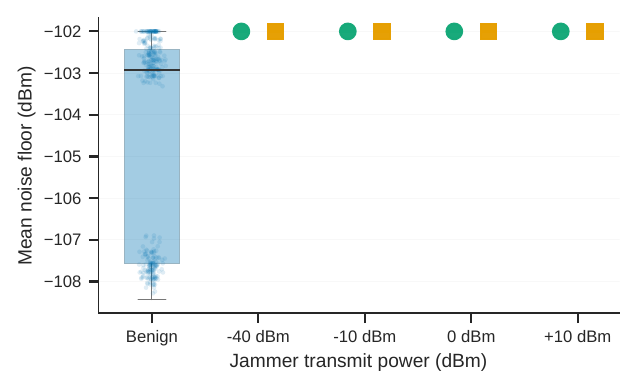}\caption{Noise floor.}\end{subfigure}\hfill
    \begin{subfigure}[t]{0.49\linewidth}\centering\includegraphics[width=\linewidth]{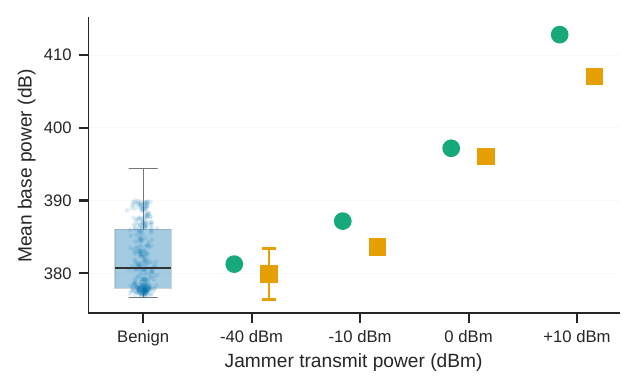}\caption{Base power.}\end{subfigure}

    \vspace{0.5em}
    \begin{subfigure}[t]{0.49\linewidth}\centering\includegraphics[width=\linewidth]{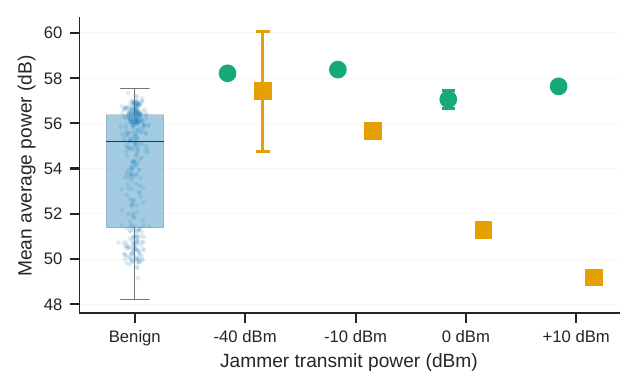}\caption{Average power.}\end{subfigure}
    \caption{Active-scan file-level feature-distribution diagnostics by waveform and transmit power. The benign active baseline is shown as a file-level distribution box plot; Gaussian-noise and single-tone malicious conditions are shown as waveform-specific mean $\pm$ standard-deviation marker/whisker summaries at each transmit power.}
    \label{fig:supp_active_weak_power_combined_additional}
\end{figure}

\paragraph{Active-scan file-level feature-distribution summaries.}
Figure~\ref{fig:supp_active_weak_power_combined_additional} collects the active-scan file-level feature-distribution diagnostics for the same file-level mean features used in the spectral validation: RSSI, maximum magnitude, noise floor, base power, and average power. The visual encoding uses a single active benign reference distribution as a box plot, while Gaussian-noise and single-tone malicious conditions are shown at each nominal transmit power as waveform-specific mean $\pm$ standard-deviation marker/whisker summaries. This avoids implying two different benign waveform conditions and makes the active-scan waveform--power comparison more compact. All values are computed from file-level aggregate features across all observations in each file, not from a single centre frequency.

\begin{figure}[!p]
    \centering
    \begin{subfigure}[t]{0.49\linewidth}\centering\includegraphics[width=\linewidth]{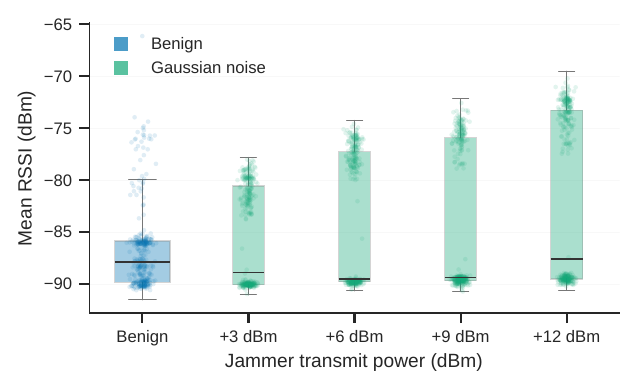}\caption{RSSI.}\end{subfigure}\hfill
    \begin{subfigure}[t]{0.49\linewidth}\centering\includegraphics[width=\linewidth]{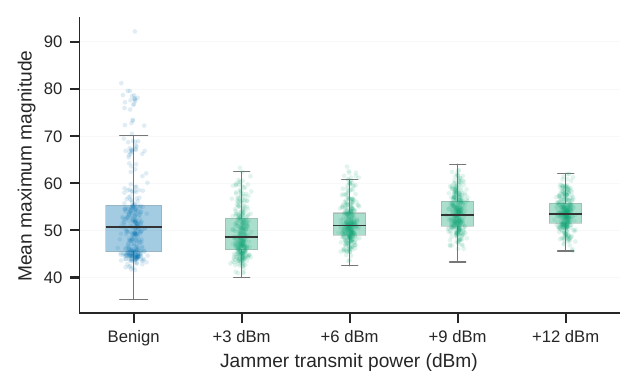}\caption{Maximum magnitude.}\end{subfigure}

    \vspace{0.5em}
    \begin{subfigure}[t]{0.49\linewidth}\centering\includegraphics[width=\linewidth]{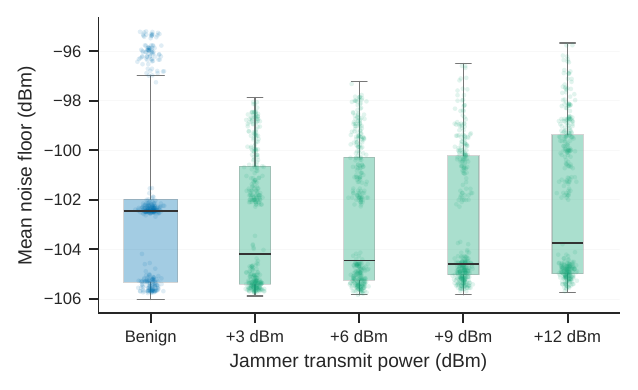}\caption{Noise floor.}\end{subfigure}\hfill
    \begin{subfigure}[t]{0.49\linewidth}\centering\includegraphics[width=\linewidth]{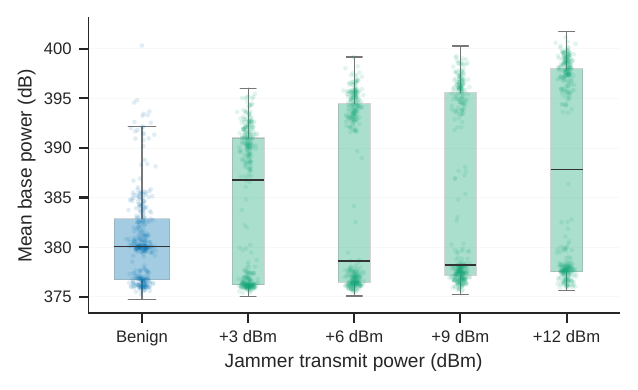}\caption{Base power.}\end{subfigure}

    \vspace{0.5em}
    \begin{subfigure}[t]{0.49\linewidth}\centering\includegraphics[width=\linewidth]{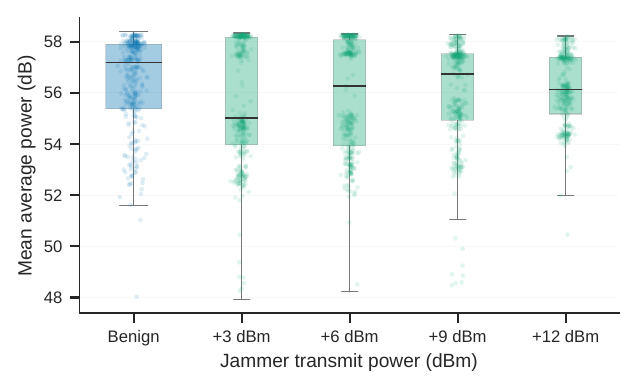}\caption{Average power.}\end{subfigure}
    \caption{Passive-scan file-level feature-distribution diagnostics by transmit power. Each box summarises a file-level mean feature, computed across all observations in the corresponding passive file.}
    \label{fig:supp_passive_weak_power_feature_distributions}
\end{figure}

\paragraph{Passive-scan file-level feature-distribution summaries.}
Figure~\ref{fig:supp_passive_weak_power_feature_distributions} collects the passive scan box-plot diagnostics for the matched file-level mean features. Passive-scan malicious captures contain Gaussian-noise jamming only, so the passive panels show benign reference files and passive malicious files stratified by transmit power. As in the active case, each box summarises a file-level aggregate feature computed across all observations in the corresponding file rather than a single centre frequency.

\subsection{Weak-condition and waveform-transfer diagnostics}
\label{sec:diagnostic_validation}

The following diagnostics expand on the held-out results in the main text. They are reported here, rather than in the main Technical Validation, because they characterise how baseline detectors behave under distribution shift rather than establishing dataset integrity; they illustrate the kind of analysis the released features support. The feature-space separability table, the matched-feature weak-condition distributions, the unresampled background-location sensitivity check, and the held-out partition sizes are collected below.

To interpret the weakest held-out settings reported in Fig.~\ref{fig:heldout_all_model_heatmap}, distinguishing conditions that are intrinsically close to benign scans from those where only the decision threshold fails to transfer, we computed feature-space separability diagnostics comparing the weakest active and passive jamming conditions against their scan-mode-matched benign baselines. Table~\ref{tab:weak_condition_diagnostics} reports standardised mean differences (SMDs), distribution-overlap coefficients, univariate AUC summaries, and a balanced cross-validated logistic diagnostic. SMD and overlap are computed feature by feature over the 63 benchmark features; the diagnostic classifier is used only as a separability probe and is not part of the baseline benchmark model selection.

\begin{table}[t]
    \centering
    \compacttable
    \caption{Feature-space separability diagnostics for weak jamming conditions relative to scan-mode-matched benign baselines. Lower overlap and higher SMD/AUC indicate stronger feature-level separation. The cross-validated diagnostic classifier is balanced and is used only to quantify separability, not as an optimized detector.}\label{tab:weak_condition_diagnostics}
    \begin{tabularx}{\linewidth}{@{}
        >{\hsize=2.5\hsize\raggedright\arraybackslash}X
        >{\hsize=0.75\hsize\centering\arraybackslash}X
        >{\hsize=0.75\hsize\centering\arraybackslash}X
        >{\hsize=0.75\hsize\centering\arraybackslash}X
        >{\hsize=0.75\hsize\centering\arraybackslash}X
        >{\hsize=0.75\hsize\centering\arraybackslash}X
        >{\hsize=0.75\hsize\centering\arraybackslash}X
    @{}}
        \toprule
        \textbf{Comparison} & \textbf{Mean abs. SMD} & \textbf{Median abs. SMD} & \textbf{Median overlap} & \textbf{Median univ. AUC} & \textbf{CV AUC} & \textbf{CV F1} \\
        \midrule
        Active $-40$\,dBm vs active benign & 1.10 & 0.97 & 0.22 & 0.76 & 1.00 & 1.00 \\
        Passive $+3$\,dBm vs passive benign & 0.50 & 0.46 & 0.54 & 0.59 & 0.997 & 0.974 \\
        \bottomrule
    \end{tabularx}
\end{table}

These diagnostics show that active $-40$\,dBm jamming is weak in nominal transmit power but not weak in feature separability: it has larger standardised feature differences, lower distribution overlap, and stronger univariate AUCs relative to active benign scans. The additional feature-level summary reinforces this distinction. For active $-40$\,dBm, 37 of the 63 benchmark features have univariate AUC at least 0.7, 31 features have absolute SMD at least 1.0, and the strongest individual feature is \texttt{rssi\_dbm\_mean} with univariate AUC 0.994. For passive $+3$\,dBm, only 13 of 63 features have univariate AUC at least 0.7 and only three have absolute SMD at least 1.0, with a substantially higher median overlap coefficient. Thus, passive $+3$\,dBm jamming is much closer to passive benign variability at the individual-feature level, even though multivariate combinations remain informative when the condition is represented during training. The passive $+3$\,dBm held-out difficulty should therefore be interpreted as a generalisation challenge to a low-power passive condition, not as evidence that the condition is undetectable in principle.

Figures~\ref{fig:supp_active_weak_power_combined_additional} and~\ref{fig:supp_passive_weak_power_feature_distributions} visualise these weak-condition diagnostics while also resolving a potential confound in the active subset. To keep the main diagnostic figure comparable across scan modes, the same two interpretable features are shown for active and passive scans: RSSI and maximum magnitude. Because active malicious files include both Gaussian-noise and single-tone captures at the same nominal transmit powers, active weak-power feature distributions are plotted with the two waveforms side by side at each power level rather than pooled into a single malicious distribution. This combined representation shows that waveform and transmit power jointly shape the active feature distributions. In RSSI, Gaussian-noise captures are generally stronger than single-tone captures at several powers, while in maximum magnitude the single-tone waveform produces the clearest peak-like separation. This behaviour is physically consistent with the measurement setup: Gaussian-noise jamming perturbs broader aggregate measures, whereas a single-tone jammer concentrates energy into a narrow spectral component and therefore appears strongly in peak-based features. The passive panels are simpler to interpret because passive malicious captures contain Gaussian-noise jamming only; showing the same RSSI and maximum-magnitude summaries prevents active--passive comparisons from being driven by different feature choices.

\subsection{Unresampled-training sensitivity analysis}

\begin{figure}[!htbp]
    \centering
    \includegraphics[width=\linewidth]{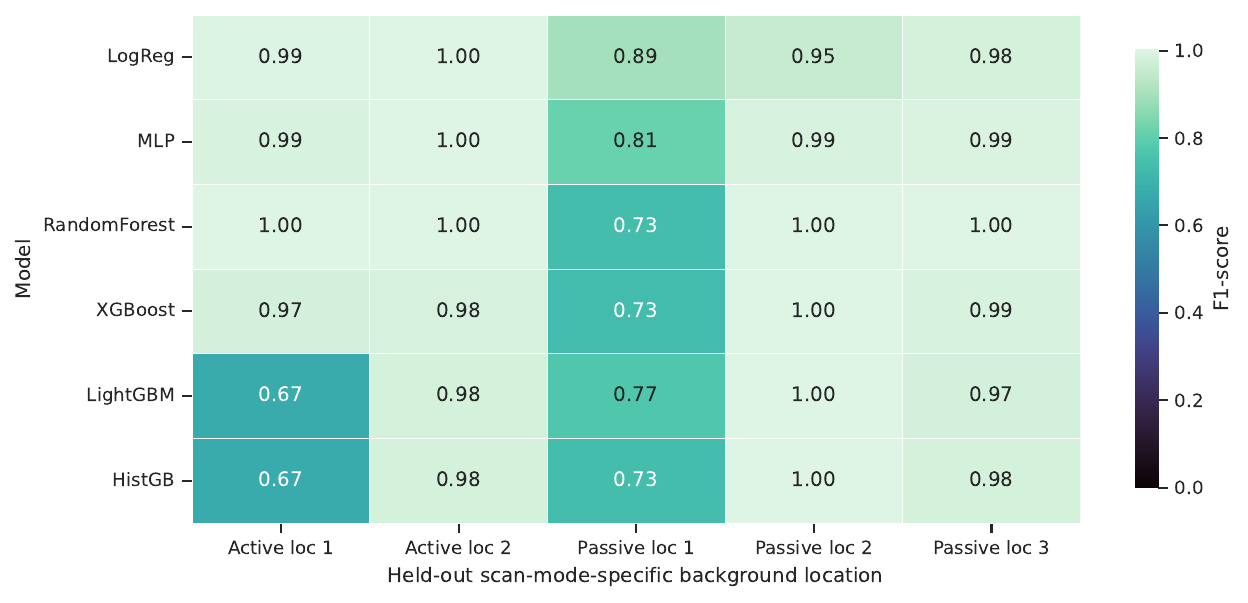}
    \caption{Supplementary per-model F1-score heatmap for the unresampled-training sensitivity analysis of scan-mode-specific held-out background-location evaluation. The held-out test sets are identical to the main balanced-training protocol, but the training set is not class-balanced.}
    \label{fig:supp_background_location_unresampled_heatmap}
\end{figure}

For the scan-mode-specific held-out background-location experiment, the main text reports the class-balanced training protocol. As a sensitivity analysis, the same held-out evaluation is repeated without training-set rebalancing and reported for all six classifiers in Fig.~\ref{fig:supp_background_location_unresampled_heatmap}. The test sets are unchanged and remain balanced between held-out benign background files and same-mode malicious files; only the training-set sampling differs. In the unresampled protocol, all available non-held-out benign training files are retained, so the active training splits contain approximately 22--24\% malicious files and the passive training splits contain approximately 13.5--13.6\% malicious files.

This sensitivity analysis is included to show whether the held-out background-location conclusions depend on class-balanced training. The qualitative pattern is stable: active held-out locations remain near saturated, passive \texttt{location2} and passive \texttt{location3} remain strong, and \texttt{passive\_location1} remains the hardest held-out background environment. The unresampled results are therefore treated as a robustness check rather than as the primary benchmark protocol.

\subsection{MLP transfer-failure diagnostic}

The all-model outputs include explicit diagnostics for degenerate MLP predictions. The clearest case occurs for Passive$\rightarrow$Active cross-scan transfer: across seeds, the MLP predicts fewer than one percent of active test files as malicious, misses all 1{,}097 malicious active files, and consequently produces F1~=~0 with ROC-AUC close to chance or worse. This differs from the held-out single-tone behaviour of several tree/boosting models, where ranking metrics remain high despite a failed fixed threshold. The MLP result is therefore treated as a genuine cross-domain optimisation/generalisation failure rather than as evidence that the underlying active malicious class is inseparable.

\subsection*{Held-out partition construction}

The passive $+3$\,dBm band-specific held-out tests contain comparable numbers of files: approximately 4{,}020 balanced test files for 2.4\,GHz and 4{,}156 for 5\,GHz. The markedly poorer 5\,GHz F1 therefore cannot be attributed to a smaller held-out sample.

Table~\ref{tab:heldout_partitions} reports the train/test sizes and class priors for the held-out malicious-condition protocols. Test partitions are balanced by construction to make threshold-dependent metrics interpretable across held-out power and waveform conditions. The scan-mode-specific background-location protocol is reported separately in Fig.~\ref{fig:background_location_withinmode_heatmap}.

\begin{table}[!htbp]
    \centering
    \widetable
    \caption{Train and test partition sizes for the held-out malicious-condition experiments reported in Fig.~\ref{fig:heldout_all_model_heatmap}. Positive rate denotes the fraction of malicious files in each partition; test partitions are balanced by design (positive rate $\approx 0.50$). Power values are in dBm.}\label{tab:heldout_partitions}
    \begin{tabularx}{\linewidth}{@{}>{\hsize=0.50\hsize}L>{\hsize=0.55\hsize}L>{\hsize=0.92\hsize}L>{\hsize=0.55\hsize}C>{\hsize=0.55\hsize}C>{\hsize=0.55\hsize}C>{\hsize=0.55\hsize}C@{}}
        \toprule
        \textbf{Condition} & \textbf{Scope} & \textbf{Held-out value} & \textbf{Train files} & \textbf{Test files} & \textbf{Train pos.} & \textbf{Test pos.} \\
        \midrule
        \multirow{16}{*}{\makecell{Power\\(dBm)}}
         & \multirow{4}{*}{Active} & $-40$ & 3{,}954 & 404 & 0.226 & 0.500 \\
         &                         & $-10$ & 3{,}744 & 614 & 0.211 & 0.500 \\
         &                         & $0$ & 3{,}844 & 514 & 0.219 & 0.500 \\
         &                         & $+10$ & 3{,}696 & 662 & 0.207 & 0.500 \\
        \cmidrule(lr){2-7}
         & \multirow{4}{*}{Passive} & $3$ & 83{,}556 & 8{,}176 & 0.162 & 0.500 \\
         &                          & $6$ & 83{,}310 & 8{,}422 & 0.161 & 0.500 \\
         &                          & $9$ & 81{,}276 & 10{,}456 & 0.152 & 0.500 \\
         &                          & $12$ & 83{,}616 & 8{,}116 & 0.162 & 0.500 \\
        \cmidrule(lr){2-7}
         & \multirow{8}{*}{Pooled} & $-40$ & 95{,}686 & 404 & 0.193 & 0.500 \\
         &                           & $-10$ & 95{,}476 & 614 & 0.192 & 0.500 \\
         &                           & $0$ & 95{,}576 & 514 & 0.193 & 0.500 \\
         &                           & $+10$ & 95{,}428 & 662 & 0.192 & 0.500 \\
         &                           & $3$ & 87{,}914 & 8{,}176 & 0.166 & 0.500 \\
         &                           & $6$ & 87{,}668 & 8{,}422 & 0.165 & 0.500 \\
         &                           & $9$ & 85{,}634 & 10{,}456 & 0.157 & 0.500 \\
         &                           & $12$ & 87{,}974 & 8{,}116 & 0.166 & 0.500 \\
        \midrule
        \multirow{2}{*}{Waveform}
         & \multirow{2}{*}{Active} & Gaussian noise & 3{,}444 & 914 & 0.186 & 0.500 \\
         &                         & Single tone & 3{,}078 & 1{,}280 & 0.148 & 0.500 \\
        \bottomrule
    \end{tabularx}
\end{table}
\FloatBarrier

\subsection{Active-scan waveform metric contrasts}

Table~\ref{tab:supp_active_waveform_contrasts} quantifies the active waveform comparison at matched transmit powers. For each metric and power level, it reports Gaussian-noise and single-tone means, their raw difference, a pooled-standard-deviation-normalised difference, and a direction-adjusted univariate AUC. This table supports the main active waveform figures by showing which metrics separate the two waveform families beyond visual inspection.

\begin{landscape}
\begin{table}[!p]
    \centering
    \scriptsize
    \setlength{\tabcolsep}{2.2pt}
    \renewcommand{\arraystretch}{0.86}
    \caption{Power-matched waveform contrasts within the active-scan malicious subset for file-level aggregate metrics. Difference is single-tone mean minus Gaussian-noise mean. The standardised difference divides this contrast by the pooled standard deviation. Direction-adjusted univariate AUC reports single-feature separability regardless of which waveform has the larger feature value. The final column, $n_G/n_S$, reports the number of Gaussian-noise files ($n_G$) and single-tone files ($n_S$) included at each transmit-power setting.}
    \label{tab:supp_active_waveform_contrasts}

    \begin{tabularx}{\linewidth}{@{}>{\hsize=1.05\hsize}L>{\hsize=0.62\hsize}C>{\hsize=0.78\hsize}R>{\hsize=0.78\hsize}R>{\hsize=0.75\hsize}R>{\hsize=0.70\hsize}R>{\hsize=0.62\hsize}R>{\hsize=0.70\hsize}C@{}}
        \toprule
        \textbf{Metric} & \textbf{Power (dBm)} & \textbf{Gaussian mean} &
        \textbf{Single-tone mean} & \textbf{Difference} &
        \textbf{Std. diff.} & \textbf{Adj. AUC} & \textbf{$n_G/n_S$} \\
        \midrule

        RSSI (dBm) & $-40$ & -85.412 & -83.334 & 2.078 & 4.489 & 1.000 & 101/101 \\
         & $-10$ & -79.305 & -82.743 & -3.439 & -7.738 & 1.000 & 102/205 \\
         & 0 & -71.924 & -72.214 & -0.290 & -2.108 & 0.927 & 153/104 \\
         & $+10$ & -52.813 & -61.700 & -8.887 & -44.919 & 1.000 & 101/230 \\

        \midrule
        \addlinespace[2pt]

        Max magnitude (linear) & $-40$ & 37.630 & 71.986 & 34.355 & 4.970 & 0.990 & 101/101 \\
         & $-10$ & 39.351 & 165.214 & 125.863 & 47.283 & 1.000 & 102/205 \\
         & 0 & 39.963 & 177.629 & 137.666 & 173.874 & 1.000 & 153/104 \\
         & $+10$ & 41.161 & 180.735 & 139.574 & 234.044 & 1.000 & 101/230 \\

        \midrule
        \addlinespace[2pt]

        Avg power (dB) & $-40$ & 58.210 & 57.415 & -0.794 & -0.424 & 0.802 & 101/101 \\
         & $-10$ & 58.374 & 55.656 & -2.719 & -12.452 & 1.000 & 102/205 \\
         & 0 & 57.060 & 51.282 & -5.778 & -17.450 & 1.000 & 153/104 \\
         & $+10$ & 57.632 & 49.182 & -8.450 & -76.642 & 1.000 & 101/230 \\

        \midrule
        \addlinespace[2pt]

        Base power (dB) & $-40$ & 381.266 & 379.911 & -1.355 & -0.543 & 0.823 & 101/101 \\
         & $-10$ & 387.171 & 383.590 & -3.581 & -8.312 & 0.995 & 102/205 \\
         & 0 & 397.176 & 396.064 & -1.112 & -2.340 & 0.933 & 153/104 \\
         & $+10$ & 412.783 & 407.036 & -5.747 & -21.639 & 1.000 & 101/230 \\

        \midrule
        \addlinespace[2pt]

        Noise floor (dBm) & $-40$ & -102.000 & -102.000 & 0.000 & -- & 0.500 & 101/101 \\
         & $-10$ & -102.000 & -102.000 & 0.000 & -- & 0.500 & 102/205 \\
         & 0 & -102.000 & -102.000 & 0.000 & -- & 0.500 & 153/104 \\
         & $+10$ & -102.000 & -102.000 & 0.000 & -- & 0.500 & 101/230 \\

        \bottomrule
    \end{tabularx}
\end{table}
\end{landscape}
\clearpage


\begin{thebibliography}{99}

\bibitem{pelechrinis2011dos}
Pelechrinis, K., Iliofotou, M. and Krishnamurthy, S.~V. Denial of Service Attacks in Wireless Networks: The Case of Jammers. \emph{IEEE Communications Surveys \& Tutorials} \textbf{13}(2), 245--257 (2011).

\bibitem{pirayesh2022survey}
Pirayesh, H. and Zeng, H. Jamming Attacks and Anti-Jamming Strategies in Wireless Networks: A Comprehensive Survey. \emph{IEEE Communications Surveys \& Tutorials} \textbf{24}(2), 767--809 (2022).

\bibitem{xu2005feasibility}
Xu, W., Trappe, W., Zhang, Y. and Wood, T. The Feasibility of Launching and Detecting Jamming Attacks in Wireless Networks. In \emph{Proceedings of the 6th ACM International Symposium on Mobile Ad Hoc Networking and Computing (MobiHoc)}, 46--57 (2005).

\bibitem{wildpackets2002rssi}
WildPackets, Inc. \emph{Converting Signal Strength Percentage to dBm Values}. Technical note (2002). \url{https://d2cpnw0u24fjm4.cloudfront.net/wp-content/uploads/Converting_Signal_Strength.pdf}.

\bibitem{ali2024rfjammingdataset}
Ali, A.~S., Lunardi, W.~T., Singh, G., Bariah, L., Baddeley, M., Lopez, M.~A., Giacalone, J.-P. and Muhaidat, S. RF Jamming Dataset: A Wireless Spectral Scan Approach for Malicious Interference Detection. \emph{IEEE Communications Magazine} \textbf{62}(11), 114--120 (2024).

\bibitem{ali2022rfjammingdataport}
Ali, A.~S. RF Jamming Dataset Using CM4 and JamRF enabled HackRF. \emph{IEEE DataPort} (2022). doi:10.21227/mekj-vw56.

\bibitem{ali2022jamrf}
Ali, A.~S. \textit{et al.} JamRF: Performance Analysis, Evaluation, and Implementation of RF Jamming Over Wi-Fi. \emph{IEEE Access} \textbf{10}, 133370--133384 (2022).

\bibitem{panitsas2025jamshield}
Panitsas, I., Yigit, Y., Tassiulas, L., Maglaras, L. and Canberk, B. JamShield: A Machine Learning Detection System for Over-the-Air Jamming Attacks. In \emph{ICC 2025, IEEE International Conference on Communications}, 1067--1072 (2025).

\bibitem{punal2014jamming}
Pu\~nal, O., Akta\c{s}, I., Schnelke, C.-J., Abidin, G., Wehrle, K. and Gross, J. Machine Learning-based Jamming Detection for IEEE 802.11: Design and Experimental Evaluation. In \emph{2014 IEEE 15th International Symposium on a World of Wireless, Mobile and Multimedia Networks (WoWMoM)}, 1--10 (2014).

\bibitem{punal2012crawdad}
Pu\~nal, O., Pereira, C., Aguiar, A. and Gross, J. CRAWDAD dataset uportorwthaachen/vanetjamming2012 (v.~2014-05-12). \emph{IEEE DataPort / CRAWDAD} (2014). doi:10.15783/C7F31Z.

\bibitem{punal2014crawdad}
Pu\~nal, O., Pereira, C., Aguiar, A. and Gross, J. CRAWDAD dataset uportorwthaachen/vanetjamming2014 (v.~2014-05-12). \emph{IEEE DataPort / CRAWDAD} (2014). doi:10.15783/C7Q306.

\bibitem{alhazbi2023indoorjammingdataset}
Alhazbi, S., Sciancalepore, S. and Oligeri, G. A Dataset of Physical-Layer Measurements in Indoor Wireless Jamming Scenarios. \emph{Data in Brief} \textbf{46}, 108773 (2023).

\bibitem{hussain2022edgeaijamming}
Hussain, A., Abughanam, N., Qadir, J. and Mohamed, A. Jamming Detection in IoT Wireless Networks: An Edge-AI Based Approach. In \emph{Proceedings of the 12th International Conference on the Internet of Things (IoT~'22)}, 57--64, ACM (2022).


\bibitem{li2023uavjammingdetection}
Li, Y., Pawlak, J., Price, J., Al Shamaileh, K., Niyaz, Q., Paheding, S. and Devabhaktuni, V. Jamming Detection and Classification in OFDM-Based UAVs via Feature- and Spectrogram-Tailored Machine Learning. \emph{IEEE Access} \textbf{10}, 16859--16870 (2022). doi:10.1109/ACCESS.2022.3150020.

\bibitem{davaslioglu2019deepwifi}
Davaslioglu, K., Soltani, S., Erpek, T. and Sagduyu, Y.~E. DeepWiFi: Cognitive WiFi With Deep Learning. \emph{IEEE Transactions on Mobile Computing} \textbf{20}(2), 429--444 (2021).

\bibitem{linux_ath11k}
Linux Wireless. About ath11k. \url{https://wireless.docs.kernel.org/en/latest/en/users/drivers/ath11k.html}. Accessed 2026-06-13.

\bibitem{linux_ath11k_spectral}
Linux Kernel Configuration. \texttt{CONFIG\_ATH11K\_SPECTRAL}: QCA ath11k spectral scan support. \url{https://www.kernelconfig.io/CONFIG_ATH11K_SPECTRAL}. Accessed 2026-06-13.

\bibitem{linux_ath12k}
Linux Wireless. About ath12k. \url{https://wireless.docs.kernel.org/en/latest/en/users/drivers/ath12k.html}. Accessed 2026-06-13.

\bibitem{WiFiSpectralJam_kaggle}
Herzalla, D. \emph{WiFiSpectralJam: A Large-Scale Open Wi-Fi Spectral-Scan Dataset with Controlled RF Jamming}. Kaggle, \url{https://www.kaggle.com/datasets/daniaherzalla/radio-frequency-jamming} (2026).

\end{thebibliography}
\end{document}